\def\DIRvalue{Pasquetti}
\def\IDvalue{PA}
\def\titlevalue{Holomorphic blocks and the 5d AGT correspondence}
\def\authorvalue{Sara Pasquetti}
\def\shortauthorvalue{\authorvalue}
\def\addressvalue{Dipartimento di Fisica, Universit\`a di Milano-Bicocca,
Piazza della Scienza 3, I-20126 Milano, Italy\\
  \tt sara.pasquetti@gmail.com}
\def\abstractvalue{
We review the holomorphic block factorisation of partition functions of supersymmetric theories on  compact manifolds in various dimensions.   We then show how to interpret 3d and 5d partition functions as correlation
functions with underlying symmetry given by a deformation of the Virasoro algebra.
}
\def\preprintvalue{.}

\ifx\ifLONG\undefined
\documentclass[11pt]{article}
\newcommand{\chapterauthor}[1]{
\begin{center}
{\bf \normalsize  #1}
\end{center}
}

\newcommand{\chapteraddress}[1]{
\begin{center}
{ \small \it \addressvalue}
\end{center}
}

\newcommand{\chapterabstract}[1]{
\vspace{\baselineskip}
\begin{center}
\textbf{\small Abstract}
\end{center}
#1}

\newcommand{\chapterheader}{

\chapter[\titlevalue{}  (by \shortauthorvalue)]{\titlevalue}
\label{Chapter\IDvalue}
\chapterauthor{\authorvalue}
\chapteraddress{\addressvalue}
\chapterabstract{\abstractvalue}
\tightmtctrue
\minitoc
}

\newcommand{\documentheader}{
\begin{flushright} \small
  \preprintvalue
 \end{flushright}

\begin{center}
{\bf \Large \titlevalue}
\end{center}

\chapterauthor{\authorvalue}
\chapteraddress{\addressvalue}
\chapterabstract{\abstractvalue}

\medskip

This is a contribution to the review volume ``Localization techniques
in quantum field theories'' (eds. V.~Pestun and M.~Zabzine) which
contains 17 Chapters available at \cite{ContributionSummary}

\tableofcontents
}

\newcommand{\ifvolume}[2]{\ifx\ifLONG\undefined#2\else#1\fi}

\newcommand{\documentfinish}{
\ifx\ifLONG\undefined
\bibliographystyle{bibreview} 
\bibliography{\IDvalue,review}  
\end{document}
\else
\addcontentsline{toc}{section}{References}
\providecommand{\href}[2]{#2}\begingroup\raggedright\endgroup

\fi
}

\newcommand{\documentfinishBBL}{
\addcontentsline{toc}{section}{References}
\ifx\ifLONG\undefined
\input{\IDvalue.separate.bbl}
\end{document}
\else
\input{\DIRvalue/\IDvalue.volume.bbl}
\fi
}

\ifx\ifLONG\undefined
\def\volcite#1{Contribution \cite{Contribution#1}}
\else 
\def\volcite#1{Chapter \ref{Chapter#1}}
\fi

\usepackage{amsmath,amssymb,amsthm,amscd,pstricks}
\usepackage{graphicx}

\usepackage{amsmath}

\usepackage{hyperref}

\begin{document}
\thispagestyle{empty}
\documentheader
\else \chapterheader \fi

\numberwithin{equation}{section}

\newcommand{\tq}{{\tilde q}}

\newcommand{\PAFP}[1]{{\color [rgb]{0,0.6,0} [FP: #1]}}
\newcommand{\PASP}[1]{{\color [rgb]{0.6,0.0,0.5} [SP: #1]}}

%%%%%%%%%%%%%%%%%%%%%%%%%%%%%%%%%%%%%%%%%%%%%%%%%%%%%%%%%%%%%%%%

\section{Introduction}

Over the last 10 years, starting from the seminal work by Pestun \cite{PAPestun:2007rz},  Witten's localisation has been extensively applied to supersymmetric  theories defined on compact manifolds of various dimensions. This has led to the derivation of a large number  of exact results such as the evaluation of partition functions and vevs of various BPS observables like Wilson-loops and surface operators.

Thanks to these results it has been possible to perform impressive large $N$ tests of various holographic dualities as  summarised in the contributions \volcite{MA}, \volcite{ZA}, \volcite{MI} of this review.
It  has also been  possible to do precision checks of various non-perturbative dualities such as Seiberg-like dualities and 3d mirror symmetry. For example, as reviewed in  \volcite{WI}, 3d partition functions are protected under RG flow hence one can explicitly compute partition functions of  pairs of UV  Lagrangian  supposed to flow to the same SCFT in the IR, and show that they are indeed equal.

Localisation has played a key role also in the discovery of new surprising correspondences relating  QFTs in different dimensions and with different types of symmetries.
This is the case of the celebrated Alday-Gaiotto-Tachikawa (AGT) correspondence \cite{PAAlday:2009aq} relating $S^4$ partition functions of $\mathcal{N}=2$ theories to Toda CFT correlators. This  correspondence together with its variation involving the superconformal index 
or $S^3\times S^1$ partition functions of $\mathcal{N}=2$ theories is reviewed in 
\volcite{TA}.  A similar correspondence  relating  3d $\mathcal{N}=2$ theories to complex Chern-Simons theories is reviewed   in \volcite{DI}.

The interest in studying SUSY theories on compact manifolds has led to the development of a comprehensive approach to the formulation of supersymmetric theories  on curved space initiated by Festuccia and Seiberg \cite{PAFestuccia:2011ws} and reviewed in   \volcite{DU}.
In particular it has been possible to derive general theorems to determine the amount of supersymmetry preserved by a given background and  the dependence of partition functions on the  data specifying the background.

It has also been observed  that  if the manifold $M$ on which the theory is formulated can be decomposed into simpler building blocks, as for example in the case of  the solid tori decomposition of a three-sphere, then also the partition function $Z_M$  can be expressed in terms of  the partition functions of the building blocks,  the so-called holomorphic-blocks.
In the first part of this review article we will  illustrate several examples of this block decomposition in 3d, 4d and 5d.

Holomorphic blocks  in various dimensions are interesting mathematical objects with intricate transformation properties under dualities which often involve  non-trivial Stokes phenomena, for an extensive analysis of the properties of the 3d holomorphic blocks see \cite{PAhb}. 3d and 5d blocks  are related to open and closed topological string amplitudes while  3d blocks appear also as Chern-Simons wave functions in the context of the 3d-3d correspondence discussed in  \volcite{DI}.

In this review article we focus instead on the  interpretation of the holomorphic blocks in the context of  AGT-like correspondences.
Via the AGT correspondence   partition functions of  $\mathcal{N}=2$ theories on $S^4$ can be mapped to Toda/CFT correlators and the   holomorphic blocks, which   in this case coincide with  the two hemi-sphere partition functions,  are mapped to the Toda chiral  conformal blocks.
Correlators involving degenerate operators are instead mapped  to  $\mathcal{N}=2$ theories on $S^4$  with surface operators inserted on a codimension-two $S^2$ and  the holomorphic blocks of the codimension-two defect theory  correspond to degenerate chiral conformal blocks.

In the second  part  of this review we will  argue that a similar correspondence can be established between  $\mathcal{N}=1$ theories on a large class of 5-manifolds and correlators with underlying symmetry given by a deformation of the Virasoro algebra.
Also in this case codimension-two defect theories and 3d holomorphic blocks can be mapped to degenerate deformed Virasoro correlators.

The plan of the review is the following: we discuss  the holomorphic block factorisation in 3d and 4d  in section \ref{PA3d4dsection}
and in 5d in section \ref{PA5dfactos}. In section \ref{PAsecdeg} we consider  the insertion of codimension-2 defect operators via Higgsing in 5d theories focusing on some simple cases.
We then move to the discussion of the dual deformed Virasoro side. After introducing the deformed Virasoro algebra in section \ref{PAdefalg}, we collect some of the evidences of the mapping of degenerate and non-degenerate  deformed  Virasoro chiral blocks to vortex and instanton partition functions in  section \ref{PAqvirinstvo}.
Finally in section \ref{PA3d5dmapsec} we discuss how to combine deformed Virasoro blocks to construct correlators and how these can be mapped to 3d and 5d partition functions.

\section{Compact manifolds and Holomorphic Block factorisation}

\subsection{Factorisation and Holomorphic blocks in 3d and 4d}\label{PA3d4dsection}

In this section we discuss the holomorphic  block factorisation in 4d  theories defined on Hermitian manifolds of the form $M^{4d}=M^{3d}\times  S^1$, where $M^{3d}$ is a possibly non-trivial  $U(1)$ fibration over the 2-sphere, and their 3d  reductions.
More precisely we focus on  the following $\mathcal{N}=1$  backgrounds: the $S^3\times S^1$   and  lens  $L_r\times S^1$ indexes, the $S^2\times T^2$ background  and their 3d $\mathcal{N}=2$ reductions: the squashed sphere $S^3$,  the  lens space $L_r$, the $S^2\times S^1$ index and the twisted  $S^2\times S^1$ index.

The  3-manifolds listed above can all be  realised by gluing two solid tori  $D^2\times S^1$ with an  element $g\in SL(2,\mathbb{Z})$, we  call them  $M^{3d}_g$.
Similarly  all the  4-manifolds  above can be constructed from the fusion of two
 solid tori $D^2\times T^2$ with appropriate  elements in $g\in SL(3,\mathbb{Z})$ and we  call them $M^{4d}_g$.

 As reviewed  in   \volcite{DU}  partition functions on these backgrounds are metric independent, they do however depend on other data specifying the background (for example in the 4d case they depend on the complex structure and holomorphic vector bundles associated to the flavouflavourr symmetries) so they are not properly topological objects, nevertheless, as we will review,  there is  evidence that the following chain of identities  holds:
 %General results  \cite{PAClosset:2013vra}, \cite{PAClosset:2012ru} state that partition functions on these spaces 
%do not depend on the  metric but are  holomorphic functions of the complex structure parameters and of the background gauge fields. 
\begin{equation}
\label{PAthree34}
 Z^{M^{3d/4d}_g}=    \sum\!\!\!\!\!\!\!\!  \int ~
 \Big\|\Upsilon^{\rm 3d/4d}\Big\|^2_{g}= \sum_{c}\Big\|\mathcal{B}^{\rm 3d/4d}_c\Big\|^2_g
  =\sum_{c}\Big\|\oint_{\Gamma_c}\Upsilon^{\rm 3d/4d}\Big\|^2_g~.
\end{equation}
The first equality states  the factorisation  into a  ``$g$-square" 
of the integrand of the Coulomb branch partition function\footnote{The symbol $\sum\!\!\!\!\!\!  \int$ indicates that the Coulomb branch partition  might  include  a sum over a discrete index besides the integration over the constant values of the fields parameterising the Coulomb branch.} 
given by a classical and 1-loop contribution: 
 \begin{equation}
{ Z}^{M^{3d/4d}_g}_{\text{cl}} ~ { Z}^{M^{3d/4d}_g}_{\text{1-loop}}  =\Big\|\Upsilon^{\rm 3d/4d}\Big\|^2_{g}\,.
\end{equation}
The functions $\Upsilon^{\rm 3d/4d}$ can be interpreted as integrands of the $D^2\times S^1$ or $D^2\times T^2$ partition functions. The data specialising the manifold $M^{3d/4d}_g$ are all encoded in the gluing rule $g$.

To explain this first equality  we  consider the case of the simplest 3d $\mathcal{N}=2$ theory: the free chiral with (minus) half Chern-Simon unit, which we add to remove the parity anomaly. This theory is often referred to as the tetrahedron theory since in the context of the 3d-3d correspondence it  computes  the quantum volume of the ideal tetrahedron  
\cite{PADimofte:2011py},\cite{PADimofte:2011ju},\cite{PADimofte:2011jd}.

If we specialise to the squashed 3-sphere $S^3_b=\{(x,y)\in \mathbb{C}^2|~ b^2|x|^2+b^{-2}|y|^2=1\}$, as reviewed in \volcite{WI}, the contribution  of a charge plus chiral multiplet with canonical R-charge and real mass $X$ for the background vector multiplet associated to the $U(1)$ flavour symmetry, is given in terms of the double-sine function $s_b$ defined in the Appendix \ref{PAspecfun}.
When combined with the Gaussian contribution of  the $-1/2$ CS unit, the partition function admits  a factorised form:
\begin{equation}\label{}
Z^{S^3_b}_{\Delta}(X)=e^{\tfrac{i\pi X^2}{2}}s_b(\tfrac{iQ}{2}-X)=\frac{(qx^{-1};q)_\infty}{(\tilde x^{-1};\tilde q^{-1})_\infty}=
\Big\|\mathcal{B}_\Delta^{\rm 3d}(x;q) \Big\|^2_S\,,
\end{equation}
where $Q=b+b^{-1}$ and the  {\it holomorphic}  variables are defined as
\begin{eqnarray}\nonumber
 &q=e^{2\pi i b^2}=e^{2\pi i \tau} \,,\qquad  \tilde q=e^{\tfrac{2\pi i}{b^2}} =e^{\tfrac{2\pi i }{\tau}}\,,&\\
 &x=e^{2\pi b X}=e^{2\pi i \chi}\,, \qquad    \tilde  x=e^{2\pi  \tfrac{X}{b}}=e^{2\pi  i\tfrac{\chi}{\tau}}\,.&%e^{\tfrac{2\pi i \chi}{\tau}}
\end{eqnarray}
The  3d holomorphic block $\mathcal{B}_\Delta^{\rm 3d}(x;q) =(q x^{-1} ;q)_\infty$ 
is the  partition function  on the solid torus or  Melvin cigar  $D^2\times S^1$ of the tetrahedron theory.
 Notice that when $|q|<1$ we have $|\tilde q|>1$ and 
\begin{align} \label{zqinf}
(x;q)_\infty &= \sum_{n=0}^\infty \frac{(-1)^n q^{\frac{n(n-1)}{2}}x^n}{(q;q)_n} =
 \begin{cases} \prod_{r=0}^\infty (1-q^rx) &{\rm if}\quad |q|<1 \\[.1cm]
 \prod_{r=0}^\infty (1-q^{-r-1}x)^{-1} &{\rm if}\quad |q|>1~. \end{cases}
\end{align}
Basically blocks in $x,q$, and $\tilde x, \tilde q$,  share the same series expansion but
they converge to different functions.
This is a key feature of holomorphic blocks which has been extensively discussed in
\cite{PAhb}.
 The two blocks are glued through the $S$-gluing    acting on $\tau$,   the modular parameter of the boundary $T^2$, and on the flavour variable $\chi$ as:
 \begin{equation}
\tau\to\tilde \tau= -S(\tau)=\frac{1}{\tau}\,,\qquad \chi\to \tilde \chi= \frac{\chi}{\tau}\,.
\end{equation}
This gluing corresponds to the element $S\in SL(2,\mathbb{Z})$ (composed with orientation inversion) realising a three-sphere  from a pair of solid tori. 

There is an similar  factorisation of the  tetrahedron theory on the lens space $L_r$.
This  smooth 3-manifold is  the free $\mathbb{Z}_r$  orbifold of the squashed 3-sphere  with the identification
\begin{equation}
(x,y) \sim (e^{\frac{2\pi  \i}{r}} x,  e^{-\frac{2\pi  \i}{r}} y)~.
\end{equation}
In this case, as reviewed in \volcite{WI}, the contribution of  the chiral multiplet is expressed in terms of the modified double-sine function which takes into account the periodicity inherited from the $\mathbb{Z}_r$ quotient. The  factorisation properties of the modified double-sine are somewhat subtle and we refer the reader to  \cite{PANieri:2015yia} for details, in the end when combined with the appropriate half Chern-Simons unit one finds:
\begin{equation} 
Z^{L_r}_\Delta(X,H)=
\Big\|\mathcal{B}_\Delta^{\rm 3d}(x;q) \Big\|^2_r~.
\end{equation}
Where we have turned on for the flavour symmetry  a continuous real mass $X$ and a discrete holonomy  $H\in Z_r$,
parameterising the  topological sectors.
The  holomorphic  variables are now defined as
\begin{equation}
\label{df}
\begin{array}{lll}x=e^{\frac{2\pi }{r}bX} e^{\frac{2\pi \i }{r}H}= e^{2\pi\i\chi} e^{\frac{2\pi \i}{r}H}
,&\quad & \tilde x=e^{\frac{2\pi }{r b}X}  e^{-\frac{2\pi \i }{r}H}
=e^{2\pi\i\frac{\chi}{r\tau-1}} e^{-\frac{2\pi \i }{r}H}~
,\\
q=e^{2\pi\i\frac{ b Q}{r}}=e^{2\pi \i \tau},&\quad & \tilde q=e^{2\pi\i\frac{ Q}{br}}=e^{2\pi\i\frac{\tau}{r\tau-1}}~.\end{array}
\end{equation}
The two blocks are glued through the $r$-pairing  acting as
\begin{equation}
\tau\to\tilde \tau=-\hat r(\tau)= \frac{\tau}{r\tau-1}~,\qquad 
\chi\to \tilde \chi= \frac{\chi}{r\tau-1}~,\quad H\to\tilde H= r-H\,.
\end{equation}
%where $\tau$ is to be identified with the modular parameter of the boundary $T^2$, while the flavour fugacity and holonomy transform as 
%\begin{equation}
%\chi\to \tilde \chi= \frac{\chi}{r\tau-1}~,\quad H\to\tilde H= r-H~.
%\end{equation}
This gluing rule as expected coincides with  the $\hat r\in SL(2,\mathbb{Z})$ element (composed with the orientation inversion) realising   the $L_r$ geometry from a pair of solid tori. The factorisation on the  more general $L(p,q)$ Lens spaces has been discussed in \cite{PADimofte:2014zga}.

The  next example we discuss is the $S^2\times S^1$ background.  Localisation on this background is reviewed in in \volcite{WI}. For the tetrahedron 
with a  fugacity $\zeta$ which we take to be a phase, and the  integer $m \in \mathbb{Z}$ for the background  magnetic flux through $S^2$, we find:
 $$
Z_\Delta^{S^2\times S^1}(\zeta,m)=\Big|\Big| (q x^{-1};q)_\infty \Big|\Big|^2_{id}  \, ,
$$
where  the holomorphic variables
\begin{equation}
\begin{array}{lll}x=\zeta q^{m/2}= e^{2\pi\i\chi} 
,&\quad & \tilde x=\zeta^{-1} q^{m/2}
=e^{2\pi\i \bar \chi} 
,\\
q=e^{2\pi \i \tau},&\quad & \tilde q=q^{-1}=e^{-2\pi \i \tau}~.\end{array}
\end{equation}
The two blocks are glued through the $id$-pairing  (combined with orientation reverse) acting as
\begin{equation}
\tau\to\tilde \tau= -id(\tau)=-\tau~,\quad \chi \to \tilde\chi=\bar \chi.
\end{equation}
This gluing rule as expected coincides with  the $id\in SL(2,\mathbb{Z})$ element (composed with orientation inversion).
Finally as discussed in  \cite{PANieri:2015yia}, there is a similar factorisation also in the case of the twisted index background \cite{PABenini:2015noa}.

Generic interacting theories, with no parity anomaly, that is with integer effective Chern-Simons couplings,
 can  be constructed by gauging products of tetrahedron theories and then adding integers units of Chern-Simons terms
and the contribution of vector multiplets. 
  This observation allow us to {\it take the square root} of the integrand of generic theories  whenever there is no parity anomaly. Indeed we have just reviewed how to take tetrahedron theories as squares of tetrahedron blocks,
thanks to these  special functions identities we can easily factorise the  matter and vector multiplet contributions.
Chern-Simons terms at integer level, can instead be dealt with by using  the properties of the  theta function:
  \begin{equation}
 \theta( x ;q ):=( - q^{1/2} x;q )_\infty (- q^{1/2} x^{-1};q )_\infty\,,
\end{equation}
which for example satisfies
\begin{eqnarray}\nonumber
&\Big|\Big| \theta\big( (-q^{1/2})^{c} x^{a};q\big)\Big|\Big|_{S}^2 =  {\rm C}^{-2} e^{-i \pi\Big(a \frac{X}{2 i\pi b}+c \frac{Q}{2}\Big)^2} \,, \qquad {\rm C}=e^{-\frac{i\pi}{12}(b^2+\frac{1}{b^2})}\,.
%&\\
%&\Big|\Big| \theta((- q^{1/2})^c x^a;q )\Big|\Big|^2_{id}=
%(- q^{1/2} )^{-(a\cdot m) c} \zeta^{-(a\cdot m) a}\,.&
\end{eqnarray}
We can use this identity to express  Chern-Simons terms on $S^3$ as squares of theta functions depending on the holomorphic variables. Chern-Simons terms on other 3-manifolds $M^{3d}_g$ are similarly factorised in terms of $g$-squares of theta functions.

To make a concrete example we consider the  SQED partition function on $S^3_b$, with masses  $\tilde m_i$ for the $N_f$ charge plus chirals,  masses $ m_i$ for the $N_f$ charge minus chirals and  an FI parameter $\xi$:
\begin{equation}
\label{PAsqedexample}
Z^{S^3_b}[SQED]=\int d\sigma ~e^{2\pi i \sigma \xi}~ \prod_{j,k=1}^{N_f}~s_b(\sigma+m_j+ i Q/2)s_b(-\sigma-\tilde m_k+ i Q/2)\,.
\end{equation} 
In this case the classical (FI term) and 1-loop term can be factorised as
\begin{equation}
\Upsilon^{3d}[SQED]= \frac{\theta(x u;q)}{\theta(u;q)\theta(x;q)} \prod_{j,k=1}^{N_f} \frac{(q x_j x^{-1};q)_\infty}{( y_k x^{-1};q)_\infty } \, ,
\end{equation}
with the following definition of  holomorphic variables: 
\begin{eqnarray}
\label{etil}\nonumber
\!\! x=e^{2\pi b \sigma}, \quad x_i=e^{2\pi b m_i}, \quad y_i=e^{2\pi b \tilde m_i}, \quad z=e^{2\pi b \xi}, \quad q=e^{2\pi i b^2}\,,\\
\!\!\! \tilde x=e^{2\pi  \sigma/b}, \quad \tilde x_i=e^{2\pi  m_i/b}, \quad \tilde y_i=e^{2\pi  \tilde m_i/b}, \quad \tilde z=e^{2\pi \xi/b}, \quad \tilde q=e^{2\pi i/ b^2} \,,
\end{eqnarray}
and
\begin{equation} 
 \prod_{j,k=1}^{N_f}x_jy_k^{-1}=r,\quad  u=(-q^\frac{1}{2})^{N_f}r^\frac{1}{2}z^{-1}\,.\\
\end{equation}

The discussion of the first equality  in the chain of identities (\ref{PAthree34}) for the 4-manifolds case is similar. Here  the factorisation of the integrand in terms of the $D^2\times T^2$ integrand $\Upsilon^{4d}$ 
again involves several non-trivial identities for the special functions appearing in the one-loop contributions but this time the necessary and sufficient condition for the factorisation is the cancellation of the cubic anomaly.
This was observed in  \cite{PANieri:2015yia}  building on the discovery of the  surprising relation between the modular properties of the superconformal index   and the appearance  of the  anomaly polynomial \cite{PASpiridonov:2012ww}.\\

The next  equality in  the chain of identities (\ref{PAthree34}) is the block-factorisation of the Coulomb branch partition function.
1-loop factors are meromorphic functions and it is possible to evaluate the integral by taking residues at their poles by choosing suitable convergent integration contours.
The result takes the form of a sum over the  supersymmetric vacua (critical points of the  effective twisted of superpotential) of the semiclassical   $(2,2)$ theory on the  $\mathbb{R}^2\times S^1$ and $\mathbb{R}^2\times T^2$ solid tori:
\begin{equation}
{ Z}^{M^{3d/4d}_g} =\sum_c \left({ Z}^{M^{3d/4d}_g}_{\text{cl}} ~ { Z}^{M^{3d/4d}_g}_{\text{1-loop}} \right)_c  \Big| \Big| {\cal Z}^{3d/4d, c}_{\text{V}}\Big| \Big|_{g}^2\,.
\end{equation}
The contribution of each vacua consists of  the product of  classical and 1-loop terms evaluated at the $c$-th vacuum and of the vortex ${\cal Z}^{3d/4d, c}_{\text{V}}$ partition function.
This is  the partition function of the theory placed on the cigars  or $\mathbb{R}^2\times S^1$ or $\mathbb{R}^2\times T^2$ with the Omega background turned on $\mathbb{R}^2$ \cite{PADimofte:2010tz} and enumerates finite energy BPS vortex configurations. Typically  vortex partition functions expressed in terms of  $q$-deformed or elliptic hypergeometric series.

We can also factorise the one-loop and classical contributions as  discussed above and  present the partition  function as a sum  of $g$-squares of 3d or 4d holomorphic blocks defined as:
\begin{equation}
\label{PA34hb}
{\cal  B}^{3d/4d}_c=  {\Upsilon}^{3d/4d}\Big|_c  ~{\cal Z}^{3d/4d, c}_{\text{V}}\,.
\end{equation}

To make a concrete example we consider again the the  SQED partition function  on $S^3_b$.
To evaluate the integral (\ref{PAsqedexample})  we can  close the contour in the upper half plane and take the contributions of poles
located at $x=-m_i + i m b + i n/b$,  see \cite{PAPasquetti:2011fj} for details.
The result reads
\begin{equation}
Z^{S^3_b}[SQED]= \sum_{i=1}^{N_f} e^{-2\pi i \xi m_i} \prod_{j,k=1}^{N_f} \frac{s_b(m_j-m_i+i Q/2)}{s_b(\tilde m_k-m_i-i Q/2)}~\Big|\Big|\mathcal{Z}^{(i)}_V\Big|\Big|^2_S\,,
\end{equation} 
where the various terms are given by the FI and 1-loop contributions with the Coulomb branch parameter fixed at $x=-m_i$ and the vortex partition function
\begin{eqnarray} 
\nonumber\mathcal{Z}^{(i)}_V&=&\sum_{n\geq 0}\prod_{j,k=1}^{N_f} \frac{( y_k x_i^{-1};q)_n}{(q x_j x_i^{-1};q)_n}  u^n=\\
&&=\phantom{|}_{N_f}\Phi_{N_f-1}(x_i^{-1}y_1,\ldots,x_i^{-1} y_{N_f};q x_i^{-1} x_1,\hat\ldots,q x_i ^{-1}x_{N_f};u)\, .\qquad
\end{eqnarray} 
where $\phantom{|}_{N_f}\Phi_{N_f-1}$ is the basic hypergeometric function. The classical and 1-loop term can also be factorised to obtain the 3d block:
\begin{eqnarray}\nonumber
{\cal  B}^{3d}_i&=& \frac{\theta(x_i u;q)}{\theta(u;q)\theta(x_i;q)} \prod_{j,k=1}^{N_f} \frac{(q x_j x_i^{-1};q)_\infty}{( y_k x_i^{-1};q)_\infty }\\ &&\!\!\!\!\!\!\!\times \phantom{|}_{N_f}\Phi_{N_f-1}(x_i^{-1}y_1,\ldots,x_i^{-1} y_{N_f};q x_i^{-1} x_1,\hat\ldots,q x_i ^{-1}x_{N_f};u)\, .
\end{eqnarray}

Explicit examples of block factorisation have been obtained for various  theories  and backgrounds including  $S^3_b$, lens space $L_r$, the superconformal index $S^2\times S^1$, the twisted index, \cite{PAPasquetti:2011fj},  \cite{PAhb},\cite{PATaki:2013opa}, \cite{PAHwang:2012jh}, \cite{PAHwang:2015wna}, \cite{PAImamura:2013qxa}, \cite{PANieri:2015yia}, \cite{PABenini:2015noa}.

Similar residue computations yield the  block factorisation of $\mathcal{N}=1$ theories on $S^3\times S^1$, $L_r\times S^1$, $S^2\times T^2 $ \cite{PAYoshida:2014qwa}, \cite{PAPeelaers:2014ima},  \cite{PANieri:2015yia},  \cite{PAChen:2014rca}  and  on the ellipsoid  \cite{PAChen:2015fta},  and,
as reviewed in  \volcite{BL}  in 2d $\mathcal{N}=(2,2)$ theories on $S^2$ \cite{PABenini:2012ui}, \cite{PADoroud:2012xw}, \cite{PAClosset:2015rna}.

The block factorisation of $M^{3d/4d}_g$  partition function can also be  interpreted as the result of an alternative localisation scheme know as Higgs branch localisation. As reviewed in  \volcite{BL}  the Higgs branch localisation was originally introduced for the (2,2) theories in  \cite{PADoroud:2012xw}, \cite{PABenini:2013yva}, 
and later   applied to other backgrounds  in \cite{PABenini:2012ui}, \cite{PAFujitsuka:2013fga},  \cite{PAPeelaers:2014ima},  \cite{PAYoshida:2014qwa}, see also the chapter \volcite{BL} in this review.

Another perspective on the factorisation in the  $S^3_b$ case was provided in \cite{PAAlday:2013lba},
 where it was shown that it is possible to deform the three-sphere  geometry into two cigars $D^2\times S^1$ connected by a long tube without changing the value of the partition function. This deformation  has  exactly the effect of projecting  down the theory into the SUSY ground states which  are the only states contributing to the overlap of the two blocks.
It should be possible to extend this argument to other 3d and 4d backgrounds. In 2d a similar proof of the block-decomposition of the two-sphere was provided in \cite{PAGomis:2012wy}. The 2d holomorphic block in this case are the 
cigar partition functions appearing in the   Cecotti-Vafa  $tt^*$ set-up \cite{PACecotti:1991me}, \cite{PACecotti:2013mba} and their overlap computes the exact K\"ahler potential as reviewed in  \volcite{MO}.\\

To explain the last equality in (\ref{PAthree34}) we begin by observing a  key property of the 3d  (or 4d) holomorphic blocks: they are annihilated by a set of difference equations which can be interpreted as Ward identities for Wilson loops (or surface operators) wrapping the circle $S^1$ (or the torus $T^2$)  and acting at the tip of the cigar. There are in fact two commuting sets of difference operators annihilating respectively the holomorphic and the anti-holomorphic blocks. 
This set of difference operators can be systematically derived from the UV Lagrangian
\cite{PADimofte:2011py},\cite{PADimofte:2011ju}, \cite{PADimofte:2011jd}.
Building on this in \cite{PAhb} it  was developed an integral formalism to compute the 3d holomorphic blocks by integrating the  meromorphic one-form $\Upsilon^{\rm 3d}$ on a  basis  of  middle-dimensional cycles in $(\mathbb{C}^{*})^{|G|}$
\begin{equation}
\label{two2}
\mathcal{B}^{\rm 3d }_c=\oint_{\Gamma_c}\Upsilon^{\rm 3d}~.
\end{equation}
 Each contour is associated to a critical point of the integrand, which in turn is related to a supersymmetric ground state
and it is chosen so that the integral converges and solves the set of difference equations.  The space of blocks can then be viewed as  the vector space of solutions to the system of  difference equations.
Closely connected to this constructions are the global transformations properties of the blocks  in parameter space. It was shown that by fixing $q$ and varying $x$  the  holomorphic blocks are subject to Stokes phenomena. 
We refer the reader to \cite{PAhb} for  a detailed discussion of the block integrals and the interplay between mirror symmetry and Stokes phenomenon. See also  \cite{PAYoshida:2014ssa} for a derivation of the block integrals  from localisation on $D^2\times S^1$.

In the context of the 3d-3d  correspondence  reviewed in the contribution  \volcite{DU}, 3d blocks are identified with  complex  Chern-Simons wave-functions. In the second part of this chapter we will instead see how 3d blocks can be
 been identified with $q$-deformed  Virasoro  correlators.

We should also mention that factorisation and the definition of the blocks suffer an intrinsic ambiguity.
By defining blocks as solutions to difference equations we have the possibility to multiply them by $q$-constant 
which satisfy $c(x q;q)= c(x;q)$. By requiring that these constants  don't modify the compact space results we restrict to $q$-constants with unit $g$-square $||c(x;q)||_g^2=1$. These are elliptic ratios of theta functions and have a trivial 
 semiclassical limit ($q=e^h$ $h\to 0$). These functions represent our ambiguity.\\

 A block integral formulation  for 4d holomorphic blocks leading to the last equality in (\ref{PAthree34})   has been proposed  in \cite{PANieri:2015yia}. The definition of the integration contours for 4d block integrals  is quite subtle and
a careful study of their properties is  still missing. For example it would  be interesting to study their
 behaviour under various 4d dualities.
It should  also  be possible  to re-derive the 4d block integrand prescription via localisation on $D^2\times T^2$.
The relation of  4d  block integrals to  free field correlators in an elliptic deformation of the Virasoro algebra has been explored in \cite{PANieri:2015dts}.

\subsection{Factorisation and Holomorphic blocks in 5d}\label{PA5dfactos}
As reviewed in \volcite{QZ}, localisation can be applied to 5d $\mathcal{N}=1$  theories formulated on a  large class of 5d manifolds. The aim of this section is to show that partition functions on these manifolds can be obtained by gluing the so-called 5d holomorphic blocks  $\mathcal{B}^{5d}$,
which are partition functions on $\mathbb{R}^{4}_{\epsilon_1,\epsilon_2}\times S^1$. The gluing rule
can be read out from the geometric data of the 5d manifolds.

\subsubsection*{Squashed $S^5$ partition functions and 5d holomorphic blocks}

 We begin our discussion  with the squashed $S^5$, the simplest example of toric Sasaki-Einstein  5-mainifold.
It is convenient to think the $S^5$ as $T^3$ fibration over the interior of a  triangle, with the fiber degenerating to a torus  on the faces and to a  circle over the  vertices.

 As reviewed  in   \volcite{QZ},    the partition function of 5d ${\cal N}=1$ theories  on this background takes the following form:
%In a series of papers \cite{PAKallen:2012cs, PAHosomichi:2012ek, PAKallen:2012va, PAKim:2012ava, PAImamura:2012bm, PALockhart:2012vp, PAKim:2012qf, PAMinahan:2013jwa} the 5d ${\cal N}=1$ supersymmetric gauge theory has been formulated on 
% $S^5$ with squashing parameters $\omega_1,\omega_2,\omega_3$,
%and the partition function has been shown to localize  to the integral over the zero-mode of the vector multiplets scalars $\vec \sigma$ which takes value in  the Cartan subalgebra of the gauge group:
%%{which we take to be $SU(N)$ with generators $T_a$ normalized as $\text{Tr}_R(T_a T_b)=C_2(R)\delta_{ab}$, with $C_2(F)=1/2$ for the fundamental}:
\begin{equation}\label{PAs5}
Z^{S^5}=\int d \vec\sigma ~{ Z}^{S^5}_{\text{cl}}(\vec \sigma) { Z}^{S^5}_{\text{1-loop}}(\vec \sigma,\vec M)  { Z}^{S^5}_{\text{inst}}(\vec \sigma,\vec M) \,.
\end{equation}
The integral is over the zero-mode of the vector multiplets scalars $\vec \sigma$  taking value in  the Cartan subalgebra of the gauge group,  $\vec M$ indicate the hypermultiplet masses.\\

The  non-perturbative ${ Z}^{S^5}_{\text{inst}}(\sigma,\vec M)$ receives  contributions 
from the three fixed points of the Hopf fibration over the $\mathbb{CP}^2$ base and takes the following factorized form \cite{PALockhart:2012vp, PAKim:2012qf}: 
\begin{equation}
\label{PAfirstk}
 Z^{S^5}_\text{inst}(\vec \sigma,\vec M) =\prod_{k=1}^3  ({\cal Z}_\text{inst})_k   =\Big|\Big|{\cal Z}_\text{inst}\Big|\Big|^3_{S} \,,
\end{equation}
where ${\cal Z}^{}_\text{inst}$ coincides with  the equivariant instanton partition function 
on ${\mathbb R}^4\times S^1$ \cite{PANekrasov:2002qd, PANekrasov:2003rj} with Coulomb and mass parameters appropriately rescaled and with equivariant parameters $\epsilon_1= \frac{e_1}{e_3}$
and $\epsilon_2= \frac{e_2}{e_3}$:
\begin{equation} 
\label{PAcinst}
 {\cal Z}_\text{inst}=  {\cal Z}^{\mathbb{R}^4\times S^1}_\text {inst}\left(\frac{i \vec \sigma}{e_3}, \frac{ \vec m}{e_3}; 
 \frac{e_1}{e_3},\frac{e_2}{e_3}\right)\,,
\end{equation}
where  $m=iM+E/2$ and  $E=\omega_1+\omega_2+\omega_3$.
The sub-index $k=1,2,3$ in eq. (\ref{PAfirstk})  refers to  the following identification of the parameters  $e_1,e_2,e_3$ to the squashing parameters $\omega_1, \omega_2,\omega_3$
in each sector:
\begin{eqnarray}\displaystyle\label{PAp123}
\begin{array}{|c|ccc|}
\hline
{\rm sector}~&\quad e_1\quad &\quad e_2\quad &\quad e_3\quad\\
\hline
1&\omega_3&\omega_2&\omega_1\\
2&\omega_1&\omega_3&\omega_2\\
3&\omega_1&\omega_2&\omega_3\\
\hline
\end{array}
\end{eqnarray}The three sectors correspond to the loci where the Reeb vector forms closed orbits (in the round $S^5$ case they close everywhere). For more general toric SE manifolds,  it is conjectured that the non-perturbative contributions are indeed localised at these isolated loci.

Actually  the instanton partition function depends on the squashing parameters   through the combination
\begin{equation}\label{PApq}
q=e^{2\pi i e_1/e_3}\,,\quad t=e^{2\pi ie_2/e_3}\,,
\end{equation}
from which we see that the product $\Big|\Big|\cdots\Big|\Big|^3_{S} $  enjoys an $SL(3,\mathbb{Z} )$ symmetry
which acts  as {\it S-dualizing} the  couplings $q$ and $t$.

The classical  Yang-Mills action can also be expressed in the $SL(3,\mathbb{Z})$ factorized form as the instanton contribution \cite{PANieri:2013vba}:
\footnote{To simplify this expression we defined $ g^2= \frac{g_{YM}^2}{4i \pi^2}$ and used that
 $2C_2(ad)\sum_\rho\rho(\sigma)^2=\sum_\alpha \alpha(\sigma)^2$.} 
  \begin{equation}\label{PAclass}
{Z}^{S^5}_{\rm cl}(\sigma)=
e^{\frac{2\pi i}{\omega_1\omega_2\omega_3 g^2}\,\text{Tr}(\sigma^2)}=
e^{-\frac{2\pi i}{\omega_1\omega_2\omega_3 g^2 2C_2(ad)} \sum_\alpha [i\alpha(\sigma)]^2}=\Big|\Big|  {\cal Z}_{\rm cl}\Big|\Big|^3_{S}\,,
\end{equation}
where
\begin{eqnarray}\displaystyle
\label{PAzcl}
{\cal Z}_{\rm cl}=\prod_\alpha \frac{\Gamma_{q,t}\left(\frac{1}{e_3}\left(i\alpha(\sigma)+\frac{1}{g^22C_2(ad)}+
\frac{E}{2}\right)\right)}{\Gamma_{q,t}\left(\frac{1}{e_3}\left(\frac{1}{g^22C_2(ad)}+\frac{E}{2}\right)\right)}\,,
\end{eqnarray}
and we denoted by $\alpha$ the roots of the gauge group Lie algebra. To arrive at this expression we first need to
write  the Gaussian  term as a combination of  Bernoulli polynomial $B_{33}$ defined in Appendix \ref{PAber}
and then use the identity  \cite{PAfv}
\begin{eqnarray}\displaystyle\label{PAnfac}
e^{-\frac{2\pi i}{3!}B_{33}(z)}&=& \prod_{k=1}^3\Gamma_{q,t}\left(\frac{z}{e_3}\right)_{k}=\Big|\Big|  \Gamma_{q,t}\left(\frac{z}{e_3}\right) \Big|\Big|^3_{S}  
\end{eqnarray}
where  the elliptic gamma function $\Gamma_{q,t}$ is defined in  the Appendix \ref{PAspecfun}. We can therefore write the  partition function as:
\begin{eqnarray}\displaystyle
\label{PAgood2cft}
Z_{S^5}= \int~d \sigma~ { Z}^{S^5}_{\text{1-loop}}(\sigma,\vec M)~ 
 \Big|\Big|{\cal Z}_{\rm cl} \, {\cal Z}_\text{inst}     \Big|\Big|^3_{S}\,,
\end{eqnarray}
where ${\cal Z}_{\rm cl} $ and ${\cal Z}_\text{inst}$ are given respectively in (\ref{PAzcl}) and (\ref{PAcinst}).\\
%In the next section we will identify $ {\cal F}$ with a chiral \textcolor{blue}{${\cal V}ir_{q,t}$} block \PAFP{New:} \cite{PANieri:2013yra}.\\

The  1-loop contribution    to the partition function is expressed in terms of the triple sine function   $S_3$  defined in appendix \ref{PAtri}:
\begin{equation}
\label{PAvecs5}
{ Z}^{{S^5}}_{\text{1-loop}}(\vec \sigma,\vec M)=\frac{\prod_{\alpha>0} S_3(i\alpha(\sigma)) S_3(E+i\alpha(\sigma))}{\prod_R
\prod_{\rho\in R} S_3\left({i\rho(\sigma)}+iM+\frac{E}{2}\right)}\,,
\end{equation} 
where $\rho$ is the weight of the representation $R$.
As suggested  in \cite{PALockhart:2012vp}, using the relation (\ref{PAs3fac}),  the  1-loop contribution to the partition function  can be factorised as:
%
% the vector  and hyper multiplet contributions can be written as 
%{\begin{eqnarray}\displaystyle
%\label{PAff1}
% Z^{{S^5}, \text{vect}}_{\text{1-loop}}(\sigma)&=&\prod_{\alpha}e^{-\frac{\pi i}{3!}B_{33}(i\alpha(\sigma))}\prod_{k=1}^3(e^{\frac{2\pi i}{e_3} {[}i\alpha(\sigma){]}};q,t)_k\,,
%\end{eqnarray}}
%\begin{eqnarray}\displaystyle
%\label{PAff2}
%{Z}^{{S^5}, \text{hyper}}_{\text{1-loop}}(\sigma,M,R)&=&\prod_{\rho\in R} e^{\frac{\pi i }{3!}
%B_{33}(i\rho(\sigma)+iM+\frac{E}{2})} \prod_{k=1}^3 
% (e^{\frac{2\pi i}{e_3} [ i\rho(\sigma)+i M+\frac{E}{2} {]} };q,t)_k^{-1}\,,
%\end{eqnarray} The full 1-loop contribution to the partition function can then be  written as 
\begin{equation}
Z^{S^5}_{ \text{1-loop}}(\vec \sigma,\vec M)=\prod_R\prod_{\substack{\alpha\\
\rho\in R}}e^{-\frac{\pi i}{3!}[B_{33}(i\alpha(\sigma))-B_{33}(i\rho(\sigma)+m)]}\prod_{k=1}^3\frac{(e^{\frac{2\pi i}{e_3} [i\alpha(\sigma)]};q,t)_k}{(e^{\frac{2\pi i}{e_3}[i\rho(\sigma)+m_R]};q,t)_k}
\end{equation}
where $(z;q,t)=\prod_{i,j\geq 0}(1-zq^i t^j)$ denotes the double $(q,t)$-factorial and  the sub-index $k$ indicates
that $q,t$ defined in (\ref{PApq}) are related to  the squashing parameters according to the $k$-th entry in  table (\ref{PAp123}).
Each factor of the $k$-product can in turn be identified with the 1-loop  contribution to  the ${\mathbb R}^4\times S^1$ theory:
\begin{equation}
 {\cal  Z}_{\text{1-loop}}={\cal Z}^{\mathbb{R}^4\times S^1}_\text {1-loop}\left(\frac{i \sigma}{e_3}, \frac{ \vec m}{e_3}; 
 \frac{e_1}{e_3},\frac{e_2}{e_3}\right)=\prod_R\prod_{\substack{\alpha\\
\rho\in R}}
 \frac{(e^{\frac{2\pi i}{e_3} [i\alpha(\sigma)]};q,t)}{(e^{\frac{2\pi i}{e_3}[i\rho(\sigma)+m_R]};q,t)}\,.
\end{equation}
If we consider (pseudo) real representations, for each weight $\rho$ there is the opposite weight $-\rho$
and the sum of the  Bernoulli is a quadratic polynomial.
% which can be easily factorized in terms of elliptic Gamma functions as we did for the classical term. In fact, in this case the Bernoulli from the 1-loop factor will amounts to a renormalization of the gauge coupling constant. 
For example consider the case with $N_f$ fundamentals of mass $M_f$ and  $N_f$ anti-fundamentals of mass $\bar M_f$, with $f=1,\ldots,N_f$, and $N_a$ adjoints of mass $M_a$, $a=1,\ldots, N_a$. The total contribution from the  Bernoulli's in the 1-loop terms is (up to a $\sigma$-independent constant)
\begin{equation}
\label{PAxox}
{\rm Ber}=e^{-\frac{2\pi i}{\omega_1\omega_2\omega_3} \frac{\sum_\alpha[i\alpha(\sigma)]^2}{2C_2(ad)}\left[\frac{E}{4}N_f-\frac{1}{4}\sum_f(m_f+\bar m_f)+C_2(ad)(\frac{E}{2}(N_a-1)-\sum_a m_a)\right]}\,,
\end{equation}
which, once combined with the classical terms,  amounts to the    shift of the coupling constant
\begin{equation}
\frac{1}{g^2}\rightarrow \frac{1}{g^2_{eff}}=\frac{1}{g^2}+\frac{E}{4}N_f-\frac{1}{4}\sum_f(m_f+\bar m_f)+C_2(ad)\left(\frac{E}{2}(N_a-1)-\sum_a m_a\right)\,.
\end{equation}
Combining all these observations one arrives at the completely factorized form \cite{PANieri:2013vba}
(up to constant prefactors):
\begin{equation}
Z^{S^5}=\int d \sigma ~\Big|\Big| {\cal  B}^{5d} \Big|\Big|^3_{S}\,,
\end{equation}
where  ${\cal  B}^{5d}$, the  5d holomorphic block, is defined as
\begin{equation}
\label{PA5hb}
{\cal  B}^{5d}= {\cal  Z}_{\text{1-loop}}~{\cal  Z}_{\text{cl}}  ~{\cal Z}_{\text{inst}}\,,
\end{equation}
with ${\cal Z}_{\text{cl}} $  defined as in eq. (\ref{PAzcl})  with $g^2\to g^2_{eff}$. As in the 3d case there is an ambiguity in the definition of the 5d blocks, this is discussed  in \cite{PANieri:2013vba}.

We will now see that  the partition functions on a large class of 5-manifolds can be expressed  in terms of the 5d blocks ${\cal  B}^{5d}$.
 
\subsubsection*{Block-factorisation of 5d  toric Sasaki-Einstein  partition functions}
As reviewed in \volcite{QZ} localisation can be performed  on general simply connected toric Sasaki-Einstein (SE) manifolds $\mathcal{M}^n$. These backgrounds preserve 2 supersymmetries. 

As in the $S^5$ case it is convenient to think of  these 5-manifolds as $T^3$ fibration over the interior of a polygon, with the fiber degenerating to a torus  on the  $n$ faces and to  circe over the $n$ vertices.

The perturbative  partition function on a SE manifold $\mathcal{M}^n$ is again a Coulomb branch integral
\begin{equation}
Z^{\mathcal{M}^n}= \int~d \vec \sigma~{Z}^{{\mathcal{M}^n}}_{\rm cl}(\vec \sigma) ~ { Z}^{{\mathcal{M}^n}}_{\text{1-loop}}(\vec \sigma,\vec M)\,.
\end{equation}

 The 1-loop contribution  ${ Z}^{{\mathcal{M}^n}}_{\text{1-loop}}(\vec \sigma,\vec M)$  takes the same form as in the $S^5$ case (\ref{PAvecs5}) with  the triple-sine functions replaced by the generalised triple-sine  function  defined as:
\begin{equation}
S^{\mathcal{M}^n}_3(X)\sim
 \prod_{\vec m \in C_\mu(\mathcal{M}^n) \cap \mathbb{Z}^3}(\vec m  \cdot \vec R +X)
 (\vec m  \cdot \vec R+\vec \xi  \cdot \vec R -X)\,.
\end{equation}
In the above expression the product is over the integers in the moment map cone:
\begin{equation}
C_\mu(\mathcal{M}^n)=\left\{\vec m \in \mathbb{R}^3 | \vec m \cdot \vec v_i \geq 0, i=1,\cdots n           \right\}\,,
\end{equation}
the three-component vector $\vec R$ parameterises the Reeb vector field,
the  vector $\vec \xi$ satisfies   $\vec \xi \cdot \vec v_i=1$  for $i=1,\cdots n$ and
 $\vec v_i$ are the inward pointing normals of the $n$ faces.
% (the SE condition ensures that such vector exists).  
The constant $E$ in (\ref{PAvecs5}) is also replaced  by $E\to \vec \xi \cdot \vec R$. \\

In \cite{PAQiu:2014oqa},\cite{PATizzano:2014roa} it was derived a factorisation formula for the generalised 3-ple sine functions:
\begin{equation}
\label{PAgdf}
S_3^{\mathcal{M}^n}(X)= {\rm Ber} \prod_{k=1}^n  \left( e^{\frac{2\pi i X}{e_3}} ;  e^{\frac{2\pi i e_1}{e_3}},
e^{\frac{2\pi i e_2}{e_3}},    \right)_k,
\end{equation}
where  we denoted by ${\rm Ber} $  the contribution  of the  exponential of the cubic  Bernoulli polynomials and 
in each sector the equivariant parameters map to the toric data via the following table
\begin{eqnarray}\displaystyle\label{PAprod}
\label{PAntable}
\begin{array}{|c|ccc|}
\hline
{\rm sector}~&\quad e_1\qquad&\quad e_2\qquad&\quad e_3\qquad\\
\hline
k~&~{\rm det}[\vec R,\vec v_{k+1},\vec n] ~&~{\rm det}[\vec v_k,\vec R,\vec n] ~&~{\rm det}[\vec v_k,\vec v_{k+1},\vec R]
\\
\hline
\end{array}
\end{eqnarray}
where  $\vec n$ is chosen to satisfy  the condition ${\rm det}[\vec v_i,\vec v_{i+1},\vec n]=1$.
Proceeding as in the $S^5$ case, we apply the identity (\ref{PAgdf}) to decompose the 1-loop part into $n$-copies of  the 1-loop  Nekrasov partition function $\mathcal{Z}_{\rm 1loop}$. \\

The classical contribution is given by
 \begin{equation}
{Z}^{{\mathcal{M}^n}}_{\rm cl}(\vec \sigma)=
e^{\frac{2\pi i  r \rho}{ g^2}\,\text{Tr}(\vec \sigma^2)}
\end{equation}
where $r$ is the overall scale of $\mathcal{M}^n$ and $\rho$ is the squashed volume normalised to  $vol(S^5)=\pi^3$. By repeatedly using the Gamma function  identities (\ref{PAgamma unit}) and (\ref{PA3gammabb})
it is possible to show that
\begin{equation}
{Z}^{{\mathcal{M}^n}}_{\rm cl}(\vec \sigma)= \prod_{k=1}^n(  {\cal  Z}_{\text{cl}})_k
\end{equation}
with $ {\cal  Z}_{\text{cl}}$ defined as in (\ref{PAzcl}) and in each sector the equivariant parameters are related to the toric data according to  the  table (\ref{PAprod}).

 By collecting the Bernoulli factors ${\rm Ber} $  from the factorisation of each generalised 3-ple sine one obtains a quadratic polynomial
which produces the usual  renormalisation of the gauge coupling. This allows us to write
\begin{equation}
Z^{{\mathcal{M}^n}}_{\rm cl}(\vec\sigma) ~ { Z}^{{\mathcal{M}^n}}_{\text{1-loop}}(\vec \sigma,\vec M)=
 \prod_{k=1}^n  \left(\mathcal{Z}_{\rm cl}\mathcal{Z}_{\rm 1loop}\right)_k
\end{equation}
where $\mathcal{Z}_{\rm cl} $ is defined as in eq. (\ref{PAzcl}) with $g^2\to g^2_{eff}$.\\

In   \cite{PAQiu:2014oqa}  it has been conjectured that the full non-perturbative partition function on $\mathcal{M}^n$ would receive contributions only from instantons solutions localised around closed Reeb orbits. Around each orbit the
instanton contribution is given by a copy of the $\mathbb{R}^4\times S^1$ instanton partition function ${\cal Z}_\text{inst}$, leading  to the following fully factorised   proposal
\begin{equation}
\label{PAfull5}
Z_{full}^{\mathcal{M}^n}=\int  \prod_{k=1}^n ( \mathcal{B}^{5d})_k\,.
\end{equation}
Proving this conjecture would  require a careful study of the contact instanton equation which are quite difficult to analyse. At the moment we cannot rule out the possibility that other solutions will contribute to the full partition function.

In 3d and 4d,  partition functions  on the lens space $S^3/\mathbb{Z}_r$ and 
on the  lens index $S^3/\mathbb{Z}_r\times S^1$ admit a block factorised form only when all the flat connections are summed over \cite{PANieri:2015yia}. The contribution of a fixed flat connection to the partition function is not factorised.   In the  $\mathcal{M}^n$ case instead, as we have just seen, the perturbative result in the trivial flat connection is already factorised. 
This fact  could be a hint that the proposal (\ref{PAfull5}) is indeed complete or perhaps just an accident.
In conclusion further studies are necessary to shed light on this point.

\subsubsection*{$S^4\times S^1$ partition functions and 5d holomorphic blocks}

The next case we consider is 5d index or $S^4\times S^1$ partition function.
${\cal N}=1$  gauge theories  on this background have been  studied  in \cite{PAKim:2012gu,PATerashima:2012ra,PAIqbal:2012xm}.
The partition function takes the form of a Coulomb branch integral 
\begin{eqnarray}\displaystyle
\label{PAs1s4}
Z^{S^4\times S^1}&=&\int d\vec\sigma ~ {Z}^{S^4\times S^1}_{\text{1-loop}}(\vec \sigma,\vec M)~  { Z}^{S^4\times S^1}_{\text{inst}}(\vec \sigma,\vec M) \,,
\end{eqnarray}
with instantons  contributions   from the fixed points at north and south poles of the
 $S^4$:
\begin{equation}
 Z^{S^4\times S^1}_\text{inst}=\prod_{k=1}^2  ({\cal Z}_\text{inst})_k=\Big|\Big|{\cal Z}_\text{inst}\Big|\Big|^2_{id} \,.
\end{equation}
Each pole contributes with a copy of the $\mathbb{R}^4\times S^1$ partition function ${\cal Z}^{}_{\rm inst}$ (\ref{PAcinst})
with $m=i M + Q_0/2$ and $Q_0=b_0+1/b_0$ and the following identification
\begin{eqnarray}\displaystyle\label{PAidqt}
\begin{array}{|c|ccc|}
\hline
{\rm sector}&\quad e_1\quad&\quad e_2\quad&\quad e_3\quad\\
\hline
1&b_0^{-1}&b_0&2\pi i/R\\
2&b_0^{-1}&b_0&-2\pi i/R\\
\hline
\end{array}\nonumber
\end{eqnarray}
where $R$ is  the circumference  of $S^1$ and  $b_0$   the squashing parameter  of  $S^4$.

Due to the property   (\ref{PAgamma unit})  of the elliptic Gamma function,  the classical term  ${\cal  Z}_{\text{cl}}$ defined in eq. (\ref{PAzcl})
{``squares"} to one $\Big|\Big| {\cal  Z}_{\text{cl}}\Big|\Big|_{id}^2=1$, we can therefore   write 
\begin{eqnarray}\displaystyle
\label{PAidcft2}
Z_{S^4\times S^1}&=&\int d \sigma ~{ Z}^{S^4\times S^1}_{\text{1-loop}}(\sigma,\vec M) ~ \Big|\Big| {\cal Z}_{\rm cl} \, {\cal Z}_\text{inst}\Big|\Big|^2_{id}\,.
\end{eqnarray}
 The {1}-loop contributions of vector   and hyper multiplets are given  by
\begin{equation}
{ Z}^{{S^4\times S^1}, \text{vect}}_{\text{1-loop}}(\vec \sigma)=\frac{\prod_{\alpha>0}\Upsilon^R\left(i\alpha(\sigma)\right)\Upsilon^R\left(-i\alpha(\sigma)\right)}{\prod_{\rho\in R}\Upsilon^R\left({i\rho(\sigma)+iM}+
\frac{Q_0}{2}\right)}\,,
\end{equation}
where the special function $\Upsilon^R$  is defined in appendix \ref{PAupb}. 
Also in this case it is possible to bring the {1}-loop term in a factorized form hence 
 the 5d index can be factorised in terms  of the same 5d blocks  ${\cal  B}^{5d}$
 we found in the $S^5$ case  \cite{PANieri:2013vba}:
 \begin{equation}
Z^{S^4\times S^1}=\int d \sigma  ~\Big|\Big| {\cal  B}^{5d}\Big|\Big|^2_{id}\,.
\end{equation}

\subsection{Codimension-two defects via Higgsing }\label{PAsecdeg}

Codimension-two half BPS defects in SUSY gauge theories, such as surface operators in 4d, are an important class of non-local operators which can be used to probe the dynamics of gauge theories. For a  recent review  on surface operators see \cite{PAGukov:2014gja} and \volcite{HO}. 
They can be defined by prescribing a singular behaviour of the fields on the codimension-two locus where  defect operators live, this has the effect of breaking  the gauge group $G$ to a Levi sub-group $L$.
Another possibility to define defects in gauge theories  is by a
coupled system with  extra degrees of freedom leaving on the defects.
A related construction,  the so called Higgsing procedure,  introduced in \cite{PAGaiotto:2012xa}  involves turning on a position dependent vev or vortex configuration in a UV theory $T'$ and following the RG flow  to an IR point described by a   theory $T$. This construction should be indeed equivalent to coupling the 4d gauge fields to a 2d sigma model with target space the vortex moduli space.

\begin{figure}[!ht]
\begin{center}
\includegraphics[width=0.8\textwidth]{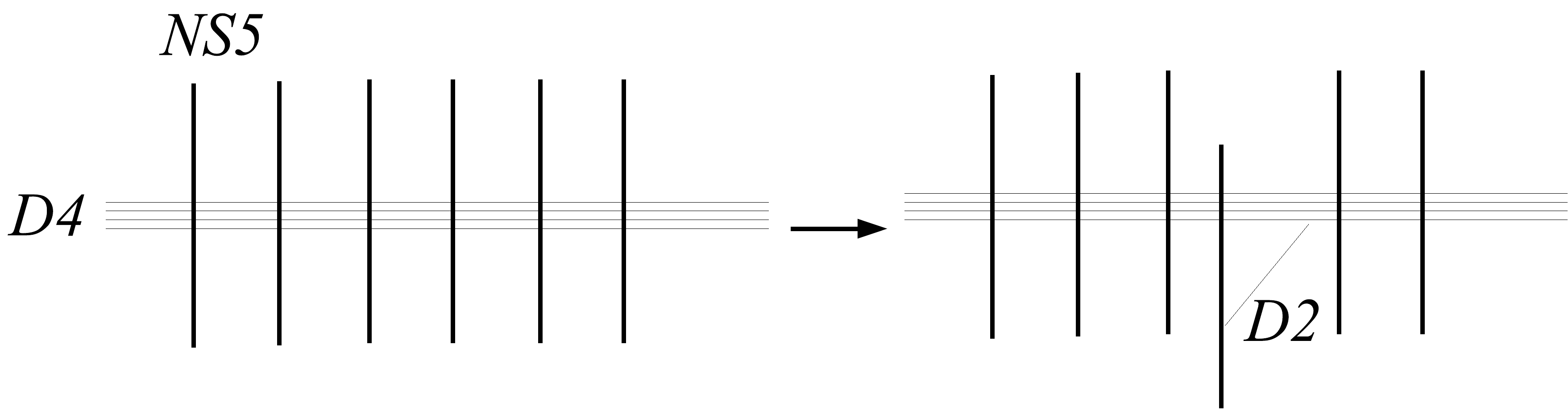}
\caption{The Hanany-Witten brane setup for a linear quiver. On the RHS a surface operator corresponding to a position-dependent vev is engineered by an extra D2 brane.}
\label{PAbarbrak}
\end{center}
\end{figure}

The Higgsing prescription gives rise to a large class of $\mathcal{N}=2$ and  $\mathcal{N}=1$ theories with  surface operators insertions. Some of these systems  can be realised in  Hanany-Witten brane set ups with  surface operators  engineered by extra branes  \cite{PAHanany:2003hp} as show in Fig. \ref{PAbarbrak} and 
admit a description in terms of a 2d GLSM coupled to the bulk theory \cite{PAHanany:1997vm}. 
Although this is not the most general type of surface operator, we will restrict to this class in the following.

In recent years there has been much progress in computing partition functions and indices of theories with the insertion of these  operators. In 4d one can compute the superconformal index ($S^3\times S^1$ partition function) of a theory $T$ enriched by a surface operator via  Higgsing by tuning the flavour fugacities of the theory $T'$ to  special values. This causes the integration contour to pinch a set of poles. The index is then evaluated by taking the residue at these poles and the result yields the index of theory $T$ enriched by a surface operator. 
If the surface operators admit a description in terms of a 2d GLSM coupled to the bulk theory \cite{PAHanany:1997vm}  the index for the coupled 2d-4d system is decorated with the elliptic genus of the  2d GLSM \cite{PAGadde:2013ftv}.

It is also possible to compute the $\mathbb{R}^4_{\epsilon_1,\epsilon_2}$ instanton partition function for 4d theory 
 in presence   of a surface defect. This takes the form of a double expansion (see for example  \cite{PADimofte:2010tz})
\begin{equation}
\sum_{n=0}^\infty \sum_{m\in \Lambda}  Q^n z^m       Z^{inst+vortex}_{n,m}\,,
\end{equation}
where $n,Q$ are  respectively  the instanton number and the instanton counting parameter, 
$m\in \Lambda\sim H_2(G/L;Z)$  (where $L$ is the Levi subgroup of the gauge group $G$) are the monopole numbers, and $z$  is the  vortex counting parameter.
 By decoupling the bulk theory (sending $Q\to 0$ and focusing on the $n=0$ sector) one gets the purely 2d vortex counting partition function.
 
\begin{figure}[!ht]
\begin{center}
\includegraphics[width=0.9\textwidth]{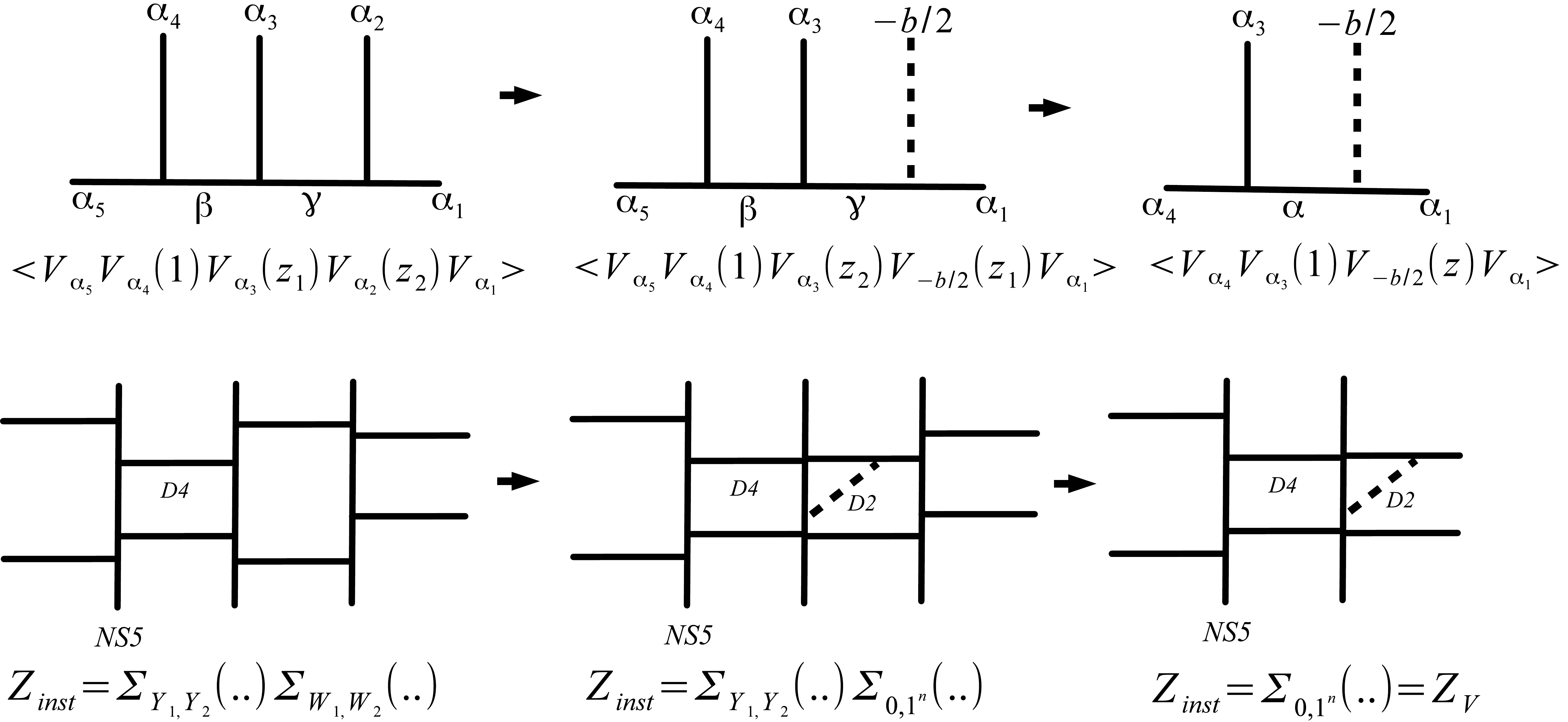}
\caption{Higgsing a linear quiver and dual CFT interpretation. The first column represents an $2+SU(2)\times SU(2)+2$ linear quiver and its dual CFT 5-point block, the instanton partition functions involves summing over two 2-vectors of Young  tableaux associated to the two $SU(2)$ nodes. The second column represents the analytic continuation of the mass/momenta to the  Higgsing/degeneration condition. Correspondently the 
sum over the second vector of Young tableaux reduces to a sum over a column-diagram. In the third column the bulk theory is decoupled and the  instanton partition  function reduces to a vortex  partition function which is mapped to the conformal block with 3 non-degenerate and one degenerate-primaries.}
\label{PAsurfopfig}
\end{center}
\end{figure}

 At the level of the Hanany-Witten brane setup, as shown in the second line of Fig. \ref{PAsurfopfig}, decoupling the bulk theory amounts to sending to infinity all the NS5 branes far  from the insertion point. 
 The 4d theory is then just a collection of free 4d hypers coupled to  the 2d theory on the stretched D2 brane.
 Also in this case the combined instanton-vortex counting partition function can be obtained via Higgsing, by
tuning the mass parameters to special values  \cite{PAMironov:2009qt}.\\

Another option to compute the  instanton-vortex partition function is  geometric engineering, where 
$\mathcal{N}=2$ gauge theories are obtained  via type II strings compactified on toric Calab-Yau threefolds. 
The refined A-model topological string partition function on these targets
\cite{PAAganagic:2003db},\cite{PAIqbal:2007ii}
 coincides with the
 $\mathbb{R}^4_{\epsilon_1,\epsilon_2}\times S^1$ instanton partition function, 
while  open topological strings amplitudes in presence of  toric branes reproduce  the instanton-vortex counting \cite{PADimofte:2010tz}, \cite{PAKozcaz:2010af}, \cite{PABonelli:2011wx}, \cite{PABonelli:2011fq}.\\

In the context of the AGT correspondence \cite{PAAlday:2009aq}, two types of surface  defects in 
class $\mathcal{S}$ theories were discussed. The first type is realised by  M2  branes with two directions extending in the 4d space-time and  sitting at a point  on the Gaiotto curve $\mathcal{C}_{g,n}$. The second type of surface operators  is instead realised   by  M5  branes
 wrapping  $\mathcal{C}_{g,n}$ and  with two directions extending in the 4d space-time.
In  \cite{PAAlday:2009fs} it was proposed  to relate the first type class of surface operators to degenerate primary operators in the dual CFT. 
Indeed it  was later shown that it is possible to match the instanton-vortex partition function for these surface operators  to the conformal blocks with degenerate insertions as sketched in Fig. \ref{PAsurfopfig}, \cite{PADimofte:2010tz}, \cite{PAKozcaz:2010af}, \cite{PABonelli:2011fq}. \\

The compact space version of the AGT+surface operators  correspondence was proposed in  \cite{PADoroud:2012xw}, \cite{PAGomis:2014eya} where it was  shown that degenerate correlators  can be mapped to  partition functions of class $\mathcal{S}$ theories  on $S^4$,  coupled to a 2d GLSM describing  the  surface defects defined on   $S^2$.

As a simple example we consider the $SU(2)$ theory with $4$ flavours on $S^4$.
 When a combination of the masses is tuned to the ``Higgsing condition"
 $m_1+m_2=- b_0$, where $b_0$ is the squashing parameter of the ellipsoid,
 the  integration contour pinches two poles. The sum of the  residue of the partition function at these two points can be identified  with the $S^2$ partition function of the $(2,2)$ SQED with  2 flavours multiplied by the
 contributions of the free 4d hypers.
 Via the AGT dictionary, which  relates the mass parameters to the four external momenta in the CFT correlator, 
 the Higgsing condition  is translated into the analytic continuation of one  momentum
 to a degenerate value $\alpha_2\to -b_0/2$.\footnote{Via the AGT dictionary
the  squashing parameter $b_0$ is identified with the parameter 
appearing in central charge $c=1+6 (b_0+b_0^{-1})^2$ of the dual CFT.} 
Summarising we have the following web of correspondences:\footnote{The expressions obtained via Higgsing/degenerations have been normalised respectively by the contribution of 4d free hypers and by the 3-point function of the  non-degenerate primaries.}
 \begin{eqnarray}\displaystyle\nonumber
&Z^{S^4}[2+SU(2)+2]~ ~\overset{\text{Higgsing}}{\longrightarrow} ~~Z^{S^2}[U(1)+2]&\\
& ~~AGT\updownarrow ~~~~~  \quad\quad \qquad\qquad ~~ ~~~AGT\updownarrow&\nonumber\\
~~&\qquad ~\quad\langle  V_{\alpha_1}V_{\alpha_2}V_{\alpha_3}V_{\alpha_4}\rangle~\overset{\text{Degeneration}}{\longrightarrow} ~\langle  V_{\alpha_1}V_{-b_0/2}
V_{\alpha_3}V_{\alpha_4}\rangle\,.&\nonumber
\end{eqnarray}

We conclude our digression on surface operators in 4d theories by mentioning that 
for surface operators realised in terms of M5 branes the standard instanton moduli space is replaced by  the Òramified instantonsÓ moduli space and the CFT duals, studied for example in \cite{PAAlday:2010vg},\cite{PAKozcaz:2010yp}   and  recently in \cite{PAFrenkel:2015rda}, have affine $sl(N)$ symmetries. More general surface operators and CFT duals have been studied in \cite{PAWyllard:2010rp} and \cite{PAWyllard:2010vi}, for a review see \cite{PATachikawa:2014dja}.\\

Codimension-two half BPS defects can be introduced  also in 5d $\mathcal{N}=1$ theories. In \cite{PAGaiotto:2014ina} the Higgsing prescription has been applied to obtain the  5d index ($S^4\times S^1$ partition function),
decorated by the  3d index ($S^2\times S^1$ partition function) of  codimension-two defects.
Here  we will consider codimension-two defects defined via Higgsing on $S^5$ and $S^4\times S^1$. We will restrict  again to the simple case of free 5d hypers  coupled to the 
codimension-two defect partition function.
% respectively on $S^3$ and $S^2\times S^1$.
We consider the case of the  the $SU(2)$ theory
with four fundamental hyper multiplets with masses $M_f$,  $f=1,\ldots ,4$ on $S^5$. 
As shown in  \cite{PANieri:2013vba}, if we tune the  masses to satisfy the condition:
\begin{equation}
M_1+M_2=i (\omega_3+E) \qquad {\rm or}  \qquad m_1+m_2=-\omega_3\,,
\label{PAdeg1}
\end{equation}
where $m_j=i M_j+E/2$, the 1-loop factor develops poles which  pinch the integration 
contour. The partition functions can then be defined via a meromorphic analytic continuation which prescribes to take the residues at the  two poles  trapped along the integration path located at 
\begin{equation}
a_1=m_{1}=-m_2-\omega_3=-a_2\,, \qquad a_1=m_{1}+\omega_3=-m_2=-a_2\,.
\label{PApol1}
\end{equation} 
By analogy with the 4d case, we expect that by taking the residues at these poles,  will reconstruct
the $S_b^3$  partition function of 3d SQED with 2 flavours.

Indeed if we evaluate the instanton  partition function (\ref{PAfirstk}) at the first pole we find  that  in the  first two sectors the sum over Young tableaux degenerates respectively to the sum over a single column and single row yielding two copies of the $q$-deformed hypergeometric $\phantom{|}_2\Phi_1$. The third sector instead  gets contribution only from the empty diagrams.
In particular all the parameters enter exactly to reconstruct  the  $S$-pairing of two vortex partition functions: 
\begin{eqnarray}\displaystyle
\Big|\Big| {\cal Z}_{\rm inst}\Big|\Big|^3_{S}\quad\xrightarrow[(a_1,a_2)\rightarrow(m_{1},m_2+\omega_3)]{}\quad 
\Big|\Big| \phantom{|}_2\Phi_1(A,B;C,e^{2\pi i\frac{\omega_2}{\omega_1}};u)   
\Big|\Big|^2_S=\Big|\Big| {\cal Z}^{(1)}_V\Big|\Big|^2_S\,.\nonumber\\
\label{PAgat}
\end{eqnarray}
The coefficients $A,B,C$ of the basic hypergeometrc function depend on the 5d mass parameters, if we identify them with those of the vortex partition
function $\mathcal{Z}^{(1)}_V$, we obtain
the following dictionary between 3d and 5d masses:
\begin{equation}
\label{PA53dictio}
m^{3d}_1=-im_1\,,\quad m^{3d}_2=-im_2\,,\quad \tilde m^{3d}_1=im_3\,,\quad \tilde m^{3d}_2=im_4\,,
\end{equation}
while by matching the expansion parameters we find $i\xi = 1/\tilde g^2$.
We also identify $
\omega_2=\frac{1}{\omega_1}=b$. In complete analogy, for the other pole, located  at 
$a_1=m_{1}+\omega_3=-m_2=-a_2$  we find
\begin{eqnarray}\displaystyle
\Big|\Big| {\cal Z}_{\rm inst}\Big|\Big|^3_{S}\quad\xrightarrow[(a_1,a_2)\rightarrow(m_{1}+\omega_3,m_2)]{}\quad 
\Big|\Big|\mathcal{Z}^{(2)}_V
\Big|\Big|^2_S\,.
\end{eqnarray}

The last step is the identification of the  residue at the $i$-th pole of the $S^5$ classical and one loop contributions with  $S^3_b$  classical and one loop contributions evaluated on the $i$-th vacuum, in the end  after normalising by the contribution of the 
free 5d hypers,  one obtains the promised result:
\begin{eqnarray}
\label{PAto31}
\!\! Z^{S^5}[2+SU(2)+2]\!\!\!\!
\quad\xrightarrow[m_{1}+m_2=-\omega_3]{}   \!\!\!\!\!\!\! &&
 \sum_{i=1,2} \left({ Z}^{S^3_b}_{\text{cl}}  { Z}^{S^3_b}_{\text{1-loop}} \right)_i \Big|\Big| {\cal Z}^{(i)}_V\Big|\Big|^2_S=\nonumber\\&&=Z^{S_b^3}[U(1)+2] \,.
\end{eqnarray}

Notice that there are two extra choices for the degeneration condition, which would have
led to the same result: 
\begin{eqnarray}\displaystyle
 m_1+m_2=-\omega_{1,2}\,, \quad {\rm with } \quad \omega_{2,1}=\omega_3^{-1}=b\,.
%&& m_1+m_2=-\omega_2\,,  \quad {\rm with } \quad \omega_1=\frac{1}{\omega_3}=b\,. 
\end{eqnarray}
The three possibilities  correspond to the three  maximal squashed 3-spheres inside the squashed 5-sphere.

In a similar manner, it is possible to show that the partition function of the SCQCD on $S^4\times S^1$, 
when two of the masses satisfy the condition
\begin{equation}
m_1+m_2=-b_0\,,
\label{PApol2}
\end{equation} 
reduces to the SQED partition function on $S^2\times S^1$:
\begin{eqnarray}
\label{PAto32}
Z^{S^4\times S^1}[2+SU(2)+2]
\quad\xrightarrow[m_{1}+m_2=-b_0]{}\!\!\!\!\!\!\!\! &&
 \sum_{i=1,2} \left({ Z}^{S^2\times S^1}_{\text{cl}}  { Z}^{S^2\times S^1}_{\text{1-loop}} \right)_i \Big|\Big| {\cal Z}^{(i)}_V\Big|\Big|^2_{id}\nonumber\\&&=Z^{S^2\times S^1}[U(1)+2]\,,
\end{eqnarray}
with  the 3d angular momentum fugacity $q$ related to the 5d parameters by
 $q=e^{R/b_0}$.
Also in this case there is another possible  degeneration condition $m_1+m_2=-\frac{1}{b_0}$, 
which leads to the same result but with the identification  $q=e^{R b_0}$.
The two choices correspond to the two  maximal $S^2$ inside the squashed $S^4$.

By paralleling the 4d case  we will reinterpret the 5d-3d Higgsing (\ref{PAto31}), (\ref{PAto32}) as 
the analytic continuation of a 4-point $q$-deformed correlator to the  $q$-correlator  of three non-degenerate and one-degenerate primaries.

Before doing so in next section  we will focus on the $q$-deformation of the  chiral  version of the AGT correspondence where we review how
$\mathbb{R}^4\times S^1$ instanton  and $\mathbb{R}^2\times S^1$ vortex counting can be mapped to chiral blocks of deformed Virasoro primaries.

 \section{Chiral 5d AGT}

\subsection{Deformed Virasoro algebras and chiral blocks}\label{PAdefalg}

Deformed Virasoro  and $\mathcal{W}_{N}$  algebras  were introduced independently in
the mid 90s by various groups \cite{PAShiraishi:1995rp, PAAwata:1995zk},\cite{PALukyanov:1994re,PALukyanov:1996qs,PALukyanov:1995gs}, \cite{PAfr,PAFeigin:1995sf},  see \cite{PAOdake:1999un} for a review.
The    $\mathcal{V}ir_{q,t}$ algebra is a deformation of the Virasoro algebra, it depends on   two complex parameters $q,t$ and is generated by an infinite set of  generators $T_n$ with $n\in \mathbb{Z}$,  satisfying  the commutation relations 
\begin{eqnarray}\displaystyle
\!\![T_n \, , \, T_m]\!=\!-\sum_{l=1}^{+\infty}f_l\left(T_{n-l}T_{m+l}-T_{m-l}T_{n+l}\right)
-\tfrac{(1-q)(1-t^{-1})  (p^{n}-p^{-n})}{1-p}\delta_{m+n,0}
\end{eqnarray}
where $p=\tfrac{q}{t}$ and  the functions  $f_{l}$  are determined by  the expansion 
\begin{eqnarray}\displaystyle
f(z)=\sum_{l=0}^{+\infty}f_l z^l
=\exp \left[\sum_{l=1}^{+\infty}\tfrac{1}{n}\tfrac{(1-q^n)(1-t^{-n})}{1+p^n}
 z^n \right]\,.
\end{eqnarray}
The algebra  $\mathcal{V}ir_{q,t}$  is invariant under $(q,t)\rightarrow (q^{-1},t^{-1})$ and $(q,t)\rightarrow (t,q)$.

As in the Virasoro algebra, the  representations of $\mathcal{V}ir_{q,t}$ can be constructed using Verma modules \cite{PAShiraishi:1995rp}. The highest weight state $|\lambda\rangle$ satisfies 
\begin{eqnarray}\displaystyle
T_0 |\lambda\rangle=\lambda |\lambda\rangle,\qquad T_n |\lambda\rangle=0\quad\text{for}\quad n>0,
\end{eqnarray}
and the Verma module is constructed acting on the highest weight state $|\lambda\rangle$ with the operators $T_{-n}$  with  $n>0$.  Singular states in the Verma module can be detected using the Kac determinant. In particular
there is a level two singular vector when  $\lambda$ takes the following values
\begin{eqnarray}\displaystyle\label{PAddeg}
\lambda_1=p^{1/2}q^{1/2}+p^{-1/2}q^{-1/2},\qquad\qquad \lambda_2=p^{1/2}t^{-1/2}+p^{-1/2}t^{1/2}.
\end{eqnarray}
The states  $\lambda_1$ and $\lambda_2$ are mapped into each other by the exchange $(q,t)\rightarrow (t,q)$ and  are left invariant by $(q,t)\rightarrow (q^{-1},t^{-1})$.\\

The algebra $\mathcal{V}ir_{q,t}$ can be related to other known algebras by taking suitable limits on parameters  $q,t$ \cite{PAAwata:1996fq}. In particular the ordinary Virasoro algebra is recovered by
setting  $t=q^{\beta}$, $q=e^{\hbar} $ and expanding 
 the deformed Virasoro current
$T(z)=\sum_{n\in \mathbb{Z}} T_n z^{-n}$:
\begin{equation}
\label{PAhc}
T(z)=2+T^{(2)}\hbar^2+T^{(4)}\hbar^4+\cdots
\end{equation}
In the second term of the expansion we recognise the Virasoro current $L(z)=\sum_{n\in \mathbb{Z}} L_n z^{-n-2}$  
\begin{equation}
T^{(2)}=\beta\left(z^2L(z)  +\frac{(1-\beta)^2}{4\beta}\right)
\end{equation}
 with the identification  $\beta=-b_0^2$.
From the expansion (\ref{PAhc}) we see that to control the deformed theory we need all the
the higher  spin currents $T^{(n)}(z)$, while in the undeformed case the current $L(z)$ constraints completely the conformal blocks.
We also notice that since in the $\hbar \to 0$ limit the  $(q,t)\rightarrow (t,q)$ symmetry of  $\mathcal{V}ir_{q,t}$  reduces to the $b_0\leftrightarrow \frac{1}{b_0}$ symmetry,  it is natural to identify the states $\lambda_1$, $\lambda_2$ (\ref{PAddeg}) as the $q$-deformation of the level-two degenerate states $\alpha^{(1,2)}=-\frac{b_0}{2}$ and $\alpha^{(2,1)}=-\frac{1}{2b_0}$ of the undeformed Virasoro case.\\

Correlation functions in  2d CFTs can be computed by decomposing them into 3-point functions
and conformal blocks, by  the insertion of complete sets of Virasoro descendants.  
3-point functions can be determined by the bootstrap approach by requiring the associativity of the OPE,
which is equivalent to the modular invariance and single valuedness  of the correlators \cite{PABelavin:1984vu}.
Conformal blocks can in turn be computed as series expansions in powers of the cross ratios, 
with coefficients obtained by repeated applications  of the commutations rules of the Virasoro generators with the  primary operators such as:
\begin{equation}
\label{PAnduse}
[L_m,V_\alpha(z)]=z^{m+1} \frac{\partial }{\partial z}V_\alpha(z)+ \Delta(\alpha) (m+1) z ^{m}V_\alpha(z)\,.
\end{equation}
There is also an alternative  approach due to Dotsenko-Fateev \cite{PADotsenko:1984nm} (see also \cite{PAFelderBRST}) 
which consists in deriving a Feigin-Fuchs integral representations for conformal blocks as $n$-point functions of  Coulomb gas vertex  operators.\\

In the deformed case, the analogue of eq. (\ref{PAnduse}) is not known and so far most of the results have been obtained via the free boson integral  approach.
 In \cite{PAShiraishi:1995rp} the deformed Virasoro algebra and its free field realisation  was 
introduced   to construct singular vectors  which are eigenvectors  of the Macdonald  operator hence proportional to the Macdonald  symmetric functions.
In ordinary CFT there is an similar relation between 
 singular vectors  of  the Virasoro algebra and the  Jack symmetric functions which in turn appears in the description of  excited states in the the Calogero-Sutherland model (CSM).
Generalising the CFT-CSM correspondence to the $q$-deformed case was indeed the motivation to  study the deformed    Virasoro algebra in  \cite{PAShiraishi:1995rp}.

The free boson integral formulation was the central tool also  in the series of works  \cite{PALukyanov:1994re,PALukyanov:1996qs,PALukyanov:1995gs} which led to an independent derivation of the deformed Virasoro algebra.
%Indeed, to realize the irreducible representations of the Virasoro algebra in the Fock space, one needs to know only explicit bosonic realization of intertwining (screening) operators. It is remarkable that the explicit form of commutation relations of the Virasoro algebra is not really used.
In these works the authors  focused on the  algebra of chiral vertex operators 
\begin{equation}
\label{PAuu}
\Phi^{\nu_3~ \nu_4}_{\Delta_1}(z_1)\Phi^{\nu_4~ \nu_1}_{\Delta_2}(z_2)\Big|_{\mathcal{L}_{\nu_1}}=
\sum_{\nu_2} W_{\Delta_1\Delta_2}\left[\frac{\nu_3\nu_2}{\nu_4\nu_1}\right]
\Phi^{\nu_3~ \nu_2}_{\Delta_2}(z_2)\Phi^{\nu_2~ \nu_1}_{\Delta_1}(z_1)\,,
\end{equation}
where the primary operators 
\begin{equation}
\Phi^{\lambda~ \nu}_\Delta(z):\quad \mathcal{L}_\lambda \otimes \mathbb{C}[z] z^{\Delta_\lambda-\Delta_\nu}\,,
\end{equation}
interpolate between irreducible representation of the algebra $\mathcal{L}_\lambda $ specified by highest weight $\Delta_\lambda$.
In the Virasoro case the  matrix $W_{\Delta_1\Delta_2}$ is a constant (independent on $z_1z_2^{-1}$)  solution of  the Yang-Baxter equation.
The idea of \cite{PALukyanov:1994re} was to consider a suitable   deformation of  the bosonization construction to realise  chiral vertex operators which satisfy the commutation relations (\ref{PAuu}) with elliptic 
 $W_{\Delta_1\Delta_2}$ matrices.  Remarkably in this way they obtained the same deformed Virasoro algebra 
 considered in \cite{PAShiraishi:1995rp}.
This deformation of the Virasoro algebra was also identified as the dynamical symmetry of the  Andrews-Baxter-Forester (ABF) model in  \cite{PALukyanov:1996qs}.\\

\begin{figure}[!ht]
\leavevmode
\begin{center}
\includegraphics[width=0.5\textwidth]{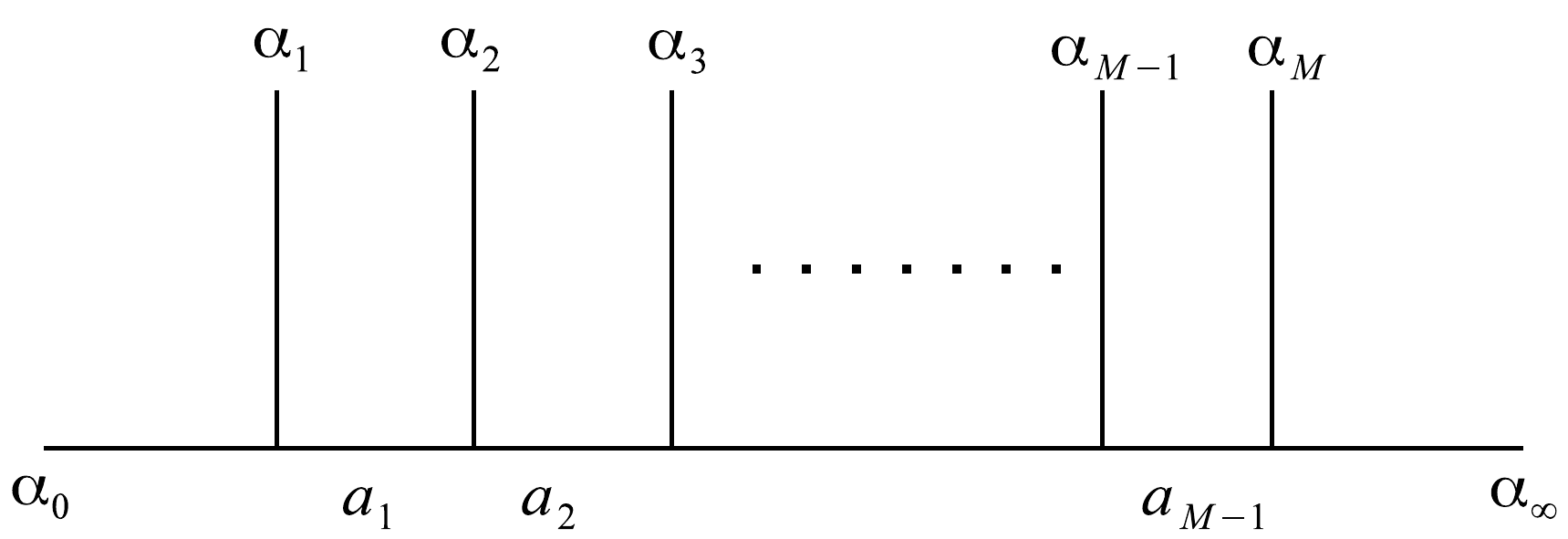}
\caption{The comb conformal diagram for the $M+2$ block.}
\label{PAcomb}
\end{center}
\end{figure}

Recently   the integral representation of the deformed Virasoro chiral blocks has been reconsidered in  \cite{PAmina1}, before reviewing this work  we record the integral representation of  ordinary Virasoro conformal blocks.
An $M+2$-point  conformal block is obtained by inserting  $M+2$ vertex operators with momenta $\alpha_a$
at positions $z_a$,  $a=0,\cdots, M+1$, with $z_{M+1}=\infty$, in the background of   $N$ screening charge
integrals
\begin{eqnarray}\displaystyle
\mathcal{B}_{M+2}=\langle \oint dw_1\cdots\oint dw_N~ V_{\alpha_0}(0)\cdots  V_{\alpha_M}(z_M)S(w_1)\cdots S(w_N)\rangle\,.
\end{eqnarray}
The vertex operators and screening charges are given in terms of the  Liouville field $\phi(z)$:
\begin{equation}
V_\alpha(z)=:~e^{-\frac{\alpha \phi(z)}{b_0}}~:\,, \qquad   S(z)=:~e^{2 b_0 \phi(z)}~:\,.
\end{equation}
It is convenient to represent conformal blocks   by means of comb diagrams as  
shown in Fig. \ref{PAcomb}.
The momentum $\alpha_{M+1}$ is determined by remaining $M+1$ momenta and by the total number of screening charges $N$
\begin{equation}
\label{PAqunatis}
\sum_{a=0}^{M+1}\alpha_a+2\beta N=2-2\beta\,.
\end{equation}
By expanding in modes the Liouville field:
\begin{equation}
\phi(z)=\frac{\phi_0}{b_0}+\frac{h_0 \log z}{b_0}+\frac{1}{b_0} \sum_{k\neq 0} h_k \frac{z^{-k}}{k}\,,
\end{equation}
and using the modes commutation rule
\begin{equation}
[h_n,h_m]=\frac{-b_0^2}{2} n \delta_{n+m,0} \,,
\end{equation}
the conformal block reduces to the following Dotsenko-Fateev (DF) integral
\begin{equation}
\label{PAunDF}
\mathcal{B}_{M+2}=\frac{1}{\prod_{a=1}^N N_a!}\oint_{\mathcal{C}_1\cdots\mathcal{C}_M} dw_1\cdots dw_N
\prod_{1\leq i\neq j\leq N} (w_i-w_j)^\beta ~\prod_{a=0}^M \prod_{i=1}^N (w_i-z_a)^{\alpha_a}\,.
\end{equation}
The integration contour is defined by splitting the $ N$ screening integrals into $M$ groups with $N_a$ screening integrals each:
\begin{equation}
N=\sum_{a=1}^N N_a\,.
\end{equation}
The contour for the $a$-th group  encircles the segments $\mathcal{C}_a$
\begin{equation}
\mathcal{C}_a: \qquad [0,z_a],\quad a=1,...,M\,.
\end{equation}
Naively the integral (\ref{PAunDF}) seems to miss some parameters. As shown in in Fig. \ref{PAcomb}, the conformal block depends in fact on the internal momenta $a_1,\cdots, a_{M-1}$.
The missing parameters are precisely the fractions of screening charges integrated along each contour
\cite{PAdv,PAItoyama:2009sc,PAEguchi:2010rf,PASchiappa:2009cc,PAmms1,PAmms2}, with 
 the identifications 
\begin{equation}
\label{PAintcon}
a_1=\alpha_0+\alpha_1+2\beta N_1\,, \qquad    a_k=a_{k-1}+\alpha_k+2\beta N_k\,,\qquad
k=2,\cdots M-1\,.
\end{equation}

In the deformed Virasoro case one  promotes the modes commutation rule to  a $q$-deformed commutation 
\begin{equation}\label{PAcomm}
[h_n,h_m]=\frac{-b_0^2}{2} n \delta_{n+m,0} \quad \to \quad [ h_n, h_m]=\frac{1}{1+(t/q)^{n}} \frac{1-t^{ n}}{(1-q^{n})}n \delta_{n+m,0}\,,
\end{equation}
and defines bosonised vertex operators and screening charges as
\begin{eqnarray}\displaystyle
\hat  V_\alpha(z)&=&:e^{\left( - \frac{\alpha}{b_0^2} \phi_0-\frac{\alpha}{b_0^2}  h_0\log z+\sum_{n\neq 0}\frac{1-q^{-\alpha n}}{n(1-t^{-n})}   h_n z^{-n} \right)}:\,\\
\hat  S(z)&=&:e^{\left( 2 \phi_0+ 2h_0 \log z+\sum_{n\neq 0}\frac{1+(t/q)^{ n}}{n} h_n z^n \right)}: \,.
\end{eqnarray}
The $q$-deformed chiral block  with  $M+2$ insertions  then   reads \cite{PAmina1},\cite{PAShiraishi:1995rp}: 
\begin{eqnarray}\displaystyle
\label{PAscc1}
\mathcal{B}^q_{M+2}=\langle \oint dw_1\cdots\oint dw_N~ \hat V_{\alpha_0}(0)\cdots  \hat V_{\alpha_M}(z_M)\hat S(w_1)\cdots \hat S(w_N)\rangle\,.
\end{eqnarray}
By using the commutation rules (\ref{PAcomm}) the  block (\ref{PAscc1}) reduces to the following 
$q$-deformed DF integral:
\begin{equation}
\label{PAmina}
\mathcal{B}^q_{M+2}=\frac{C}{\prod_{a=1}^N N_a!}\oint_{\mathcal{C}_1\cdots\mathcal{C}_M} dw_1\cdots dw_N ~\Delta_{q,t}^2(w)~\prod_{a=0}^M V_a(w,z_a)\,,
\end{equation}
where
\begin{equation}
\label{PAmcdandvec}
\Delta_{q,t}^2(w)=\prod_{1\leq i\neq j\leq N}\frac{ (w_i/w_j;q)_\infty}{(t w_i/w_j;q)_\infty}\,,\qquad
V_a(w,z_a)=\prod_{i=1}^N \frac{ (q ^{\alpha_a}z_a/w_i;q)_\infty}{(z_a/w_i;q)_\infty}\,.
\end{equation}
The  higher rank case, involving the  $q$-$\mathcal{W}_N$ algebra is studied in \cite{PAmina2}.\\

In the literature we find often another presentation of the DF integrals:
\begin{eqnarray}\displaystyle
\nonumber
\label{PAmoroDF1}
&&\mathcal{B}_{M+2}=\int_0^1 d^{N_1}w\int_0^{\Lambda_2^{-1}}d^{N_2}w\cdots\int_0^{\Lambda_2^{-1}\cdots \Lambda_M^{-1}}d^{N_M}w ~
\prod_{1\leq i\neq j\leq N} (1-\frac{w_i}{w_j})^\beta\times \\
&&\times
\prod_{i=1}^{N_1+\cdots N_M}  w_i^{\alpha_0} (1-w_i)^{v_1}(1-\Lambda_2 w_i)^{v_2} \cdots 
(1-\Lambda_2 \cdots \Lambda_M w_i)^{v_M}\,,
\end{eqnarray}
it is is easy to find a dictionary relating the parameters $v_i $ to the momenta $\alpha_i$ and the cross ratios $\Lambda_i$ to the $z_i$ in 
eq. (\ref{PAunDF}), so the two presentations are equivalent.
The DF integral  (\ref{PAmoroDF1}) can be promoted to the $q$-deformed case by replacing 
\begin{equation}
\label{PAmordef}
\prod_{1\leq i\neq j\leq N} (w_i-w_j)^\beta  \to \prod_{1\leq i\neq j\leq N}  \prod_{k=0}^{\beta}(w_i-q^k w_j)\,,
\qquad
(1-x)^c \to  \prod_{n=0}^{c-1} (1-q^n x)\,,
\end{equation}
and the integration measure by the Jackson measure $dx \to d_q x$ defined as 
\begin{equation}
\int_0^1 f(x) d_qx=(1-q)\sum_{k=0}^\infty q^k f(q^k)\,.
\end{equation}
In the  end  one arrives at the $q$-deformed DF integral  \cite{PAMironov:2011dk}:
\begin{eqnarray}\displaystyle
\label{PAmoroDF}
\nonumber
&&\!\!\!\!\!\!\!\!\!\!\mathcal{B}^q_{M+2}=\int_0^1 d^{N_1}_qw\int_0^{\Lambda_2^{-1}}d^{N_2}_qw\cdots\int_0^{\Lambda_2^{-1}\cdots \Lambda_M^{-1}}d^{N_M}_qw ~
\prod_{1\leq i\neq j\leq N} \prod_{k=0}^{\beta-1}(1-q^k\frac{w_i}{w_j})^\beta\\
&&\!\!\!\!\!\!\!\!\!\!\!\!\times 
\prod_{i=1}^{N_1+\cdots N_M}  w_i^{\alpha_0} \prod_{k=0}^{v_1-1}(1-q^k w_i)
\prod_{k=0}^{v_2-1}
(1-q^k\Lambda_2 w_i) \cdots \prod_{k=0}^{v_M-1}
(1-q^k \Lambda_2 \cdots \Lambda_M w_i)\,,\nonumber\\
\end{eqnarray}
which can be shown to be equivalent to (\ref{PAmina})  for  integer values of $\beta$ and of the  momenta.
Actually  the expressions in (\ref{PAmcdandvec}) provide the analytic continuation to non-integer values of
the products  in (\ref{PAmordef}).

\subsection{$\mathcal{V}ir_{q,t}$ chiral blocks and instanton-vortex counting}\label{PAqvirinstvo}

In this section we will discuss the map between $\mathcal{V}ir_{q,t}$ chiral blocks and  $\mathbb{R}^4\times S^1$ instantons or $\mathbb{R}^2\times S^1$  vortex partition functions.
The $q,t$ parameter appearing  in the deformed Virasoro algebra are identified with  the equivariant parameters in the instanton partition function $q=e^{R \epsilon_1}$,  $t=e^{-R \epsilon_2}$ with $R$ the radius of the $S^1$.\footnote{The parameter  $t$ in this section is the inverse of the parameter $t$ appearing in section \ref{PA5dfactos}.}

The first evidence  of this map is the very neat result of \cite{PAAwata:2009ur} (see also \cite{PAYanagida:2010vz}) where the $q$-deformed version of the so-called Gaiotto-Whittaker states \cite{PAGaiotto:2009ma}  was constructed.
The deformed analogue of the Gaiotto-Whittaker states $|G\rangle$ are states  in the 
Verma module $M_h$ satisfying:
\begin{equation}
T_1 |G\rangle =\Lambda^2 G\,, \qquad  T_n |G\rangle =0\quad (n\geq 2)\quad \Lambda^2\in \mathbb{C},
\end{equation}
and normalised such that  $|G\rangle=|h\rangle +\cdots $ where 
$|h\rangle$ is an  highest weight vector ($T_n |h\rangle=0$, for $n>0$,  $ T_0|h\rangle=h|h\rangle$, $h\in \mathbb{C}$).
 %\PASP{The coefficient $h_k$ where obtained ...bit more..}
The inner product $\langle G|G\rangle$ can be computed using the free boson realisation of $\mathcal{V}ir_{q,t}$ and it can be shown that 
\begin{equation}
\langle G|G\rangle=\mathcal{Z}^{5d}_{inst}[SU(2)]\,,
\end{equation}
where $\mathcal{Z}^{5d}_{inst}[SU(2)]$ is the  instanton partition function of the $5d$ $\mathcal{N}=1$ pure $SU(2)$
 theory and the parameter $\Lambda^2$ is identified with the instanton counting parameter.
This result has been generalised to higher rank in \cite{PATaki:2014fva}.\\

Moving to multipoint correlators, we have to rely  on the $q$-DF integral representation discussed in the previous section since, as we mentioned, the OPE approach in the $q$-deformed case is not well developed.
This fact also implies that at the moment we cannot evaluate the correlators  at generic values of external and internal momenta since  they will have to satisfy conditions like  (\ref{PAqunatis}),
hence we can only test the correspondence with 5d instanton partition functions at special points in the moduli space. 

The undeformed AGT correspondence associates to the sphere  with $M+2$ punctures the
$M+2$-point correlator on the  CFT side and the $\mathcal{N}=2$, $2+SU(2)^{M-1}+2$  linear quiver on the gauge theory side, 
one would then  expect  an analogous map between the
$M+2$-point block in $\mathcal{V}ir_{q,t}$ and the 5d $\mathcal{N}=1$, $2+SU(2)^{M-1} +2$   quiver theory.

\begin{figure}[!ht]
\leavevmode
\begin{center}
\includegraphics[width=0.8\textwidth]{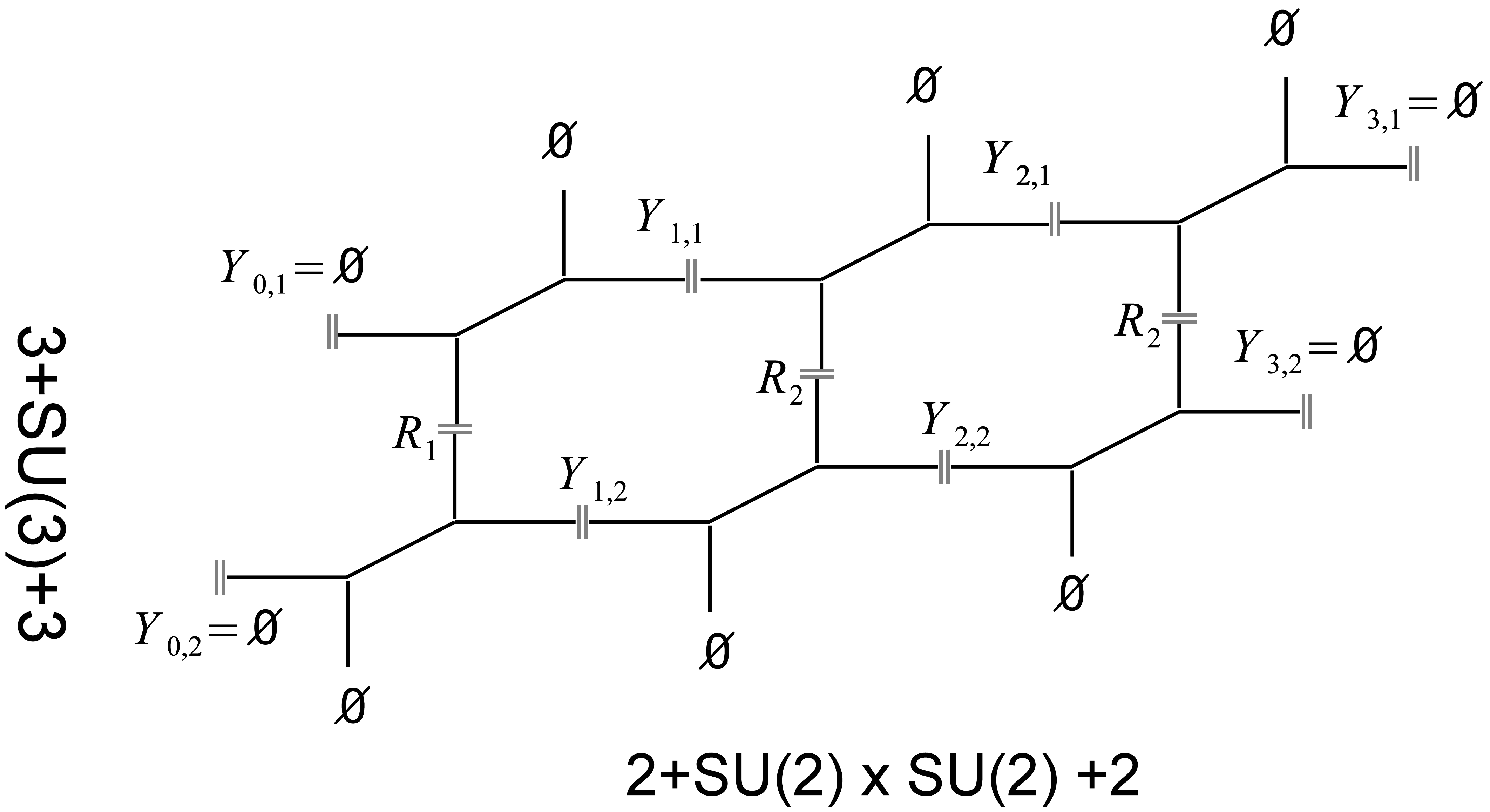}
\caption{ The toric diagram of the  CY geometry engineering either $3+SU(3)+3$ or $2+SU(2)^2+2$.}
\label{PAM3}
\end{center}
\end{figure}

However it is important to notice that  in 5d  the $2+SU(2)^{M-1} +2$ theory
is dual to the $M+SU(M)+M$  theory (more generally  the duality relates $K+SU(K)^{M-1}+K$ to $M+SU(M)^{K-1}+M$ theories) \cite{PAKatz:1997eq,PAAharony:1997bh}.
A neat way to understand this duality is to consider the geometric engineering perspective.
In fact both theories can be  engineered by a IIA compactification on the same
 toric Calabi-Yau threefold, whose toric diagram for the  $M=3$ case is depicted in Fig. \ref{PAM3}.
 The topological string partition function $Z_{top}$ on this CY  can be computed by means of the refined vertex formalism \cite{PAIqbal:2007ii}. One has to glue trivalent vertices by summing over sets of representations associated to each internal leg.
In general it is  not possible to  perform  all the sums in a closed form and one typically
gets a perturbative expansion in powers of the  K\"ahler parameters of the  legs with representations left to sum.

For example in the  case of Fig. \ref{PAM3} one can resum all the reps associated to the diagonal and vertical legs and obtain a perturbative expansion in the K\"ahler parameters of the  horizontal legs with Young tableaux $Y_{1,1},Y_{1,2}$ and $Y_{2,1},Y_{2,2}$. The result one gets in this way  can be identified
with the 5d  $2+SU(2)^{2} +2$ instanton partition function with the K\"ahler parameters of the horizontal legs mapped to 5d gauge couplings of the two nodes.

Alternatively one can resum first the diagonal and horizontal legs and obtain a perturbative expansion in the K\"ahler parameters associated to the vertical legs  with Young tableaux $R_1,R_2,R_3$.
This expansion  can be mapped to the  $3+SU(3) +3$ instanton partition function.
The fact that the two ways of performing the sum are equivalent
requires  the non-trivial  slicing invariance property  of the refined topological vertex.
The dictionary relating the topological string  K\"ahler parameters to the 
 Coulomb, gauge coupling  and mass parameters in two dual gauge theories can be found in \cite{PABao:2013pwa}. \\
 
As a result of the discussion  above, the  $M+2$-point $q$-DF block is expected to be related to both these 5d theories. As we will see it turns out that there are two distinct   evaluation methods which yield the two instanton expansions.

In  \cite{PAmina1} it was devised a procedure to evaluate the $q$-DF $M+2$-point block  (\ref{PAmina}) by residues computation. The integrand in   (\ref{PAmina}) is a meromorphic function,
poles come from the zeros of the $q$-products $\prod_{k=1}^\infty(1- q^k z)$ in the denominator located at $z=q^{-k}$.
By taking into account carefully that certain poles are cancelled by the zeros of the $q$-products in the numerator,
it can be shown  that the poles enclosed by each integration contour $\mathcal{C}_a$
where  $N_a$ screening charges are integrated, are labelled by Young tableaux $R_a$ with at most  $N_a$  rows.
Basically one finds:
\begin{equation}
\frac{1 }{N_1!}\oint_{\mathcal{C}_1} dw^{N_1}\cdots \frac{1 }{N_M!}\oint_{\mathcal{C}_M} dw^{N_M}
\rightarrow 
\sum_{R_1,\cdots, R_M}\,.
\end{equation}
By taking the residues at these sets of poles in (\ref{PAmina}) the result very nicely organises as the
5d Nekrasov partition function for the $M+SU(M)+M$ theory:
\begin{equation}
\mathcal{B}^q_{M+2}=
\sum_{R_1,\cdots, R_M}   \Lambda^{\sum_a |R_a|}  ~
\frac{z_{fund}(\vec R) z_{anti-fund}(\vec R) }{ z_{vec}(\vec R)} 
= Z_{inst}^{5d}[M+SU(M)+M].
\end{equation}
where $\vec R=(R_1,\cdots R_M)$.
For details on how the gauge theory parameters are mapped to the $q$-DF we refer the reader to \cite{PAmina1}, \cite{PAminarev}.
Notice that, since as discussed above the sets of representations are non-generic and the 
mass and Coulomb parameters  via the dictionary are identified with momenta satisfying the conditions  (\ref{PAintcon}),  the 5d theory is  at a special point in the moduli space where the Higgs branch and the Coulomb branch meet at the origin. The non-restricted 5d theory was conjectured to emerge via geometric-transition in the large N limit.

In \cite{PAmina1}, \cite{PAmina2}, it was also observed  that the $M+2$-point block (\ref{PAmina}) can be directly identified with the block integral  $Z^{3d}$, discussed in section \ref{PA3d4dsection}, of a  3d   $SU(N)$ theory  with $2M$ flavours and one adjoint.
Indeed  the screening charge contributions $\Delta_{q,t}^2(w)$ and the vertices $V_a(w,z_a)$  in eq. (\ref{PAmcdandvec})  can be respectively identified with the vector and hypermultiplets contribution to the 3d block integral. In the end one finds a triality relation:
\begin{equation}
\mathcal{B}^q_{M+2}=Z^{3d}= Z_{inst}^{5d}[M+SU(M)+M].
\end{equation}
The 3d theory is interpreted as the  theory  studied in \cite{PAHanany:2003hp}, \cite{PAHanany:2004ea}, on charge N vortices in the  5d gauge theory.

The relation between free field correlators in $q$-Virasoro  and 3d gauge theories partition functions has been recently discussed in \cite{PANedelin:2016gwu}. The authors showed that it is possible  to map 3d $\mathcal{N}=2$ partition functions on the  3-manifolds  $M^{3d}_g$   to free field correlators of the $q$-Virasoro modular double.\\

We now turn to the second evaluation method of the $q$-DF integrals
yielding   the  dual instanton   expansion. We begin by recording the schematic form of the instanton  expansion  of the $2+SU(2)^{M-1}+2$ quiver theory:\begin{eqnarray}\displaystyle
\label{PAquiver}\!\!\!\!\!\!\!\!\!\!\!\!
& Z_{inst}^{5d}[2+SU(2)^{M-1}+2]=\sum_{\vec Y_a} \Lambda_1^{|\vec Y_1|}\cdots \Lambda_{M-1}^{
 |\vec Y_{M-1}|} &\nonumber\\\!\!\!\!\!\!\!\!\!\!\!\!
&\displaystyle{ \frac{z_{fund}(\vec Y_1)        z_{bifund}(\vec Y_1,\vec Y_2) 
\cdots z_{bifund}(\vec Y_{M-1},\vec Y_{M-1})z_{anti-fund}(\vec Y_{M-1})}{z_{vec}(\vec Y_1)
z_{vec}(\vec Y_2) \dots z_{vec}(\vec Y_{M-1})}}\,,&
\end{eqnarray}
where $\vec Y_a$ is a two component Young Tableau and $\Lambda_1\cdots \Lambda_{M-1}$  the gauge couplings. In the 4d/undeformed case this structure suggested to  look for an analogous   decomposition of conformal blocks on a basis of states labelled by the Young tableaux $\vec Y_a$,  
 this is the so-called Nekrasov  decomposition of the conformal blocks:
\begin{eqnarray}\displaystyle \nonumber
\!\!\!\!\!\!&\langle V_{\alpha_0}(0)V_{\alpha_1}(1)V_{\alpha_2}(\alpha_2)\cdots
V_{\alpha_M+1}(\infty)    \rangle\sim&\\\nonumber&
\sum_{\vec Y_a} \langle V_{\alpha_0}(0)|V_{\alpha_1}(1)|  \tilde \alpha_1, \vec Y_1\rangle
\langle   \tilde \alpha_1, \vec Y_1  |  V_{\alpha_2}(\Lambda_2) |\tilde \alpha_2, \vec Y_2    \rangle \cdots&\\& \cdots
\langle   \tilde \alpha_{M-1}, \vec Y_{M-1}  |  V_{\alpha_M}(\Lambda_M) |V_{\alpha_M+1}(\infty)    \rangle\,,&\
\end{eqnarray}
where the symbol  $\sim$ is due to the omission of the so-called $U(1)$ factor that one needs to strip-off from the Nekrasov partition function in order to match with the CFT results. This factor plays a crucial role  in \cite{PAAlba:2010qc}, \cite{PABelavin:2011sw} where this basis 
was identified as a special orthogonal basis of states for the algebra  $Vir \otimes \mathcal{H}$, the  tensor product of Virasoro and Heisenberg algebras.
Besides rendering much simpler  the evaluation of the coefficients of the OPE 
this basis  has a clear interpretation as the class of fixed points in the equivariant cohomology  of the instanton moduli space.\footnote{ The action of the  $W_N$ algebra on the instanton moduli space  is discussed 
\cite{PAMO}, \cite{PASV}.}

Unfortunately lifting directly this approach to the deformed case is problematic since, as we already mentioned, the OPE approach in the deformed case is not know, however one  can try to find a similar decomposition of  the $q$-DF integrals in terms of these states.  This idea been developed  in a series of papers \cite{PAmms1,PAmms2,PAMironov:2010pi,PAMorozov:2013rma}.
The authors  initially focused on the 4-point block in the undeformed Virasoro case which can be schematically expressed as:
\begin{equation}
\mathcal{B}_{2+2}=\int_{\mathcal{C}_1} d\mu(x)\int_{\mathcal{C}_2}  d\mu(y) ~\mathbb{I}^2(x,y)\,,
\end{equation}
where  $x=(x_1,\cdots x_{N_1})$ and $y=(y_1,\cdots y_{N_2})$ indicate the two sets of screening charges variables integrated  on the first and second contours respectively.  We refer the reader to \cite{PAMorozov:2013rma} for the explicit
definition of the factors $d\mu(x)$, $d\mu(y)$, which can  be identified with the so-called
Selberg measure, and of the cross term $\mathbb{I}^2(x,y)$.
The idea was to express the cross term by means of the completeness of a set of new orthogonal  polynomials  $K_{\vec Y}(x)$ labelled by  Young tableux $\vec Y=Y_1,Y_2$.
These polynomials are  a generalisation of Jack polynomials depending on two reps which, in the 
$\beta=1$ case, reduce to the  products of two  Schur polynomials $K_{\vec Y}(x)=\chi_{Y_1} \chi_{Y_2}$.
Using the completeness of these polynomials 
\begin{equation}
 \mathbb{I}^2(x,y)=\sum_{\vec Y} \Lambda^{|\vec Y|} K_{\vec Y}(x)K^{*}_{\vec Y}(y)\,,
\end{equation}
the  4-point function can be expressed as a double Selberg average:
\begin{equation}
\mathcal{B}_{2+2}\!=\!\int_{\mathcal{C}_1} \! d\mu(x)\int_{\mathcal{C}_2}\! d\mu(y) ~\mathbb{I}^2(x,y)=\sum_{\vec Y} \Lambda^{|\vec Y|} \!
\int_{\mathcal{C}_1} d\mu(x)  K_{\vec Y}(x)\!\int_{\mathcal{C}_2}  d\mu(y)  K^{*}_{\vec Y}(y)\,.
\end{equation}
The  explicit evaluation of these Selberg averages 
remarkably yields 
\begin{equation}
\int_{\mathcal{C}_1} d\mu(x)  K_{\vec Y}(x)~\int_{\mathcal{C}_2}  d\mu(y) K^{*}_{\vec Y}(y)=
\frac{z_{fund}(\vec Y) z_{antifund}(\vec Y)}{z_{vec}(\vec Y)}\,,
\end{equation}
hence one reconstructs the  instanton expansion 
for the $2+SU(2)+2$   theory:
\begin{equation}
\mathcal{B}_{2+2}=
\sum_{\vec Y} \Lambda^{|\vec Y|} ~
\frac{z_{fund}(\vec Y) z_{antifund}(\vec Y)}{z_{vec}(\vec Y)}=Z^{4d}_{inst}[2+SU(2)+2]
\,.
\end{equation}
This result has been generalised to the $q$-deformed case for  $q=t$ in \cite{PAMironov:2011dk}
and then for generic $q,t$ in \cite{PAZenkevich:2014lca}.
In this latter case one needs to introduce a new set of polynomials, combinations of Macdonalds polynomials
to prove that 
 \begin{equation}
\mathcal{B}^q_{2,2}= Z^{5d}_{inst}[2+SU(2)+2]\,.
 \end{equation}
The generalisation to  multipoint blocks has been studied in  \cite{PAMorozov:2015xya}.
We refer the reader to the original paper, here we record only the key steps leading to the final result.
First one introduces an object called {\it generalised bifundamental  kernel}  $ \widetilde N_{\vec Y_{a-1} ,\vec Y_{a}}$.
For $q=t$ this kernel  admits a factorised form  in terms of skew Schur polynomials:
\begin{equation}
 \widetilde N_{\vec Y_{a-1} ,\vec Y_{a}}=
N_{Y_{a-1,1} Y_{a,1}} N_{Y_{a-1,2} Y_{a,2}}\,, \qquad N_{A,B}[x]=\sum_C \chi^*_{A/C}[y] \chi_{B/C}[y] \,.
\end{equation}
The first result is that one can express the $M+2$-point block in terms of $q$-Selberg averages of the generalised bifundamental kernels:
\begin{equation}
\mathcal{B}^q_{M+2}=
\sum_{\vec Y_a} \prod_{a=1}^{M} \Lambda_a^{|\vec Y_a|}
 \langle\widetilde N_{\vec Y_{a-1} ,\vec Y_{a}}\rangle\,,
\end{equation}
with $\vec Y_0=\vec Y_M=\emptyset$.
One can then  prove that:
\begin{equation}
\langle \widetilde N_{\vec Y_{a-1} ,\vec Y_{a}}\rangle\sim
\frac{z_{bif}[\vec Y_{a-1}, \vec Y_{a}]   }{   z_{vec}(\vec Y_{a-1})^{1/2} z_{vec}(\vec Y_{a})^{1/2}  }\,,
\end{equation}
which, by considering the form of the quiver instanton partition function in  (\ref{PAquiver}), leads to
\begin{eqnarray}\displaystyle
\mathcal{B}^q_{M+2}=Z_{inst}^{5d}[2+SU(2)^{M-1}+2]\,.
\end{eqnarray}
This provides a proof of the ``standard" 5d lift of the AGT correspondence. 
However \cite{PAMorozov:2015xya} managed to prove  that there is a finer structure.
They showed that the  $q$-DF integral can in fact be directly decomposed in terms of 
resolved conifold topological  string amplitudes $[Z_{top}]^{R_1,R_2}_{Y_1,Y_2}$, where $R_{1,2}$ and $Y_{1,2}$ are the representations respectively  carried by the external  vertical and horizontal legs. In particular the average of the bifundamental kernel decomposes as:

\vspace{.80cm}

\begin{eqnarray}\displaystyle
\hspace{-1.8cm}   \nonumber\langle \widetilde N_{\vec Y_{a-1} ,\vec Y_{a}}\rangle\!
\sim\!
\sum_R Q_f^{|R|} [Z_{top}]^{0,R^T}_{Y_{a-1,1},Y_{a,1}}
 [ Z_{top}]^{R,0}_{Y_{a-1,2},Y_{a,2}} =
\end{eqnarray}
\vspace{-3.30cm}
\begin{figure}[!ht]
\leavevmode
\begin{center}
\hspace{9.2cm}\includegraphics[width=0.25\textwidth]{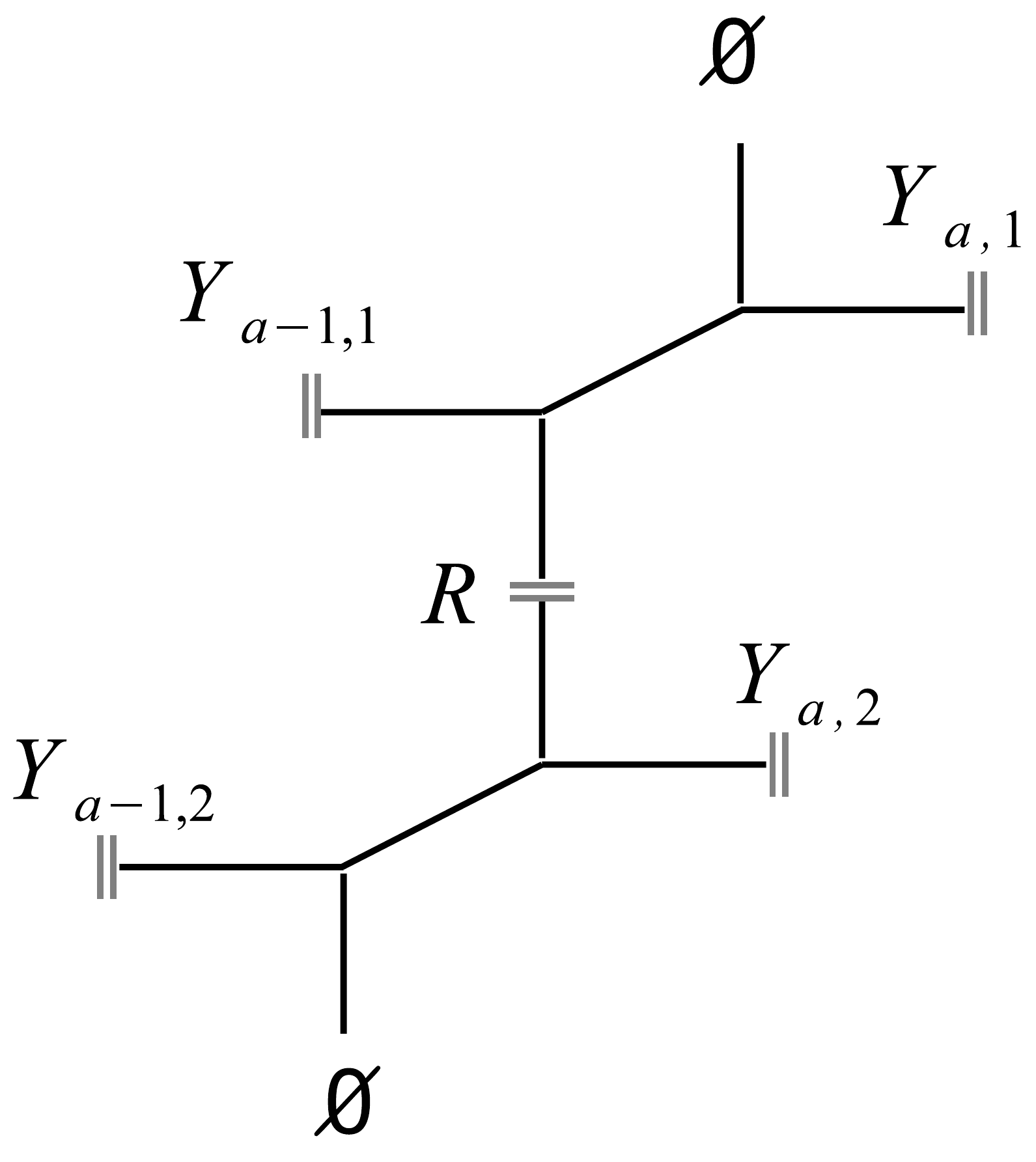}
\end{center}
\end{figure}

\vspace{-.40cm}

\noindent 
The sum over the representation $R$ is the result of the $q$ average. As  we have already mentioned
the integrands are typically  meromorphic functions and integrals  are evaluated by  summing sets of poles labelled by Young tableaux (with finite number of rows).
The K\"ahler parameter $Q_f$, associated to the internal vertical leg carrying the representation $R$, is related to Coulomb branch parameter.
In conclusion   the $M+2$-point  block can be directly mapped to the topological string 
partition function for the CY geometry  depicted (for  $M=3$)  in Fig. \ref{PAM3}:
\begin{equation}
\mathcal{B}_{M+2}=\sum_{\vec Y_a} \sum_{R_a}  \prod_{a=1}^M\Lambda_a^{|\vec Y_a|}
Q_{f_a}^{|R_a|}  [Z_{top}]^{0,R_a^T}_{ Y_{a-1,1},Y_{a,1} }
[Z_{top}]^{R_a,0}_{Y_{a-1,2},Y_{a,2}}\,,
\end{equation}
with $\vec Y_0=\vec Y_M=\emptyset$. 
This is the most fundamental decomposition of the $q$-deformed  DF blocks.

Recently it has been observed  that DF and gauge theory matrix integrals can be considered as special cases of a more  general  class of matrix models,  the so-called network matrix models which have a direct topological string interpretation.
The symmetry of these matrix models is the  Ding-Iohara-Miki algebra which has been conjectured to be 
the underlying symmetry of the AGT correspondence,  see for example \cite{PAAwata:2016riz} and references therein.\\

We close this section by mentioning  a further approach to the evaluation of (deformed) DF-integrals.
In \cite{PASchiappa:2009cc} it was shown that  $q$-deformed   blocks involving 3-generic primaries operators 
plus any number of operators  with specialised  momentum (corresponding to  level-2 degenerates) are given by multivariate basic hypergeometric series.
For example it  is easy to see that  if in the $q$-DF integral  (\ref{PAmina}) we take  arbitrary $\alpha_0,\alpha_1,\alpha_{M+1}$
and specialise  $\alpha_i=-1$ for $i=2,\cdots, M$ such that the corresponding vertices become
\begin{equation}
V_i(w,x_i)=\prod_{j=1}^N \frac{ (q ^{\alpha_i}x_i/w_j;q)_\infty}{(x_i/w_j;q)_\infty}\,
\to  \prod_{j=1}^N (1-q^{-1}x_i/w_j),
\end{equation}
the integral  can be mapped to the Jackson integral studied in \cite{PAkane2}.\footnote{The measure in \cite{PAkane2} is different from the Macdonald measure in (\ref{PAmina}) but they actually give the same results up to prefactors independent on $x$, see for example the discussion in Appendix C of \cite{PAAwata:2010yy}. } 
This integral can be exactly evaluated and yields  (up to prefactors)
a  basic hypergeometric functions of $M$ variables:
$$
\mathcal{B}_{M+2}(x_1,\cdots,x_M)\sim
{}_{2}\Phi_{1}(A,B;C;x_1,\cdots,x_M)\,,
$$
where the coefficients $A,B,C$ are functions of the 3 generic momenta.
For example, in the case of a 4-point block with 3 generic and one degenerate insertion one finds:
\begin{equation}
\label{PAb22phi}
\mathcal{B}_{2+2}(z)= \phantom{|}_2\Phi_1(A,B;C;z)\,,
\end{equation}
which is annihilated by the difference operator
\begin{eqnarray}\displaystyle
 D(A,B;C; q; z)\mathcal{B}_{\alpha_0,\alpha_1,\alpha_2}(z)=0\,,
 \end{eqnarray}
\begin{equation} 
 D(A,B;C;q;z)= h_2\,\frac{\partial_q^2}{\partial_q z^2}+h_1\frac{\partial_q}{\partial_q z}+h_0\,,
\end{equation}
where  $\frac{\partial_q}{\partial_q z}\, f(z)=\frac{f(qz)-f(z) }{z(q-1)}$ and the coefficients  $h_2,h_1,h_0$  are defined by
\begin{eqnarray}\displaystyle\nonumber
h_2&=&z(C-AB q z),\\ \nonumber
h_1&=&\frac{1-C}{1-q}+\frac{(1-A)(1-B)-(1-AB q) }{(1-q)}z,\\
h_0&=&-\frac{(1-A)(1-B)}{(1-q)^2}.
\end{eqnarray}
By removing the $q$-deformation this difference operator reduces to the familiar hypergeometric differential operator acting on level-two degenerate  4-point correlators.

As we discussed in section \ref{PAsecdeg}, we expect a map between  $\mathcal{V}ir_{q,t}$  blocks with some of the momenta analytically continued to degenerate values and  instanton-vortex partition functions associated to linear quivers with defects obtained via Higgsing. In particular the block $\mathcal{B}_{2+2}$ in (\ref{PAb22phi}), in analogy with the undeformed case, is expected  to be mapped to the vortex partition function of the 3d 
 $\mathcal{N}=2$ QED with 2 flavours, describing the codimension-two defect  theory obtained  by Higgsing the 5d $2+SU(2)+2$ theory (after normalizing by the contributions of 3-point functions and 5d free hypers).
Indeed we see that  $\mathcal{B}_{2+2}$ is given by a $q$-deformed hypergeometric series as the  $\mathcal{N}=2$ QED vortex partition function, so the two quantities can be mapped with a suitable dictionary.

\section{3d \& 5d partition functions as $q$-deformed  correlators}\label{PA3d5dmapsec}

Having reviewed   the identification of instanton/vortex partition functions with
non-degenerate/degenerate chiral  $\mathcal{V}ir_{q,t}$ blocks we now  move to the construction of $ \mathcal{V}ir_{q,t}$  correlators  and discuss their map   to  3d and 5d partition functions.

The first question one needs to address is  how to combine $\mathcal{V}ir_{q,t}$  chiral blocks into  a correlator. In the undeformed   2d CFT  case, the underlying symmetry  is the product of the holomorphic and anti-holomorphic copies of the Virasoro algebra and  correlators are constructed by taking the modulus square of the holomorphic and anti-holomorphic  conformal blocks.  This ensures that the monodromies acquired by the  chiral blocks
under change of channel (or  ordering of the OPE) cancel out in the  physical correlators which are
 modular invariant and single-value objects.
$\mathcal{V}ir_{q,t}$  chiral blocks don't have monodromies since, they have lines of poles rather than brach-cuts. This is clear if we consider for example the  degenerate chiral blocks discussed in the previous section, given by  $q$-deformed hypergeometric series. 

In \cite{PANieri:2013yra},\cite{PANieri:2013vba} it was proposed to define   deformed correlators by combining $ \mathcal{V}ir_{q,t} $ chiral blocks with the $SL(3,\mathbb{Z})$ gluing rules described in section \ref{PA5dfactos}.
In particular the authors  considered the  $S^4\times S^1$ and $S^5$ gluings and defined two families of correlators with respectively   $\mathcal{V}ir_{q,t}\otimes \mathcal{V}ir_{q,t} $ and $\mathcal{V}ir_{q,t}\otimes \mathcal{V}ir_{q,t}\otimes \mathcal{V}ir_{q,t} $ symmetry. In the first case the blocks 
are glued with the $id$-gluing  (\ref{PAidqt}), while in the second case with the $S$-gluing  (\ref{PAp123}).
Correspondingly these two families  were called $id$- and $S$-correlators and
it was proposed the following 5d-lift of the AGT correspondence relating $M+2$-point $id$- and $S$-correlators to
 $S^4\times S^1$ and $S^5$ partition functions of the  $2+SU(2)^{M-1}+2$ linear quiver:
\begin{eqnarray}\displaystyle  \label{PAcoremia1}
&&Z^{S^5}[2+SU(2)^{M-1}+2]=\int d \sigma~ { Z}^{S^5}_{\text{1-loop}}(\sigma,\vec M)~ 
 \Big|\Big| \mathcal{Z}_{cl}\mathcal{Z}^{5d}_{inst}\Big|\Big|^3_{S}=\\
 &&\!\!= \!\!\int\! \! d   a~  C^S \cdots C^S~ 
 \Big|\Big| {\cal C}_q  \mathcal{B}^q_{M+2} \Big|\Big|^3_{S}
 =\langle V_{\alpha_\infty}(\infty) V_{\alpha_M}(z_M)\cdots
 V_{\alpha_1}(z_1) V_{\alpha_0}(0) \rangle_S\,,
\nonumber
\end{eqnarray}

\begin{eqnarray}\displaystyle  \label{PAcoremia2}
&&Z^{S^4\times S^1}[2+SU(2)^{M-1}+2]=\int   d \sigma~ { Z}^{S^4\times S^1}_{\text{1-loop}}(\sigma,\vec M)~ 
 \Big|\Big|\mathcal{Z}_{cl}\mathcal{Z}^{5d}_{inst}\Big|\Big|^2_{id}=\\
  &&\!\!=\!\!\int \!\!d   a ~ C^{id} \cdots C^{id}~ 
 \Big|\Big| {\cal C}_q  \mathcal{B}^q_{M+2} \Big|\Big|^3_{id}
 =\langle V_{\alpha_\infty}(\infty) V_{\alpha_M}(z_M)\cdots
 V_{\alpha_1}(z_1) V_{\alpha_0}(0) \rangle_{id}\,.
\nonumber
\end{eqnarray}
In detail the map goes as follows. The instanton and  classical contributions are identified  with the $\mathcal{V}ir_{q,t} $ chiral blocks:
\begin{equation}
\mathcal{Z}_{cl}\mathcal{Z}^{5d}_{inst}= \mathcal{C}_{q}  \mathcal{B}_{M+2}\,.
\end{equation}
In the previous section we have discussed the map $\mathcal{Z}^{5d}_{inst}=\mathcal{B}_{M+2}$ for special values of the momenta and mentioned  that the match should extend to  generic momenta via geometric transition.

The classical terms are instead conjectured to map  to the factors
 $ \mathcal{C}_{q}$,  which were interpreted as  the deformed version of the  conformal factors, to which they reduce in the $q\to 1$ limit.
In the undeformed case  these factors  follow from the conformal Ward identities. In the $q$-deformed case at present it is not know how to derive the analogue of these identities, however an interesting discussion on the $q$-deformation of the $SU(1,1)$ Ward identities   can be found in \cite{PABernard:1989jq}.

The  key point is then to map the 1-loop factors to the 3-point functions contribution which we schematically 
indicated as $C^S \cdots C^S$ or $C^{id} \cdots C^{id}$. We will discuss this point  in a moment, before doing so
it is useful to  note that the correspondences (\ref{PAcoremia1}),  (\ref{PAcoremia2}), generate  
via analytic continuation/Higgsing, a series of  secondary  correspondences between   $id,S$-correlators with degenerate insertions and $S^4\times S^1$, $S^5$ partition functions decorated by  the $S^2\times S^1$, $S^3$ partition functions describing the  codimension-two  defects. 
For example one expects  the  map of the $id$ and $S$-correlators of 3 generic  primaries and one level-two degenerate primary $V_{\alpha_2}(z) =V_{deg}(z) $
(normalised by the 3-point function of the non-degenerate primaries)  to the partition functions of the $\mathcal{N}=2$ SQED with 2 flavours on $S^2\times S^1$, $S^3$ (normalised by the contribution
of the free 5d hypers):
\begin{eqnarray}\displaystyle
\label{PAcan1}
Z^{SQED}_{S^3}
=\sum_i ^2 G^{(i)}_{\rm 1-loop}\Big|\Big| \mathcal{G}^{(i)}_{\rm cl}{\cal Z}^{(i)}_V\Big|\Big|^2_{S}=
\langle V_{\alpha_4}(\infty) V_{\alpha_3}(1)
 V_{deg}(z) V_{\alpha_0}(0) \rangle_S\,,
\end{eqnarray}
\begin{eqnarray}\displaystyle
\label{PAcan2}
Z^{SQED}_{S^2\times S^1}=
  \sum_i ^2 G^{(i)}_{\rm 1-loop}\Big|\Big| \mathcal{G}^{(i)}_{\rm cl}{\cal Z}^{(i)}_V\Big|\Big|^2_{id}
  =\langle V_{\alpha_4}(\infty) V_{\alpha_3}(1)
 V_{deg}(z) V_{\alpha_0}(0) \rangle_{id}
  \,.
\end{eqnarray}
The  correspondences (\ref{PAcan1}) and (\ref{PAcan2})  indicate that the degenerate 4-point correlators will be  $SL(2,\mathbb{Z})$ $id$- or $S$-squares of degenerate blocks hence,  they will be annihilated by two hypergeometric difference operators (see also the discussion on the evaluation of the degenerate block  integral (\ref{PAb22phi})):
  \begin{eqnarray}\displaystyle
\langle V_{\alpha_4}(\infty) V_{\alpha_3}(1) V_{deg}(z) V_{\alpha_1}(0) \rangle_*\sim G(z,\tilde z),
\end{eqnarray}
\begin{eqnarray}\displaystyle\label{PAqdif}
 D(A,B;C; q; z)G(z, \tilde z)=0\,,\qquad
 D(\tilde A,\tilde B;\tilde C;\tilde q; \tilde z)G(z,\tilde z)=0\,.
\end{eqnarray}
The parameters $A,B,C$ are functions of the momenta $\alpha_1,\alpha_3,\alpha_4$,
un-tilded and tilded variables denote the parameters in the two chiral blocks and the subscript $*$ indicate either the $id$ or $S$ gluing.

\begin{figure}[!ht]
\leavevmode
\begin{center}
\includegraphics[width=0.6\textwidth]{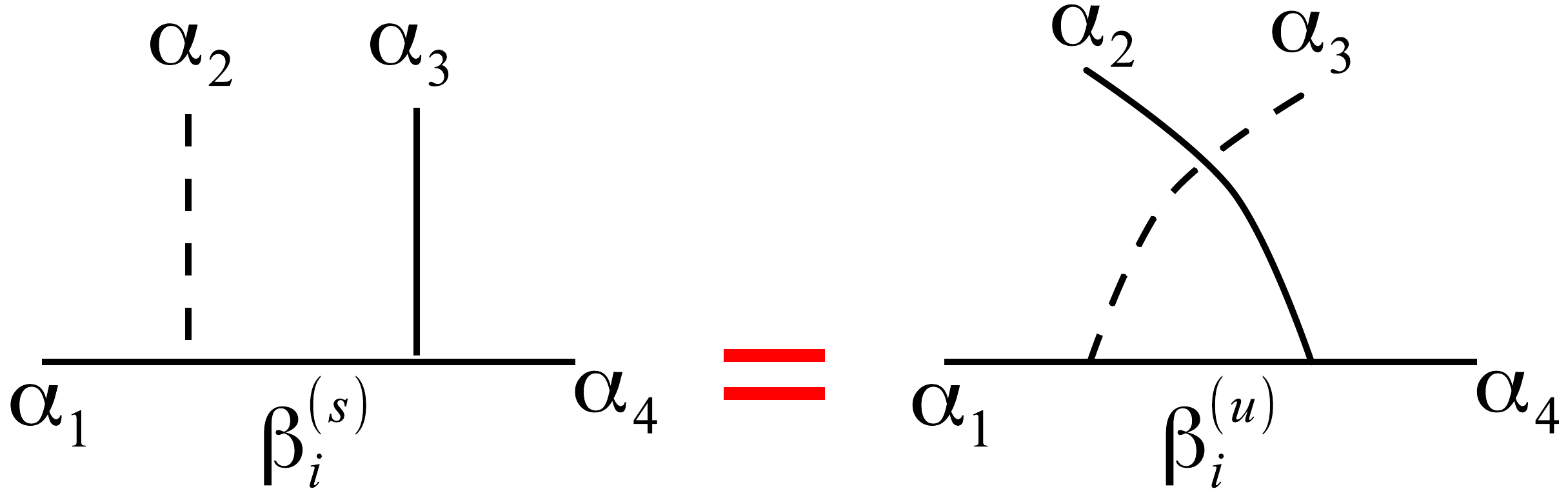}
\end{center}
\caption{Crossing symmetry requires the equality of correlations involving chiral blocks
in the s-channel (on the LHS) and in the u-channel (on the RHS).}
\label{PAcross}
\end{figure}

This observation was used in \cite{PANieri:2013yra} to derive 3-point functions by means of  the bootstrap approach \cite{PABelavin:1984vu},\cite{PATeschner:1995yf}. 
Eqs. (\ref{PAqdif})  imply that  $G(z,\tilde z)$ can be expressed as a bi-linear combination of solutions of the $q$-hypergeometric difference equation.  Let $I^{(s)}_{1,2}$ be a   basis of two  linearly independent  solutions    in the neighbourhood of $z=0$, then we can write:
\begin{eqnarray}\displaystyle\label{PAscorreq}
\langle V_{\alpha_4}(\infty) V_{\alpha_3}(1) V_{deg}(z) V_{\alpha_1}(0) \rangle_*\sim\sum_{i=1}^{2}K_{i}^{(s)} \Big| \Big| I_i^{(s)}(z;q) \Big| \Big|_{*} ^2\,,
\end{eqnarray}
where the coefficients $K_{i}^{(s)}$ are related  to the 3-point function factor associated to the diagram on the left of Fig. \ref{PAcross}\,.

Similarly in the $u$-channel the correlation function is a bilinear combination of solutions 
 $I^{(u)}_{1,2}$  in the neighbourhood of $z=\infty$:
  \begin{eqnarray}\displaystyle\label{PAucorreq}
\langle V_{\alpha_4}(\infty) V_{\alpha_3}(r) V_{deg}(z) V_{\alpha_1}(0) \rangle&\sim&\sum_{i=1}^{2}K_{i}^{(u)} \Big| \Big| I_i^{(u)}(z;q) \Big| \Big|_{*} ^2\,,
\end{eqnarray}
with coefficient $K^{(u)}_{i}$ related to the 3-point functions factor associated to the diagram on the right of Fig. \ref{PAcross}.

To bootstrap  the 3-point functions we now  impose crossing symmetry  requiring the equality of  $s$- and $u$-channel correlators:
 \begin{equation}
 \label{PAqcs}
K_{1}^{(s)} \Big| \Big| I^{(s)}_1  \Big| \Big|_{*}^2+
K_{2}^{(s)}
\Big| \Big| I^{(s)}_2  \Big| \Big|_{*}^2=
K_{1}^{(u)}
\Big| \Big| I^{(u)}_1  \Big| \Big|_{*}^2+
K_{2}^{(u)}
\Big| \Big| I^{(u)}_2  \Big| \Big|_{*}^2\,.
\end{equation}
We then take the analytic continuation to the neighbourhood of  $\infty$ of  the solutions $I^{(s)}_i$  and express  them as linear combination of $u$-channel solutions.    At this point eq. (\ref{PAqcs}) yields a set of  non-trivial equations for the coefficients $K_{i}^{(s)}$ and   $K_{i}^{(u)}$ which determine the 3-point functions  uniquely once the gluing rule is specified.

In the  case  of $id$-gluing there are two types of level-two degenerate primaries:  $\alpha_2=-\frac{b_0^{\pm 1}}{2}$  (corresponding to the two $S^2\times S^1$ defects inside $S^4\times S^1$) and we can write two sets of  equations for the 3-point function. The unique solution (up to $q$-constants) to these equations is given by:
 \begin{equation}\label{PAqdozz}
C^{id}(\alpha_3,\alpha_2,\alpha_1)=\frac{1}{\Upsilon^R(2 \alpha_T-Q_0)}\prod_{i=1}^3 \frac{\Upsilon^R(2 \alpha_i)}{\Upsilon^R(2\alpha_T-2 \alpha_i)}\,,
\end{equation}
where $2\alpha_T=\alpha_1+\alpha_2+\alpha_3$  while the parameter $R$ is related to the deformation parameter of the algebra and to the $S^1$ radius of the $S^4\times S^1$ geometry on the gauge theory side.
The definition and useful properties of the   $\Upsilon^R(X)$ function are collected in the Appendix \ref{PAspecfun}.

Similarly, for $S$-correlators   there are three  types of degenerations $\alpha_2=-\frac{\omega_i}{2}$ for  $i=1,2,3$   (corresponding to the  three large  $S^3$ inside $S^5$)   and we can write  three sets of equations with unique solution  given by:
\begin{equation}
\label{PA4p1l}
C^S(\alpha_3,\alpha_2,\alpha_1)=\frac{1}{S_3(2 \alpha_T-E)}\prod_{i=1}^3 \frac{S_3(2 \alpha_i)}{S_3(2\alpha_T-2 \alpha_i)}\,.
\end{equation}

We can now complete the map between $q$-deformed correlators and partition functions.
For example it is easy to check that the 3-point functions factor in the 4-point $S$-correlator
can be mapped to  the $S^5$  one-loop contribution   of the 
$2+SU(2)+2$ theory:
\begin{eqnarray}\displaystyle
\nonumber C^S(\alpha_1,\alpha_2,\alpha)C^S(E-\alpha,\alpha_3,\alpha_4)=
{ Z}^{S^5,\text{vect}}_{\text{1-loop}}(\sigma)\prod_{i=1}^4{ Z}^{S^5, \text{hyper}}_{\text{1-loop}}(\sigma,m_i)\,,
\end{eqnarray}
where the 1-loop factors for the vector multiplet and for the  fundamental hypers are defined in eqs. (\ref{PAvecs5}),  and masses and momenta are mapped by the following dictionary:
\begin{equation}
\alpha= i\sigma+\frac{E}{2}\,,\quad\alpha_1\pm\alpha_2=im_{1,2}+E\, ,\quad \alpha_3\pm\alpha_4= im_{3,4}+E\,.
\end{equation}
Similarly for $id$-correlators one can show that:
\begin{eqnarray}\displaystyle
C_{id}(\alpha_1,\alpha_2,\alpha)C_{id}(Q_0-\alpha,\alpha_3,\alpha_4)=
{ Z}^{S^4\times S^1,\text{vect}}_{\text{1-loop}}(\sigma)\prod_{i=1}^4{ Z}^{S^4\times S^1,\text{hyper}}_{\text{1-loop}}(\sigma,m_i)\,,\nonumber\\
\end{eqnarray}

\vspace{-.5cm}
\noindent
with the following dictionary:
\begin{equation}
\alpha=i\sigma+\frac{Q_0}{2}\,,\quad\alpha_1\pm\alpha_2=im_{1,2}+Q_0\,,\quad\alpha_3\pm\alpha_4=im_{3,4}+Q_0\,.
\end{equation}
3-point functions factor in higher point $S,id$-correlators can be similarly mapped to 1-loop contributions in
$S^5$ and $S^4\times S^1$ linear quiver partition functions. In \cite{PANieri:2013vba} it was also shown that the
3-point function contribution to the 1-punctured torus $S,id$-correlators can be mapped to the 
1-loop factor of the $SU(2)$ theory  with  a massive adjoint hyper on $S^5$ and $S^4\times S^1$. \\

It is possible to take a smooth  limit which removes the $q$-deformation and reduces $id$-correlators  to Liouville correlators.  On the gauge theory side this limit corresponds to  shrinking the $S^1$ radius and reducing  to the $S^4$ partition function. Indeed for $R\to 0$, the 3-point function (\ref{PAqdozz})  smoothly reduces to the familiar DOZZ formula for the Liouville 3-point function \cite{PADorn:1994xn}, \cite{PAZamolodchikov:1995aa}, \cite{PATeschner:1995yf}, which via AGT is mapped to 1-loop factors on $S^4$.
$S$-correlators instead don't admit a smooth undeformed limit.  
In \cite{PANieri:2013vba}  reflection coefficients,  constructed from $id$ and $S$ 3-point functions,
 were given a  geometric interpretation  as Harish-Chandra $c$-functions for certain quantum symmetric spaces.
 These $c$-functions were in turn related to the  Jost  functions  describing  scattering processes  in two  different limits of the XYZ spin chain.\\

The 3-point function (\ref{PAqdozz}) was earlier derived in \cite{PAKozcaz:2010af}, building on the results of  \cite{PABenini:2009gi}, by using the  topological string partition function on  a particular  toric CY threefold,  the blow up of the $\mathbb{C}^2/[\mathbb{Z}_2\times   \mathbb{Z}_2 ]$ orbifold. In \cite{PABenini:2009gi} it was proposed that a five-dimensional version of the $T_N$ theory  could be obtained in a  IIB  setup  in  terms  of  a  junction of   N D5-branes, N NS5-branes and N (1,1) 5-branes which realises the blow up of the $\mathbb{C}^2/[\mathbb{Z}_N\times   \mathbb{Z}_N ]$ orbifold.
 The 4d $T_N$ theory is the non-Lagrangian theory of  N M5 branes on the sphere with 3 full punctures and it is mapped via AGT to the Toda 3-point function. For this reason one expects that the topological string partition function on the $T_N$ geometry maps to the $q$-deformed   Toda  3-point function  with  3 full primaries.
The determination  of  the  Toda  3-point function  with  3 full primaries is  a long standing  open problem and  the possibility  of extracting the answer from the  $T_N$ geometry  has been explored in  \cite{PAMitev:2014isa}, \cite{PAIsachenkov:2014eya}.\\

The work done on  the correspondence between deformed $\mathcal{V}ir_{q,t}$ correlators and partition functions
has focused on establishing a direct map between the  terms contributing to the partition functions 
and  the  terms contributing to the  $\mathcal{V}ir_{q,t}$ correlators.
 At a deeper level one would like to be able to  identify how symmetries are mapped across the correspondence.
 In the  AGT case the  generalised S-duality of  4d $\mathcal{N}=2$ gauge theories  was beautifully identified with the Moore-Seiberg duality groupoid acting on 2d CFT correlators. This observation for example made it possible to borrow from the CFT the sophisticated machinery of Verlinde-loop operators and apply it to  the exact computation of vevs of line operators on the gauge theory side \cite{PAAlday:2009fs}, \cite{PADrukker:2009id}.
  More recently in \cite{PAGomis:2014eya} the map of symmetries has been understood also in the case  where  surface operators engineered by M2 branes are included. Remarkably in this case all the gauge theory dualities of the combined 4d-2d system describing the surface operator,   can  be identified with symmetries of the Toda correlators involving extra degenerate primaries. 
 It would be very interesting to  establish  an analogous complete  map between  3d/5d  gauge theory dualities and symmetries of $q$-correlators. An encouraging step in this direction was taken in \cite{PANieri:2013vba} where it was pointed out that the crossing symmetry of the degenerate  4-point $\mathcal{V}ir_{q,t}$ correlator, which was used to derive the 3-point functions, can be identified with the flop symmetry of the 3d SQED, which acts by swapping the sign of the Fayet-Iliopoulos and exchanging charge plus with charge minus chiral multiplets. 
Understanding the map of symmetries should  open up the possibility to retrace in $q$-deformed case, the various applications of the gauge/CFT correspondence to the study of defects operators in gauge theories.

\section*{Acknowledgments}
I would like to thank Fabrizio Nieri, Filippo Passerini and Alessandro Torrielli for enjoyable collaborations on 
these topics. S.P. is partially  supported by the  ERC-STG grant 637844-HBQFTNCER. 
%\appendix

\section{Appendix}\label{PAspecfun}
In this appendix we collect the definition and some properties of the special functions used in the main text.
\subsection*{Bernoulli polynomials}\label{PAber}
The  Bernoulli polynomials $B_{rr}(z|\vec{\omega})$ are defined by \cite{PAnaru}
\begin{eqnarray}\displaystyle
B_{11}(z|\vec{\omega})&=&\frac{z}{\omega_1}-\frac{1}{2}\nonumber\\
B_{22}(z|\vec{\omega})&=&\frac{z^2}{\omega_1\omega_2}-\frac{\omega_1+\omega_2}{\omega_1\omega_2}z+\frac{\omega_1^2+\omega_2^2+3\omega_1\omega_2}{\omega_1\omega_2}\nonumber\\
B_{33}(z|\vec{\omega})&=& \frac{z^3}{\omega_1 \omega_2 \omega_3}
- \frac{3\,(\omega_1 + \omega_2 + \omega_3)}
{2\, \omega_1 \omega_2 \omega_3} z^2 
 + \frac{\omega_1^2 + \omega_2^2 + \omega_3^2
+ 3(\omega_1 \omega_2 + \omega_2 \omega_3 + \omega_3 \omega_1)}
{2\, \omega_1 \omega_2 \omega_3} z\nonumber \\ 
&& - \frac{(\omega_1 + \omega_2 + \omega_3)
(\omega_1 \omega_2 + \omega_2 \omega_3 + \omega_3 \omega_1)}
{4\, \omega_1 \omega_2 \omega_3}\,,
\end{eqnarray}
where $\vec{\omega}:=(\omega_1,\ldots,\omega_r) $. We will use the shorthand $B_{rr}(z):=B_{rr}(z|\vec{\omega})$.
\subsection*{Multiple Gamma and Sine functions}\label{PAtri}

The Barnes $r$-Gamma function $\Gamma_r(z|\vec{\omega})$ can be defined as the $\zeta$-regularized infinite product \cite{PAnaru}
\begin{equation}
\Gamma_r(z|\vec{\omega})\sim \prod_{\vec{n}\in \mathbb{Z}^+_0}\frac{1}{\left(z+\vec{\omega}\cdot\vec{n}\right)}.
\end{equation}
When there is no possibility of confusion, we will simply set $\Gamma_r(z):=\Gamma_r(z|\vec{\omega})$.\\

The $r$-Sine function is defined according to \cite{PAnaru}
\begin{equation}
S_r(z|\vec{\omega})=\frac{\Gamma_r(E_r-z)^{(-1)^r}}{\Gamma_r(z)}
\end{equation}
where we defined $E_r:=\sum_i\omega_i$. We will also denote $S_r(z):=S_r(z|\vec{\omega})$ when there is no confusion. Also, introducing the multiple $q$-shifted factorial
\begin{equation}
\label{PAmulq}
\left(z;q_1,\ldots q_r\right):=\prod_{k_1,\ldots, k_r\geq 0}\left(1-zq_1^{k_1}\cdots q_r^{k_r}\right)
\end{equation}
the $r$-sine function has the following product representation ($r\geq 2$) \cite{PAnaru}
\begin{equation}\label{PAs3fac}
S_r(z)=e^{(-1)^r\frac{i \pi }{r!}B_{rr}(z)}\prod_{k=1}^r\left(e^{\frac{2\pi i}{\omega_k}z};e^{2\pi i\frac{\omega_1}{\omega_k}},\ldots,e^{2\pi i\frac{\omega_{k-1}}{\omega_k}},e^{2\pi i\frac{\omega_{k+1}}{\omega_k}},\ldots,e^{2\pi i\frac{\omega_{r}}{\omega_k}}\right)
\end{equation}
whenever ${\rm Im}\left( \omega_j/\omega_k\right) \neq 0$ (for $j\neq k$). General useful identities are 
\begin{equation}
S_r(z)S_r(E_r-z)^{(-1)^r}=1 
\end{equation}
\begin{equation}
S_r(\lambda z|\lambda\vec{\omega})=S_r(z|\vec{\omega});\quad \lambda \in \mathbb{C}/\{0\}
\end{equation}
\begin{equation}
\label{PAprorat}
\frac{S_r(z+\omega_i)}{S_r(z)}=\frac{1}{S_{r-1}(z|\omega_1,\ldots,\omega_{i-1},\omega_{i+1},\ldots,\omega_r)}\,.
\end{equation}
%Notice for $r=3$ we can write
%\begin{equation}
%S_3(z)=e^{-\frac{i \pi }{3!}B_{33}(z)}\left(e^{\frac{2\pi i}{e_3}z};q,t\right)_1\left(e^{\frac{2\pi i}{e_3}z};q,t\right)_2\left(e^{\frac{2\pi i}{e_3}z};q,t\right)_3
%\end{equation}
%where $q$, $t$ are expressed via the $e_1$, $e_2$, $e_3$ parameters as described in (\ref{PAp123}), (\ref{PApq}), and it is customary to denote $E=\omega_1+\omega_2+\omega_3$. 
For $r=2$ the multi-sine function is related to the double sine function
\begin{equation}
s_b(z)=S_2(Q/2-iz|b,b^{-1})\sim \prod_k\frac{n_1\omega_1+n_2\omega_2+Q/2-iz}{n_1\omega_1+n_2\omega_2+Q/2+iz}
\end{equation}
where we take $Q=\omega_1+\omega_2$ and $b=\omega_1=\omega_2^{-1}$.

\subsection*{$\Upsilon^R$ function}\label{PAupb}

The $\Upsilon^R$ function  is defined as the regularized infinite product
\begin{eqnarray}\displaystyle\nonumber
\Upsilon^R(z)
\sim \prod_{n_1,n_2\geq0}\sinh\left[\frac{R}{2}\left(z+n_1 b_0+n_2 b_0^{-1}\right)\right]\sinh\left[\frac{R}{2}\left(Q_0-z+n_1b_0+n_2b_0^{-1}\right)\right]\,.
\end{eqnarray}
Important defining properties are 
\begin{equation}
\Upsilon^R(z)=\Upsilon^R(Q_0-z)
\end{equation}
\begin{equation}
\frac{\Upsilon^R(z+b_0^{\pm 1})}{\Upsilon^R(z)}\sim \frac{\left(e^{R(b_0^{\mp 1}- z)};e^{R b_0^{\mp 1}}\right)}{\left(e^{R z};e^{R b_0^{\mp 1}}\right)}\,.
%=\frac{1}{\left(e^{R z};e^{R b_0^{\mp 1}}\right)\left(e^{-R z};e^{-R b_0^{\mp 1}}\right)}.
\end{equation}
In the $R\to 0$ limit it reduces to the $\Upsilon(z)$ function appearing in Liouville field theory
\begin{equation}
\Upsilon(z)=\Gamma_2(z|b_0,b_0^{-1})^{-1}\Gamma_2(Q_0-z|b_0,b_0^{-1})^{-1}
\end{equation}
where $Q_0:=b_0+b_0^{-1}$.

\subsection*{Jacobi Theta and elliptic Gamma functions}
The Jacobi $\Theta$ function is defined by \cite{PAgrahman}
\begin{equation}
\Theta(z;\tau)=\left(e^{2\pi i z};e^{2\pi i\tau}\right)\left(e^{2\pi i\tau}e^{-2\pi i z};e^{2\pi i\tau}\right)
\end{equation}
and satisfies the functional relation
\begin{equation}
\frac{\Theta(\tau+z;\tau)}{\Theta(z;\tau)}=-e^{-2\pi i z}.
\end{equation}
Another relevant property is \cite{PAnaru}
\begin{equation}
\Theta\left(\frac{z}{\omega_1};\frac{\omega_2}{\omega_1}\right)\Theta\left(\frac{z}{\omega_2};\frac{\omega_1}{\omega_2}\right)=e^{-i\pi B_{22}(z)}.
\end{equation}
\\

The elliptic Gamma function $\Gamma_{q,t}$ is defined by \cite{PAgrahman}
\begin{equation}
\Gamma_{q,t}(z)=\frac{(qt\;e^{-2\pi i z};q,t)}{(e^{2\pi i z};q,t)};\quad q=e^{2\pi i\tau};\quad t=e^{2\pi i\sigma}
\end{equation}
and satisfies the functional relations
\begin{equation}
\frac{\Gamma_{q,t}(\tau+z)}{\Gamma_{q,t}(z)}=\Theta(z;\sigma);\quad \frac{\Gamma_{q,t}(\sigma+z)}{\Gamma_{q,t}(z)}=\Theta(z;\tau)\,.
\end{equation}
Other relevant properties are \cite{PAfv}
\begin{equation}
\label{PA3gammabb}
\Gamma_{q,t}\left(\frac{z}{e_3}\right)_1\Gamma_{q,t}\left(\frac{z}{e_3}\right)_2\Gamma_{q,t}\left(\frac{z}{e_3}\right)_3=e^{-\frac{i\pi}{3}B_{33}(z)}
\end{equation}
where $q$, $t$ are expressed via the $e_1$, $e_2$, $e_3$ parameters as described in (\ref{PAp123}), (\ref{PApq})
and 
\begin{equation}
\label{PAgamma unit}
\Gamma_{q,t}\left(\frac{z}{e_3}\right)\Gamma_{q,t}\left(\frac{e_1+e_2-z}{e_3}\right)=1\,.
\end{equation}

\documentfinish
\begin{thebibliography}{100}

\bibitem{ContributionSummary}
V.~Pestun and M.~Zabzine, eds., {\em Localization techniques in quantum field
  theory}, vol.~xx.
\newblock Journal of Physics A, 2016.
\newblock \href{http://arxiv.org/abs/1608.02952}{{\tt 1608.02952}}.
\newblock \url{https://arxiv.org/src/1608.02952/anc/LocQFT.pdf},
  \url{http://pestun.ihes.fr/pages/LocalizationReview/LocQFT.pdf}.

\bibitem{PAPestun:2007rz}
V.~Pestun, ``{Localization of gauge theory on a four-sphere and supersymmetric
  Wilson loops},'' \href{http://dx.doi.org/10.1007/s00220-012-1485-0}{{\em
  Commun. Math. Phys.} {\bf 313} (2012)  71--129},
\href{http://arxiv.org/abs/0712.2824}{{\tt arXiv:0712.2824 [hep-th]}}.
%%CITATION = ARXIV:0712.2824;%%.

\bibitem{ContributionMA}
M.~Mari{\~n}o, ``Localization at large $N$ in Chern-Simons-matter theories,''
  {\em Journal of Physics A} {\bf xx} (2016)  000,
  \href{http://arxiv.org/abs/1608.02959}{{\tt 1608.02959}}.

\bibitem{ContributionZA}
K.~Zarembo, ``Localization and AdS/CFT Correspondence,'' {\em Journal of
  Physics A} {\bf xx} (2016)  000, \href{http://arxiv.org/abs/1608.02963}{{\tt
  1608.02963}}.

\bibitem{ContributionMI}
J.~Minahan, ``Matrix models for 5D super Yang-Mills,'' {\em Journal of Physics
  A} {\bf xx} (2016)  000, \href{http://arxiv.org/abs/1608.02967}{{\tt
  1608.02967}}.

\bibitem{ContributionWI}
B.~Willett, ``Localization on three-dimensional manifolds,'' {\em Journal of
  Physics A} {\bf xx} (2016)  000, \href{http://arxiv.org/abs/1608.02958}{{\tt
  1608.02958}}.

\bibitem{PAAlday:2009aq}
L.~F. Alday, D.~Gaiotto, and Y.~Tachikawa, ``{Liouville Correlation Functions
  from Four-dimensional Gauge Theories},''
  \href{http://dx.doi.org/10.1007/s11005-010-0369-5}{{\em Lett.Math.Phys.} {\bf
  91} (2010)  167--197},
\href{http://arxiv.org/abs/0906.3219}{{\tt arXiv:0906.3219 [hep-th]}}.
%%CITATION = ARXIV:0906.3219;%%.

\bibitem{ContributionTA}
Y.~Tachikawa, ``A brief review of the 2d/4d correspondences,'' {\em Journal of
  Physics A} {\bf xx} (2016)  000, \href{http://arxiv.org/abs/1608.02964}{{\tt
  1608.02964}}.

\bibitem{ContributionDI}
T.~Dimofte, ``Perturbative and nonperturbative aspects of complex Chern-Simons
  Theory,'' {\em Journal of Physics A} {\bf xx} (2016)  000,
  \href{http://arxiv.org/abs/1608.02961}{{\tt 1608.02961}}.

\bibitem{PAFestuccia:2011ws}
G.~Festuccia and N.~Seiberg, ``{Rigid Supersymmetric Theories in Curved
  Superspace},'' \href{http://dx.doi.org/10.1007/JHEP06(2011)114}{{\em JHEP}
  {\bf 06} (2011)  114},
\href{http://arxiv.org/abs/1105.0689}{{\tt arXiv:1105.0689 [hep-th]}}.
%%CITATION = ARXIV:1105.0689;%%.

\bibitem{ContributionDU}
T.~Dumitrescu, ``An Introduction to Supersymmetric Field Theories in Curved
  Space,'' {\em Journal of Physics A} {\bf xx} (2016)  000,
  \href{http://arxiv.org/abs/1608.02957}{{\tt 1608.02957}}.

\bibitem{PAhb}
C.~Beem, T.~Dimofte, and S.~Pasquetti, ``{Holomorphic Blocks in Three
  Dimensions},''
\href{http://arxiv.org/abs/1211.1986}{{\tt arXiv:1211.1986 [hep-th]}}.
%%CITATION = ARXIV:1211.1986;%%.

\bibitem{PADimofte:2011py}
T.~Dimofte, D.~Gaiotto, and S.~Gukov, ``{3-Manifolds and 3d Indices},''
  \href{http://dx.doi.org/10.4310/ATMP.2013.v17.n5.a3}{{\em Adv. Theor. Math.
  Phys.} {\bf 17} (2013) no.~5, 975--1076},
\href{http://arxiv.org/abs/1112.5179}{{\tt arXiv:1112.5179 [hep-th]}}.
%%CITATION = ARXIV:1112.5179;%%.

\bibitem{PADimofte:2011ju}
T.~Dimofte, D.~Gaiotto, and S.~Gukov, ``{Gauge Theories Labelled by
  Three-Manifolds},'' \href{http://dx.doi.org/10.1007/s00220-013-1863-2}{{\em
  Commun. Math. Phys.} {\bf 325} (2014)  367--419},
\href{http://arxiv.org/abs/1108.4389}{{\tt arXiv:1108.4389 [hep-th]}}.
%%CITATION = ARXIV:1108.4389;%%.

\bibitem{PADimofte:2011jd}
T.~Dimofte and S.~Gukov, ``{Chern-Simons Theory and S-duality},''
  \href{http://dx.doi.org/10.1007/JHEP05(2013)109}{{\em JHEP} {\bf 05} (2013)
  109},
\href{http://arxiv.org/abs/1106.4550}{{\tt arXiv:1106.4550 [hep-th]}}.
%%CITATION = ARXIV:1106.4550;%%.

\bibitem{PANieri:2015yia}
F.~Nieri and S.~Pasquetti, ``{Factorisation and holomorphic blocks in 4d},''
\href{http://arxiv.org/abs/1507.00261}{{\tt arXiv:1507.00261 [hep-th]}}.
%%CITATION = ARXIV:1507.00261;%%.

\bibitem{PADimofte:2014zga}
T.~Dimofte, ``{Complex Chern-Simons theory at level k via the 3d-3d
  correspondence},''
\href{http://arxiv.org/abs/1409.0857}{{\tt arXiv:1409.0857 [hep-th]}}.
%%CITATION = ARXIV:1409.0857;%%.

\bibitem{PABenini:2015noa}
F.~Benini and A.~Zaffaroni, ``{A topologically twisted index for
  three-dimensional supersymmetric theories},''
\href{http://arxiv.org/abs/1504.03698}{{\tt arXiv:1504.03698 [hep-th]}}.
%%CITATION = ARXIV:1504.03698;%%.

\bibitem{PASpiridonov:2012ww}
V.~P. Spiridonov and G.~S. Vartanov, ``{Elliptic hypergeometric integrals and
  't Hooft anomaly matching conditions},''
  \href{http://dx.doi.org/10.1007/JHEP06(2012)016}{{\em JHEP} {\bf 06} (2012)
  016},
\href{http://arxiv.org/abs/1203.5677}{{\tt arXiv:1203.5677 [hep-th]}}.
%%CITATION = ARXIV:1203.5677;%%.

\bibitem{PADimofte:2010tz}
T.~Dimofte, S.~Gukov, and L.~Hollands, ``{Vortex Counting and Lagrangian
  3-manifolds},'' \href{http://dx.doi.org/10.1007/s11005-011-0531-8}{{\em
  Lett.Math.Phys.} {\bf 98} (2011)  225--287},
\href{http://arxiv.org/abs/1006.0977}{{\tt arXiv:1006.0977 [hep-th]}}.
%%CITATION = ARXIV:1006.0977;%%.

\bibitem{PAPasquetti:2011fj}
S.~Pasquetti, ``{Factorisation of ${\cal{N}} = 2$ Theories on the Squashed
  3-Sphere},'' \href{http://dx.doi.org/10.1007/JHEP04(2012)120}{{\em JHEP} {\bf
  1204} (2012)  120},
\href{http://arxiv.org/abs/1111.6905}{{\tt arXiv:1111.6905 [hep-th]}}.
%%CITATION = ARXIV:1111.6905;%%.

\bibitem{PATaki:2013opa}
M.~Taki, ``{Holomorphic Blocks for 3d Non-abelian Partition Functions},''
\href{http://arxiv.org/abs/1303.5915}{{\tt arXiv:1303.5915 [hep-th]}}.
%%CITATION = ARXIV:1303.5915;%%.

\bibitem{PAHwang:2012jh}
C.~Hwang, H.-C. Kim, and J.~Park, ``{Factorization of the 3d superconformal
  index},'' \href{http://dx.doi.org/10.1007/JHEP08(2014)018}{{\em JHEP} {\bf
  1408} (2014)  018},
\href{http://arxiv.org/abs/1211.6023}{{\tt arXiv:1211.6023 [hep-th]}}.
%%CITATION = ARXIV:1211.6023;%%.

\bibitem{PAHwang:2015wna}
C.~Hwang and J.~Park, ``{Factorization of the 3d superconformal index with an
  adjoint matter},''
\href{http://arxiv.org/abs/1506.03951}{{\tt arXiv:1506.03951 [hep-th]}}.
%%CITATION = ARXIV:1506.03951;%%.

\bibitem{PAImamura:2013qxa}
Y.~Imamura, H.~Matsuno, and D.~Yokoyama, ``{Factorization of the
  $S^3/\mathbb{Z}_n$ partition function},''
  \href{http://dx.doi.org/10.1103/PhysRevD.89.085003}{{\em Phys.Rev.} {\bf D89}
  (2014) no.~8, 085003},
\href{http://arxiv.org/abs/1311.2371}{{\tt arXiv:1311.2371 [hep-th]}}.
%%CITATION = ARXIV:1311.2371;%%.

\bibitem{PAYoshida:2014qwa}
Y.~Yoshida, ``{Factorization of 4d N=1 superconformal index},''
\href{http://arxiv.org/abs/1403.0891}{{\tt arXiv:1403.0891 [hep-th]}}.
%%CITATION = ARXIV:1403.0891;%%.

\bibitem{PAPeelaers:2014ima}
W.~Peelaers, ``{Higgs branch localization of $ \mathcal{N} $ = 1 theories on
  S$^{3}$ x S$^{1}$},'' \href{http://dx.doi.org/10.1007/JHEP08(2014)060}{{\em
  JHEP} {\bf 1408} (2014)  060},
\href{http://arxiv.org/abs/1403.2711}{{\tt arXiv:1403.2711 [hep-th]}}.
%%CITATION = ARXIV:1403.2711;%%.

\bibitem{PAChen:2014rca}
H.-Y. Chen and H.-Y. Chen, ``{Heterotic Surface Defects and Dualities from
  2d/4d Indices},'' \href{http://dx.doi.org/10.1007/JHEP10(2014)004}{{\em JHEP}
  {\bf 10} (2014)  004},
\href{http://arxiv.org/abs/1407.4587}{{\tt arXiv:1407.4587 [hep-th]}}.
%%CITATION = ARXIV:1407.4587;%%.

\bibitem{PAChen:2015fta}
H.-Y. Chen and T.-H. Tsai, ``{On Higgs Branch Localization of Seiberg-Witten
  Theories on Ellipsoid},''
\href{http://arxiv.org/abs/1506.04390}{{\tt arXiv:1506.04390 [hep-th]}}.
%%CITATION = ARXIV:1506.04390;%%.

\bibitem{ContributionBL}
F.~Benini and B.~{Le Floch}, ``Supersymmetric localization in two dimensions,''
  {\em Journal of Physics A} {\bf xx} (2016)  000,
  \href{http://arxiv.org/abs/1608.02955}{{\tt 1608.02955}}.

\bibitem{PABenini:2012ui}
F.~Benini and S.~Cremonesi, ``{Partition functions of ${\cal{N}}=(2,2)$ gauge
  theories on $S^2$ and vortices},''
\href{http://arxiv.org/abs/1206.2356}{{\tt arXiv:1206.2356 [hep-th]}}.
%%CITATION = ARXIV:1206.2356;%%.

\bibitem{PADoroud:2012xw}
N.~Doroud, J.~Gomis, B.~Le~Floch, and S.~Lee, ``{Exact Results in D=2
  Supersymmetric Gauge Theories},''
  \href{http://dx.doi.org/10.1007/JHEP05(2013)093}{{\em JHEP} {\bf 05} (2013)
  093},
\href{http://arxiv.org/abs/1206.2606}{{\tt arXiv:1206.2606 [hep-th]}}.
%%CITATION = ARXIV:1206.2606;%%.

\bibitem{PAClosset:2015rna}
C.~Closset, S.~Cremonesi, and D.~S. Park, ``{The equivariant A-twist and gauged
  linear sigma models on the two-sphere},''
  \href{http://dx.doi.org/10.1007/JHEP06(2015)076}{{\em JHEP} {\bf 1506} (2015)
   076},
\href{http://arxiv.org/abs/1504.06308}{{\tt arXiv:1504.06308 [hep-th]}}.
%%CITATION = ARXIV:1504.06308;%%.

\bibitem{PABenini:2013yva}
F.~Benini and W.~Peelaers, ``{Higgs branch localization in three dimensions},''
  \href{http://dx.doi.org/10.1007/JHEP05(2014)030}{{\em JHEP} {\bf 1405} (2014)
   030},
\href{http://arxiv.org/abs/1312.6078}{{\tt arXiv:1312.6078 [hep-th]}}.
%%CITATION = ARXIV:1312.6078;%%.

\bibitem{PAFujitsuka:2013fga}
M.~Fujitsuka, M.~Honda, and Y.~Yoshida, ``{Higgs branch localization of 3d
  $\mathcal{N}$ = 2 theories},''
  \href{http://dx.doi.org/10.1093/ptep/ptu158}{{\em PTEP} {\bf 2014} (2014)
  no.~12, 123B02},
\href{http://arxiv.org/abs/1312.3627}{{\tt arXiv:1312.3627 [hep-th]}}.
%%CITATION = ARXIV:1312.3627;%%.

\bibitem{PAAlday:2013lba}
L.~F. Alday, D.~Martelli, P.~Richmond, and J.~Sparks, ``{Localization on
  Three-Manifolds},'' \href{http://dx.doi.org/10.1007/JHEP10(2013)095}{{\em
  JHEP} {\bf 1310} (2013)  095},
\href{http://arxiv.org/abs/1307.6848}{{\tt arXiv:1307.6848 [hep-th]}}.
%%CITATION = ARXIV:1307.6848;%%.

\bibitem{PAGomis:2012wy}
J.~Gomis and S.~Lee, ``{Exact Kahler Potential from Gauge Theory and Mirror
  Symmetry},'' \href{http://dx.doi.org/10.1007/JHEP04(2013)019}{{\em JHEP} {\bf
  04} (2013)  019},
\href{http://arxiv.org/abs/1210.6022}{{\tt arXiv:1210.6022 [hep-th]}}.
%%CITATION = ARXIV:1210.6022;%%.

\bibitem{PACecotti:1991me}
S.~Cecotti and C.~Vafa, ``{Topological antitopological fusion},''
\href{http://dx.doi.org/10.1016/0550-3213(91)90021-O}{{\em Nucl. Phys.} {\bf
  B367} (1991)  359--461}.
%%CITATION = NUPHA,B367,359;%%.

\bibitem{PACecotti:2013mba}
S.~Cecotti, D.~Gaiotto, and C.~Vafa, ``{$tt^*$ geometry in 3 and 4
  dimensions},'' \href{http://dx.doi.org/10.1007/JHEP05(2014)055}{{\em JHEP}
  {\bf 1405} (2014)  055},
\href{http://arxiv.org/abs/1312.1008}{{\tt arXiv:1312.1008 [hep-th]}}.
%%CITATION = ARXIV:1312.1008;%%.

\bibitem{ContributionMO}
D.~Morrison, ``Gromov-Witten invariants and localization,'' {\em Journal of
  Physics A} (2016)  , \href{http://arxiv.org/abs/1608.02956}{{\tt
  1608.02956}}.

\bibitem{PAYoshida:2014ssa}
Y.~Yoshida and K.~Sugiyama, ``{Localization of 3d $\mathcal{N}=2$
  Supersymmetric Theories on $S^1 \times D^2$},''
\href{http://arxiv.org/abs/1409.6713}{{\tt arXiv:1409.6713 [hep-th]}}.
%%CITATION = ARXIV:1409.6713;%%.

\bibitem{PANieri:2015dts}
F.~Nieri, ``{An elliptic Virasoro symmetry in 6d},''
\href{http://arxiv.org/abs/1511.00574}{{\tt arXiv:1511.00574 [hep-th]}}.
%%CITATION = ARXIV:1511.00574;%%.

\bibitem{ContributionQZ}
J.~Qiu and M.~Zabzine, ``Review of localization for 5D supersymmetric gauge
  theories,'' {\em Journal of Physics A} {\bf xx} (2016)  000,
  \href{http://arxiv.org/abs/1608.02966}{{\tt 1608.02966}}.

\bibitem{PALockhart:2012vp}
G.~Lockhart and C.~Vafa, ``{Superconformal Partition Functions and
  Non-perturbative Topological Strings},''
\href{http://arxiv.org/abs/1210.5909}{{\tt arXiv:1210.5909 [hep-th]}}.
%%CITATION = ARXIV:1210.5909;%%.

\bibitem{PAKim:2012qf}
H.-C. Kim, J.~Kim, and S.~Kim, ``{Instantons on the 5-sphere and M5-branes},''
\href{http://arxiv.org/abs/1211.0144}{{\tt arXiv:1211.0144 [hep-th]}}.
%%CITATION = ARXIV:1211.0144;%%.

\bibitem{PANekrasov:2002qd}
N.~A. Nekrasov, ``{Seiberg-Witten prepotential from instanton counting},'' {\em
  Adv.Theor.Math.Phys.} {\bf 7} (2004)  831--864,
\href{http://arxiv.org/abs/hep-th/0206161}{{\tt arXiv:hep-th/0206161
  [hep-th]}}.
%%CITATION = HEP-TH/0206161;%%.

\bibitem{PANekrasov:2003rj}
N.~Nekrasov and A.~Okounkov, ``{Seiberg-Witten theory and random partitions},''
\href{http://arxiv.org/abs/hep-th/0306238}{{\tt arXiv:hep-th/0306238
  [hep-th]}}.
%%CITATION = HEP-TH/0306238;%%.

\bibitem{PANieri:2013vba}
F.~Nieri, S.~Pasquetti, F.~Passerini, and A.~Torrielli, ``{5D partition
  functions, q-Virasoro systems and integrable spin-chains},''
  \href{http://dx.doi.org/10.1007/JHEP12(2014)040}{{\em JHEP} {\bf 12} (2014)
  040},
\href{http://arxiv.org/abs/1312.1294}{{\tt arXiv:1312.1294 [hep-th]}}.
%%CITATION = ARXIV:1312.1294;%%.

\bibitem{PAfv}
G.~Felder and A.~Varchenko, ``{The elliptic gamma function and
  $SL(3,\mathbb{Z}) \ltimes \mathbb{Z}^3$},''
  \href{http://dx.doi.org/10.1006/aima.2000.1951}{{\em Adv. Math.} {\bf 156}
  (2000)  44--76}.

\bibitem{PAQiu:2014oqa}
J.~Qiu, L.~Tizzano, J.~Winding, and M.~Zabzine, ``{Gluing Nekrasov partition
  functions},'' \href{http://dx.doi.org/10.1007/s00220-015-2351-7}{{\em Commun.
  Math. Phys.} {\bf 337} (2015) no.~2, 785--816},
\href{http://arxiv.org/abs/1403.2945}{{\tt arXiv:1403.2945 [hep-th]}}.
%%CITATION = ARXIV:1403.2945;%%.

\bibitem{PATizzano:2014roa}
L.~Tizzano and J.~Winding, ``{Multiple sine, multiple elliptic gamma functions
  and rational cones},''
\href{http://arxiv.org/abs/1502.05996}{{\tt arXiv:1502.05996 [math.CA]}}.
%%CITATION = ARXIV:1502.05996;%%.

\bibitem{PAKim:2012gu}
H.-C. Kim, S.-S. Kim, and K.~Lee, ``{5-dim Superconformal Index with Enhanced
  $E_n$ Global Symmetry},''
  \href{http://dx.doi.org/10.1007/JHEP10(2012)142}{{\em JHEP} {\bf 1210} (2012)
   142},
\href{http://arxiv.org/abs/1206.6781}{{\tt arXiv:1206.6781 [hep-th]}}.
%%CITATION = ARXIV:1206.6781;%%.

\bibitem{PATerashima:2012ra}
S.~Terashima, ``{On Supersymmetric Gauge Theories on $S^4\times S^1$},''
\href{http://arxiv.org/abs/1207.2163}{{\tt arXiv:1207.2163 [hep-th]}}.
%%CITATION = ARXIV:1207.2163;%%.

\bibitem{PAIqbal:2012xm}
A.~Iqbal and C.~Vafa, ``{BPS Degeneracies and Superconformal Index in Diverse
  Dimensions},''
\href{http://arxiv.org/abs/1210.3605}{{\tt arXiv:1210.3605 [hep-th]}}.
%%CITATION = ARXIV:1210.3605;%%.

\bibitem{PAGukov:2014gja}
S.~Gukov, ``{Surface Operators},''
  \href{http://dx.doi.org/10.1007/978-3-319-18769-3_8}{{\em Math. Phys. Stud.}
  {\bf 9783319187693} (2016)  223--259},
\href{http://arxiv.org/abs/1412.7127}{{\tt arXiv:1412.7127 [hep-th]}}.
%%CITATION = ARXIV:1412.7127;%%.

\bibitem{ContributionHO}
K.~Hosomichi, ``$\mathcal{N}=2$ SUSY gauge theories on $S^4$,'' {\em Journal of
  Physics A} {\bf xx} (2016)  000, \href{http://arxiv.org/abs/1608.02962}{{\tt
  1608.02962}}.

\bibitem{PAGaiotto:2012xa}
D.~Gaiotto, L.~Rastelli, and S.~S. Razamat, ``{Bootstrapping the superconformal
  index with surface defects},''
  \href{http://dx.doi.org/10.1007/JHEP01(2013)022}{{\em JHEP} {\bf 01} (2013)
  022},
\href{http://arxiv.org/abs/1207.3577}{{\tt arXiv:1207.3577 [hep-th]}}.
%%CITATION = ARXIV:1207.3577;%%.

\bibitem{PAHanany:2003hp}
A.~Hanany and D.~Tong, ``{Vortices, instantons and branes},''
  \href{http://dx.doi.org/10.1088/1126-6708/2003/07/037}{{\em JHEP} {\bf 07}
  (2003)  037},
\href{http://arxiv.org/abs/hep-th/0306150}{{\tt arXiv:hep-th/0306150
  [hep-th]}}.
%%CITATION = HEP-TH/0306150;%%.

\bibitem{PAHanany:1997vm}
A.~Hanany and K.~Hori, ``{Branes and N=2 theories in two-dimensions},''
  \href{http://dx.doi.org/10.1016/S0550-3213(97)00754-2}{{\em Nucl. Phys.} {\bf
  B513} (1998)  119--174},
\href{http://arxiv.org/abs/hep-th/9707192}{{\tt arXiv:hep-th/9707192
  [hep-th]}}.
%%CITATION = HEP-TH/9707192;%%.

\bibitem{PAGadde:2013ftv}
A.~Gadde and S.~Gukov, ``{2d Index and Surface operators},''
  \href{http://dx.doi.org/10.1007/JHEP03(2014)080}{{\em JHEP} {\bf 03} (2014)
  080},
\href{http://arxiv.org/abs/1305.0266}{{\tt arXiv:1305.0266 [hep-th]}}.
%%CITATION = ARXIV:1305.0266;%%.

\bibitem{PAMironov:2009qt}
A.~Mironov and A.~Morozov, ``{The Power of Nekrasov Functions},''
  \href{http://dx.doi.org/10.1016/j.physletb.2009.08.061}{{\em Phys. Lett.}
  {\bf B680} (2009)  188--194},
\href{http://arxiv.org/abs/0908.2190}{{\tt arXiv:0908.2190 [hep-th]}}.
%%CITATION = ARXIV:0908.2190;%%.

\bibitem{PAAganagic:2003db}
M.~Aganagic, A.~Klemm, M.~Marino, and C.~Vafa, ``{The Topological vertex},''
  \href{http://dx.doi.org/10.1007/s00220-004-1162-z}{{\em Commun. Math. Phys.}
  {\bf 254} (2005)  425--478},
\href{http://arxiv.org/abs/hep-th/0305132}{{\tt arXiv:hep-th/0305132
  [hep-th]}}.
%%CITATION = HEP-TH/0305132;%%.

\bibitem{PAIqbal:2007ii}
A.~Iqbal, C.~Kozcaz, and C.~Vafa, ``{The Refined topological vertex},''
  \href{http://dx.doi.org/10.1088/1126-6708/2009/10/069}{{\em JHEP} {\bf 10}
  (2009)  069},
\href{http://arxiv.org/abs/hep-th/0701156}{{\tt arXiv:hep-th/0701156
  [hep-th]}}.
%%CITATION = HEP-TH/0701156;%%.

\bibitem{PAKozcaz:2010af}
C.~Kozcaz, S.~Pasquetti, and N.~Wyllard, ``{A and B model approaches to surface
  operators and Toda theories},''
  \href{http://dx.doi.org/10.1007/JHEP08(2010)042}{{\em JHEP} {\bf 1008} (2010)
   042},
\href{http://arxiv.org/abs/1004.2025}{{\tt arXiv:1004.2025 [hep-th]}}.
%%CITATION = ARXIV:1004.2025;%%.

\bibitem{PABonelli:2011wx}
G.~Bonelli, A.~Tanzini, and J.~Zhao, ``{The Liouville side of the Vortex},''
  \href{http://dx.doi.org/10.1007/JHEP09(2011)096}{{\em JHEP} {\bf 1109} (2011)
   096},
\href{http://arxiv.org/abs/1107.2787}{{\tt arXiv:1107.2787 [hep-th]}}.
%%CITATION = ARXIV:1107.2787;%%.

\bibitem{PABonelli:2011fq}
G.~Bonelli, A.~Tanzini, and J.~Zhao, ``{Vertices, Vortices and Interacting
  Surface Operators},'' \href{http://dx.doi.org/10.1007/JHEP06(2012)178}{{\em
  JHEP} {\bf 06} (2012)  178},
\href{http://arxiv.org/abs/1102.0184}{{\tt arXiv:1102.0184 [hep-th]}}.
%%CITATION = ARXIV:1102.0184;%%.

\bibitem{PAAlday:2009fs}
L.~F. Alday, D.~Gaiotto, S.~Gukov, Y.~Tachikawa, and H.~Verlinde, ``{Loop and
  surface operators in ${\cal{N}}=2$ gauge theory and Liouville modular
  geometry},'' \href{http://dx.doi.org/10.1007/JHEP01(2010)113}{{\em JHEP} {\bf
  1001} (2010)  113},
\href{http://arxiv.org/abs/0909.0945}{{\tt arXiv:0909.0945 [hep-th]}}.
%%CITATION = ARXIV:0909.0945;%%.

\bibitem{PAGomis:2014eya}
J.~Gomis and B.~Le~Floch, ``{M2-brane surface operators and gauge theory
  dualities in Toda},''
\href{http://arxiv.org/abs/1407.1852}{{\tt arXiv:1407.1852 [hep-th]}}.
%%CITATION = ARXIV:1407.1852;%%.

\bibitem{PAAlday:2010vg}
L.~F. Alday and Y.~Tachikawa, ``{Affine SL(2) conformal blocks from 4d gauge
  theories},'' \href{http://dx.doi.org/10.1007/s11005-010-0422-4}{{\em Lett.
  Math. Phys.} {\bf 94} (2010)  87--114},
\href{http://arxiv.org/abs/1005.4469}{{\tt arXiv:1005.4469 [hep-th]}}.
%%CITATION = ARXIV:1005.4469;%%.

\bibitem{PAKozcaz:2010yp}
C.~Kozcaz, S.~Pasquetti, F.~Passerini, and N.~Wyllard, ``{Affine sl(N)
  conformal blocks from N=2 SU(N) gauge theories},''
  \href{http://dx.doi.org/10.1007/JHEP01(2011)045}{{\em JHEP} {\bf 01} (2011)
  045},
\href{http://arxiv.org/abs/1008.1412}{{\tt arXiv:1008.1412 [hep-th]}}.
%%CITATION = ARXIV:1008.1412;%%.

\bibitem{PAFrenkel:2015rda}
E.~Frenkel, S.~Gukov, and J.~Teschner, ``{Surface Operators and Separation of
  Variables},''
\href{http://arxiv.org/abs/1506.07508}{{\tt arXiv:1506.07508 [hep-th]}}.
%%CITATION = ARXIV:1506.07508;%%.

\bibitem{PAWyllard:2010rp}
N.~Wyllard, ``{W-algebras and surface operators in N=2 gauge theories},''
  \href{http://dx.doi.org/10.1088/1751-8113/44/15/155401}{{\em J. Phys.} {\bf
  A44} (2011)  155401},
\href{http://arxiv.org/abs/1011.0289}{{\tt arXiv:1011.0289 [hep-th]}}.
%%CITATION = ARXIV:1011.0289;%%.

\bibitem{PAWyllard:2010vi}
N.~Wyllard, ``{Instanton partition functions in N=2 SU(N) gauge theories with a
  general surface operator, and their W-algebra duals},''
  \href{http://dx.doi.org/10.1007/JHEP02(2011)114}{{\em JHEP} {\bf 02} (2011)
  114},
\href{http://arxiv.org/abs/1012.1355}{{\tt arXiv:1012.1355 [hep-th]}}.
%%CITATION = ARXIV:1012.1355;%%.

\bibitem{PATachikawa:2014dja}
Y.~Tachikawa, ``{A review on instanton counting and W-algebras},''
  \href{http://dx.doi.org/10.1007/978-3-319-18769-3_4}{{\em Math. Phys. Stud.}
  {\bf 9783319187693} (2016)  79--120},
\href{http://arxiv.org/abs/1412.7121}{{\tt arXiv:1412.7121 [hep-th]}}.
%%CITATION = ARXIV:1412.7121;%%.

\bibitem{PAGaiotto:2014ina}
D.~Gaiotto and H.-C. Kim, ``{Surface defects and instanton partition
  functions},''
\href{http://arxiv.org/abs/1412.2781}{{\tt arXiv:1412.2781 [hep-th]}}.
%%CITATION = ARXIV:1412.2781;%%.

\bibitem{PAShiraishi:1995rp}
J.~Shiraishi, H.~Kubo, H.~Awata, and S.~Odake, ``{A Quantum deformation of the
  Virasoro algebra and the Macdonald symmetric functions},''
  \href{http://dx.doi.org/10.1007/BF00398297}{{\em Lett.Math.Phys.} {\bf 38}
  (1996)  33--51},
\href{http://arxiv.org/abs/q-alg/9507034}{{\tt arXiv:q-alg/9507034 [q-alg]}}.
%%CITATION = Q-ALG/9507034;%%.

\bibitem{PAAwata:1995zk}
H.~Awata, H.~Kubo, S.~Odake, and J.~Shiraishi, ``{Quantum W(N) algebras and
  Macdonald polynomials},'' \href{http://dx.doi.org/10.1007/BF02102595}{{\em
  Commun.Math.Phys.} {\bf 179} (1996)  401--416},
\href{http://arxiv.org/abs/q-alg/9508011}{{\tt arXiv:q-alg/9508011 [q-alg]}}.
%%CITATION = Q-ALG/9508011;%%.

\bibitem{PALukyanov:1994re}
S.~L. Lukyanov and Y.~Pugai, ``{Bosonization of ZF algebras: Direction toward
  deformed Virasoro algebra},'' {\em J.Exp.Theor.Phys.} {\bf 82} (1996)
  1021--1045,
\href{http://arxiv.org/abs/hep-th/9412128}{{\tt arXiv:hep-th/9412128
  [hep-th]}}.
%%CITATION = HEP-TH/9412128;%%.

\bibitem{PALukyanov:1996qs}
S.~L. Lukyanov and Y.~Pugai, ``{Multipoint local height probabilities in the
  integrable RSOS model},''
  \href{http://dx.doi.org/10.1016/0550-3213(96)00221-0}{{\em Nucl.Phys.} {\bf
  B473} (1996)  631--658},
\href{http://arxiv.org/abs/hep-th/9602074}{{\tt arXiv:hep-th/9602074
  [hep-th]}}.
%%CITATION = HEP-TH/9602074;%%.

\bibitem{PALukyanov:1995gs}
S.~L. Lukyanov, ``{A Note on the deformed Virasoro algebra},''
  \href{http://dx.doi.org/10.1016/0370-2693(95)01410-1}{{\em Phys.Lett.} {\bf
  B367} (1996)  121--125},
\href{http://arxiv.org/abs/hep-th/9509037}{{\tt arXiv:hep-th/9509037
  [hep-th]}}.
%%CITATION = HEP-TH/9509037;%%.

\bibitem{PAfr}
E.~Frenkel and N.~Reshetikhin, ``{Quantum Affine Algebras and Deformations of
  the Virasoro and W-algebras},''
  \href{http://dx.doi.org/10.10072FBF02104917}{{\em Commun.Math.Phys.} {\bf
  178} (1996)  237--264}, \href{http://arxiv.org/abs/q-alg/9505025}{{\tt
  arXiv:q-alg/9505025 [q-alg]}}.

\bibitem{PAFeigin:1995sf}
B.~Feigin and E.~Frenkel, ``{Quantum W algebras and elliptic algebras},''
  \href{http://dx.doi.org/10.1007/BF02108819}{{\em Commun.Math.Phys.} {\bf 178}
  (1996)  653--678},
\href{http://arxiv.org/abs/q-alg/9508009}{{\tt arXiv:q-alg/9508009 [q-alg]}}.
%%CITATION = Q-ALG/9508009;%%.

\bibitem{PAOdake:1999un}
S.~Odake, ``{Beyond CFT: Deformed Virasoro and elliptic algebras},'' in {\em
  {Theoretical physics at the end of the twentieth century. Proceedings, Summer
  School, Banff, Canada, June 27-July 10, 1999}}, pp.~307--449.
\newblock 1999.
\newblock \href{http://arxiv.org/abs/hep-th/9910226}{{\tt arXiv:hep-th/9910226
  [hep-th]}}.
\newblock
\url{http://alice.cern.ch/format/showfull?sysnb=0332345}.
\newblock
%%CITATION = HEP-TH/9910226;%%.

\bibitem{PAAwata:1996fq}
H.~Awata, H.~Kubo, S.~Odake, and J.~Shiraishi, ``{Virasoro type symmetries in
  solvable models},''
\href{http://arxiv.org/abs/hep-th/9612233}{{\tt arXiv:hep-th/9612233
  [hep-th]}}.
%%CITATION = HEP-TH/9612233;%%.

\bibitem{PABelavin:1984vu}
A.~Belavin, A.~M. Polyakov, and A.~Zamolodchikov, ``{Infinite Conformal
  Symmetry in Two-Dimensional Quantum Field Theory},''
\href{http://dx.doi.org/10.1016/0550-3213(84)90052-X}{{\em Nucl.Phys.} {\bf
  B241} (1984)  333--380}.
%%CITATION = NUPHA,B241,333;%%.

\bibitem{PADotsenko:1984nm}
V.~Dotsenko and V.~Fateev, ``{Conformal Algebra and Multipoint Correlation
  Functions in Two-Dimensional Statistical Models},''
\href{http://dx.doi.org/10.1016/0550-3213(84)90269-4}{{\em Nucl.Phys.} {\bf
  B240} (1984)  312}.
%%CITATION = NUPHA,B240,312;%%.

\bibitem{PAFelderBRST}
G.~Felder, ``{BRST approach to minimal model},'' {\em J.Exp.Theor.Phys.} {\bf
  317} (1989)  215--236.

\bibitem{PAmina1}
M.~Aganagic, N.~Haouzi, C.~Kozcaz, and S.~Shakirov, ``{Gauge/Liouville
  Triality},''
\href{http://arxiv.org/abs/1309.1687}{{\tt arXiv:1309.1687 [hep-th]}}.
%%CITATION = ARXIV:1309.1687;%%.

\bibitem{PAdv}
R.~Dijkgraaf and C.~Vafa, ``{Toda Theories, Matrix Models, Topological Strings,
  and N=2 Gauge Systems},''
\href{http://arxiv.org/abs/0909.2453}{{\tt arXiv:0909.2453 [hep-th]}}.
%%CITATION = ARXIV:0909.2453;%%.

\bibitem{PAItoyama:2009sc}
H.~Itoyama, K.~Maruyoshi, and T.~Oota, ``{The Quiver Matrix Model and 2d-4d
  Conformal Connection},'' \href{http://dx.doi.org/10.1143/PTP.123.957}{{\em
  Prog.Theor.Phys.} {\bf 123} (2010)  957--987},
\href{http://arxiv.org/abs/0911.4244}{{\tt arXiv:0911.4244 [hep-th]}}.
%%CITATION = ARXIV:0911.4244;%%.

\bibitem{PAEguchi:2010rf}
T.~Eguchi and K.~Maruyoshi, ``{Seiberg-Witten theory, matrix model and AGT
  relation},'' \href{http://dx.doi.org/10.1007/JHEP07(2010)081}{{\em JHEP} {\bf
  1007} (2010)  081},
\href{http://arxiv.org/abs/1006.0828}{{\tt arXiv:1006.0828 [hep-th]}}.
%%CITATION = ARXIV:1006.0828;%%.

\bibitem{PASchiappa:2009cc}
R.~Schiappa and N.~Wyllard, ``{An $A(r)$ threesome: Matrix models, 2d CFTs and
  4d ${\cal{N}}=2$ gauge theories},''
  \href{http://dx.doi.org/10.1063/1.3449328}{{\em J.Math.Phys.} {\bf 51} (2010)
   082304},
\href{http://arxiv.org/abs/0911.5337}{{\tt arXiv:0911.5337 [hep-th]}}.
%%CITATION = ARXIV:0911.5337;%%.

\bibitem{PAmms1}
A.~Mironov, A.~Morozov, and S.~Shakirov, ``{Matrix Model Conjecture for Exact
  BS Periods and Nekrasov Functions},''
  \href{http://dx.doi.org/10.1007/JHEP02(2010)030}{{\em JHEP} {\bf 1002} (2010)
   030},
\href{http://arxiv.org/abs/0911.5721}{{\tt arXiv:0911.5721 [hep-th]}}.
%%CITATION = ARXIV:0911.5721;%%.

\bibitem{PAmms2}
A.~Mironov, A.~Morozov, and S.~Shakirov, ``{Conformal blocks as Dotsenko-Fateev
  Integral Discriminants},''
  \href{http://dx.doi.org/10.1142/S0217751X10049141}{{\em Int.J.Mod.Phys.} {\bf
  A25} (2010)  3173--3207},
\href{http://arxiv.org/abs/1001.0563}{{\tt arXiv:1001.0563 [hep-th]}}.
%%CITATION = ARXIV:1001.0563;%%.

\bibitem{PAmina2}
M.~Aganagic, N.~Haouzi, and S.~Shakirov, ``{$A_n$-Triality},''
\href{http://arxiv.org/abs/1403.3657}{{\tt arXiv:1403.3657 [hep-th]}}.
%%CITATION = ARXIV:1403.3657;%%.

\bibitem{PAMironov:2011dk}
A.~Mironov, A.~Morozov, S.~Shakirov, and A.~Smirnov, ``{Proving AGT conjecture
  as HS duality: extension to five dimensions},''
  \href{http://dx.doi.org/10.1016/j.nuclphysb.2011.09.021}{{\em Nucl.Phys.}
  {\bf B855} (2012)  128--151},
\href{http://arxiv.org/abs/1105.0948}{{\tt arXiv:1105.0948 [hep-th]}}.
%%CITATION = ARXIV:1105.0948;%%.

\bibitem{PAAwata:2009ur}
H.~Awata and Y.~Yamada, ``{Five-dimensional AGT Conjecture and the Deformed
  Virasoro Algebra},'' \href{http://dx.doi.org/10.1007/JHEP01(2010)125}{{\em
  JHEP} {\bf 1001} (2010)  125},
\href{http://arxiv.org/abs/0910.4431}{{\tt arXiv:0910.4431 [hep-th]}}.
%%CITATION = ARXIV:0910.4431;%%.

\bibitem{PAYanagida:2010vz}
S.~Yanagida, ``{Five-dimensional SU(2) AGT conjecture and recursive formula of
  deformed Gaiotto state},'' \href{http://dx.doi.org/10.1063/1.3505826}{{\em
  J.Math.Phys.} {\bf 51} (2010)  123506},
\href{http://arxiv.org/abs/1005.0216}{{\tt arXiv:1005.0216 [math.QA]}}.
%%CITATION = ARXIV:1005.0216;%%.

\bibitem{PAGaiotto:2009ma}
D.~Gaiotto, ``{Asymptotically free $\mathcal{N} = 2$ theories and irregular
  conformal blocks},''
  \href{http://dx.doi.org/10.1088/1742-6596/462/1/012014}{{\em
  J.Phys.Conf.Ser.} {\bf 462} (2013) no.~1, 012014},
\href{http://arxiv.org/abs/0908.0307}{{\tt arXiv:0908.0307 [hep-th]}}.
%%CITATION = ARXIV:0908.0307;%%.

\bibitem{PATaki:2014fva}
M.~Taki, ``{On AGT-W Conjecture and q-Deformed W-Algebra},''
\href{http://arxiv.org/abs/1403.7016}{{\tt arXiv:1403.7016 [hep-th]}}.
%%CITATION = ARXIV:1403.7016;%%.

\bibitem{PAKatz:1997eq}
S.~Katz, P.~Mayr, and C.~Vafa, ``{Mirror symmetry and exact solution of 4-D N=2
  gauge theories: 1.},'' {\em Adv.Theor.Math.Phys.} {\bf 1} (1998)  53--114,
\href{http://arxiv.org/abs/hep-th/9706110}{{\tt arXiv:hep-th/9706110
  [hep-th]}}.
%%CITATION = HEP-TH/9706110;%%.

\bibitem{PAAharony:1997bh}
O.~Aharony, A.~Hanany, and B.~Kol, ``{Webs of (p,q) five-branes,
  five-dimensional field theories and grid diagrams},''
  \href{http://dx.doi.org/10.1088/1126-6708/1998/01/002}{{\em JHEP} {\bf 9801}
  (1998)  002},
\href{http://arxiv.org/abs/hep-th/9710116}{{\tt arXiv:hep-th/9710116
  [hep-th]}}.
%%CITATION = HEP-TH/9710116;%%.

\bibitem{PABao:2013pwa}
L.~Bao, V.~Mitev, E.~Pomoni, M.~Taki, and F.~Yagi, ``{Non-Lagrangian Theories
  from Brane Junctions},''
\href{http://arxiv.org/abs/1310.3841}{{\tt arXiv:1310.3841 [hep-th]}}.
%%CITATION = ARXIV:1310.3841;%%.

\bibitem{PAminarev}
M.~Aganagic and S.~Shakirov, ``{Gauge/Vortex duality and AGT},''
  \href{http://dx.doi.org/10.1007/978-3-319-18769-3_13}{{\em Math. Phys. Stud.}
  {\bf 9783319187693} (2016)  419--448},
\href{http://arxiv.org/abs/1412.7132}{{\tt arXiv:1412.7132 [hep-th]}}.
%%CITATION = ARXIV:1412.7132;%%.

\bibitem{PAHanany:2004ea}
A.~Hanany and D.~Tong, ``{Vortex strings and four-dimensional gauge
  dynamics},'' \href{http://dx.doi.org/10.1088/1126-6708/2004/04/066}{{\em
  JHEP} {\bf 04} (2004)  066},
\href{http://arxiv.org/abs/hep-th/0403158}{{\tt arXiv:hep-th/0403158
  [hep-th]}}.
%%CITATION = HEP-TH/0403158;%%.

\bibitem{PANedelin:2016gwu}
A.~Nedelin, F.~Nieri, and M.~Zabzine, ``{$q$-Virasoro modular double and 3d
  partition functions},''
\href{http://arxiv.org/abs/1605.07029}{{\tt arXiv:1605.07029 [hep-th]}}.
%%CITATION = ARXIV:1605.07029;%%.

\bibitem{PAAlba:2010qc}
V.~A. Alba, V.~A. Fateev, A.~V. Litvinov, and G.~M. Tarnopolskiy, ``{On
  combinatorial expansion of the conformal blocks arising from AGT
  conjecture},'' \href{http://dx.doi.org/10.1007/s11005-011-0503-z}{{\em
  Lett.Math.Phys.} {\bf 98} (2011)  33--64},
\href{http://arxiv.org/abs/1012.1312}{{\tt arXiv:1012.1312 [hep-th]}}.
%%CITATION = ARXIV:1012.1312;%%.

\bibitem{PABelavin:2011sw}
A.~A. Belavin, M.~A. Bershtein, B.~L. Feigin, A.~V. Litvinov, and G.~M.
  Tarnopolsky, ``{Instanton moduli spaces and bases in coset conformal field
  theory},'' \href{http://dx.doi.org/10.1007/s00220-012-1603-z}{{\em Commun.
  Math. Phys.} {\bf 319} (2013)  269--301},
\href{http://arxiv.org/abs/1111.2803}{{\tt arXiv:1111.2803 [hep-th]}}.
%%CITATION = ARXIV:1111.2803;%%.

\bibitem{PAMO}
D.~Maulik and A.~Okounkov, ``{Quantum Groups and Quantum Cohomology},''
  \href{http://arxiv.org/abs/1211.1287}{{\tt arXiv:1211.1287 [math.AG]}}.

\bibitem{PASV}
O.~Schiffmann and E.~Vasserot, ``{Cherednik algebras, W algebras and the
  equivariant cohomology of the moduli space of instantons on $A^2$},''
  \href{http://arxiv.org/abs/1202.2756}{{\tt arXiv:1202.2756 [math.QA]}}.

\bibitem{PAMironov:2010pi}
A.~Mironov, A.~Morozov, and S.~Shakirov, ``{A direct proof of AGT conjecture at
  beta = 1},'' \href{http://dx.doi.org/10.1007/JHEP02(2011)067}{{\em JHEP} {\bf
  1102} (2011)  067},
\href{http://arxiv.org/abs/1012.3137}{{\tt arXiv:1012.3137 [hep-th]}}.
%%CITATION = ARXIV:1012.3137;%%.

\bibitem{PAMorozov:2013rma}
A.~Morozov and A.~Smirnov, ``{Towards the Proof of AGT Relations with the Help
  of the Generalized Jack Polynomials},''
  \href{http://dx.doi.org/10.1007/s11005-014-0681-6}{{\em Lett.Math.Phys.} {\bf
  104} (2014) no.~5, 585--612},
\href{http://arxiv.org/abs/1307.2576}{{\tt arXiv:1307.2576 [hep-th]}}.
%%CITATION = ARXIV:1307.2576;%%.

\bibitem{PAZenkevich:2014lca}
Y.~Zenkevich, ``{Generalized Macdonald polynomials, spectral duality for
  conformal blocks and AGT correspondence in five dimensions},''
\href{http://arxiv.org/abs/1412.8592}{{\tt arXiv:1412.8592 [hep-th]}}.
%%CITATION = ARXIV:1412.8592;%%.

\bibitem{PAMorozov:2015xya}
A.~Morozov and Y.~Zenkevich, ``{Decomposing Nekrasov Decomposition},''
\href{http://arxiv.org/abs/1510.01896}{{\tt arXiv:1510.01896 [hep-th]}}.
%%CITATION = ARXIV:1510.01896;%%.

\bibitem{PAAwata:2016riz}
H.~Awata, H.~Kanno, T.~Matsumoto, A.~Mironov, A.~Morozov, A.~Morozov,
  Y.~Ohkubo, and Y.~Zenkevich, ``{Explicit examples of DIM constraints for
  network matrix models},''
\href{http://arxiv.org/abs/1604.08366}{{\tt arXiv:1604.08366 [hep-th]}}.
%%CITATION = ARXIV:1604.08366;%%.

\bibitem{PAkane2}
J.~Kaneko, ``{Constant term identities of Forrester-Zeilberger-Cooper},''
  \href{http://dx.doi.org/10.1016/S0012-365X(96)00098-2}{{\em Discrete
  Mathematics} {\bf 173} (1997)  79--90}.

\bibitem{PAAwata:2010yy}
H.~Awata and Y.~Yamada, ``{Five-dimensional AGT Relation and the Deformed
  beta-ensemble},'' \href{http://dx.doi.org/10.1143/PTP.124.227}{{\em
  Prog.Theor.Phys.} {\bf 124} (2010)  227--262},
\href{http://arxiv.org/abs/1004.5122}{{\tt arXiv:1004.5122 [hep-th]}}.
%%CITATION = ARXIV:1004.5122;%%.

\bibitem{PANieri:2013yra}
F.~Nieri, S.~Pasquetti, and F.~Passerini, ``{3d \& 5d gauge theory partition
  functions as q-deformed CFT correlators},''
\href{http://arxiv.org/abs/1303.2626}{{\tt arXiv:1303.2626 [hep-th]}}.
%%CITATION = ARXIV:1303.2626;%%.

\bibitem{PABernard:1989jq}
D.~Bernard and A.~LeClair, ``{$q$ deformation of $SU(1,1)$ conformal Ward
  identities and $q$-strings},''
  \href{http://dx.doi.org/10.1016/0370-2693(89)90953-2}{{\em Phys.Lett.} {\bf
  B227} (1989)  417--423}.

\bibitem{PATeschner:1995yf}
J.~Teschner, ``{On the Liouville three point function},''
  \href{http://dx.doi.org/10.1016/0370-2693(95)01200-A}{{\em Phys.Lett.} {\bf
  B363} (1995)  65--70},
\href{http://arxiv.org/abs/hep-th/9507109}{{\tt arXiv:hep-th/9507109
  [hep-th]}}.
%%CITATION = HEP-TH/9507109;%%.

\bibitem{PADorn:1994xn}
H.~Dorn and H.~Otto, ``{Two and three point functions in Liouville theory},''
  \href{http://dx.doi.org/10.1016/0550-3213(94)00352-1}{{\em Nucl.Phys.} {\bf
  B429} (1994)  375--388},
\href{http://arxiv.org/abs/hep-th/9403141}{{\tt arXiv:hep-th/9403141
  [hep-th]}}.
%%CITATION = HEP-TH/9403141;%%.

\bibitem{PAZamolodchikov:1995aa}
A.~B. Zamolodchikov and A.~B. Zamolodchikov, ``{Structure constants and
  conformal bootstrap in Liouville field theory},''
  \href{http://dx.doi.org/10.1016/0550-3213(96)00351-3}{{\em Nucl.Phys.} {\bf
  B477} (1996)  577--605},
\href{http://arxiv.org/abs/hep-th/9506136}{{\tt arXiv:hep-th/9506136
  [hep-th]}}.
%%CITATION = HEP-TH/9506136;%%.

\bibitem{PABenini:2009gi}
F.~Benini, S.~Benvenuti, and Y.~Tachikawa, ``{Webs of five-branes and N=2
  superconformal field theories},''
  \href{http://dx.doi.org/10.1088/1126-6708/2009/09/052}{{\em JHEP} {\bf 09}
  (2009)  052},
\href{http://arxiv.org/abs/0906.0359}{{\tt arXiv:0906.0359 [hep-th]}}.
%%CITATION = ARXIV:0906.0359;%%.

\bibitem{PAMitev:2014isa}
V.~Mitev and E.~Pomoni, ``{Toda 3-Point Functions From Topological Strings},''
  \href{http://dx.doi.org/10.1007/JHEP06(2015)049}{{\em JHEP} {\bf 06} (2015)
  049},
\href{http://arxiv.org/abs/1409.6313}{{\tt arXiv:1409.6313 [hep-th]}}.
%%CITATION = ARXIV:1409.6313;%%.

\bibitem{PAIsachenkov:2014eya}
M.~Isachenkov, V.~Mitev, and E.~Pomoni, ``{Toda 3-Point Functions From
  Topological Strings II},''
\href{http://arxiv.org/abs/1412.3395}{{\tt arXiv:1412.3395 [hep-th]}}.
%%CITATION = ARXIV:1412.3395;%%.

\bibitem{PADrukker:2009id}
N.~Drukker, J.~Gomis, T.~Okuda, and J.~Teschner, ``{Gauge Theory Loop Operators
  and Liouville Theory},''
  \href{http://dx.doi.org/10.1007/JHEP02(2010)057}{{\em JHEP} {\bf 02} (2010)
  057},
\href{http://arxiv.org/abs/0909.1105}{{\tt arXiv:0909.1105 [hep-th]}}.
%%CITATION = ARXIV:0909.1105;%%.

\bibitem{PAnaru}
A.~Narukawa, ``{The modular properties and the integral representations of the
  multiple elliptic gamma functions},''
  \href{http://arxiv.org/abs/math/0306164}{{\tt arXiv:math/0306164 [Math.QA]}}.

\bibitem{PAgrahman}
G.~Gasper and M.~Rahman, ``{Basic Hypergeometric Series},'' {\em 2nd edition,
  Cambridge University Press} (2004)  .

\end{thebibliography}
